\documentclass[prd,eqsecnum,twocolumn,nofootinbib,superscriptaddress]{revtex4-2}

\usepackage{amsfonts}
\usepackage{ulem} 
\usepackage{mathtools}
\usepackage{amssymb}
\usepackage{amsthm}
\usepackage{bm}
\usepackage{dcolumn}
\usepackage{epsfig}
\usepackage{graphicx}
\usepackage{graphics}
\usepackage[latin1]{inputenc}
\usepackage{latexsym}
\usepackage{rotating}
\usepackage{xcolor}
\usepackage{hyperref}
\usepackage{calligra} \DeclareMathAlphabet{\mathcalligra}{T1}{calligra}{m}{n} \DeclareFontShape{T1}{calligra}{m}{n}{<->s*[2.2]callig15}{}

\renewcommand{\sp}{\epsilon} 
\newcommand{\mr}{\varepsilon} 

\newcommand{\avg}[1]{\left\langle #1 \right\rangle}

\newcommand{\osc}[1]{\Breve{#1}}
\newcommand{\nit}[1]{\Tilde{#1}}
\newcommand{\PD}[2]{\frac{\partial #1}{\partial #2}}
\newcommand{\HOT}[1]{\mathcal{O}( \mr ^{ #1 } )} 

\begin{document}
\title{Extreme mass-ratio inspiral and waveforms for a spinning body into a Kerr black hole\\
via osculating geodesics and near-identity transformations}
\author{Lisa V. Drummond}
\affiliation{Department of Physics and MIT Kavli Institute, MIT, Cambridge, MA 02139 USA}
\author{Philip Lynch}
\affiliation{Max Planck Institute for Gravitational Physics (Albert Einstein Institute), Am M\"uhlenberg 1, 14476 Potsdam, Germany}
\affiliation{School of Mathematics and Statistics , University College Dublin, Belfield, Dublin 4, Ireland}
\author{Alexandra G. Hanselman}
\affiliation{Department of Physics, University of Chicago, Chicago, IL 60637 USA}
\author{Devin R. Becker}
\affiliation{Department of Physics and MIT Kavli Institute, MIT, Cambridge, MA 02139 USA}
\author{Scott A. Hughes}
\affiliation{Department of Physics and MIT Kavli Institute, MIT, Cambridge, MA 02139 USA}

\begin{abstract}
Understanding the orbits of spinning bodies in curved spacetime is important for modeling binary black hole systems with small mass ratios.  At zeroth order in mass ratio and ignoring its size, the smaller body moves on a geodesic of the larger body's spacetime.  Post-geodesic effects, driving motion away from geodesics, are needed to model the system accurately.  One very important post-geodesic effect is the gravitational self-force, which describes the small body's interaction with its own contribution to a binary's spacetime.  The self-force includes the backreaction of gravitational-wave emission driving inspiral.  Another post-geodesic effect, the {\it spin-curvature force}, is due to the smaller body's spin coupling to spacetime curvature.  In this paper, we combine the leading orbit-averaged backreaction of point-particle gravitational-wave emission with the spin-curvature force to construct the worldline and associated gravitational waveform for a spinning body spiraling into a Kerr black hole.  We use an osculating geodesic integrator, which treats the worldline as evolution through a sequence of geodesic orbits, as well as near-identity (averaging) transformations, which eliminate dependence on orbital phases, allowing for very fast computation of generic spinning body inspirals.  The resulting inspirals and waveforms include all critical dynamical effects which govern such systems (orbit and precession frequencies, inspiral, strong-field gravitational-wave amplitudes), and as such form an effective first model for the inspiral of spinning bodies into Kerr black holes.  We emphasize that our present calculation is not self consistent, since we neglect effects which enter at the same order as effects we include.  However, our analysis demonstrates that the impact of spin-curvature forces can be incorporated into EMRI waveform tools with relative ease, making it possible to augment these models with this important aspect of source physics.  The calculation is sufficiently modular that it should not be difficult to include neglected post-geodesic effects as efficient tools for computing them become available.
\end{abstract}
\maketitle

\section{Introduction and Motivation}
\label{sec:intro}

Binary systems with very small mass ratios that inspiral due to gravitational wave (GW) backreaction are known as extreme mass-ratio inspirals (EMRIs).  Such systems are formally interesting and important, as they represent a limit of the binary problem in general relativity that can be solved precisely, providing important input for modeling the relativistic two-body problem.  They are also expected to be important sources of low-frequency GWs.  Binaries consisting of stellar-mass compact objects (mass $\mu \sim 1-100\,M_\odot$) in strong-field orbits of massive black holes (mass $M \sim 10^6\,M_\odot$) produce GWs in the sensitive band of the planned Laser Interferometer Space Antenna (LISA) \cite{eLISA2013,Barausse2020}.  The detection of GWs from EMRI sources will enable precise measurements of properties of massive black holes, and robustly probe the Kerr nature of the spacetime \cite{Collins2004, Glampedakis2006, Barack2007, Amaro-Seoane2007, Vigeland2010, Gair2013,Babak2017}.  This will be achieved by matching the phase of theoretical waveforms with observed GW data over thousands to millions of orbits.  Making such measurements will require precise, long-duration waveform models.

At ``zeroth'' order in mass ratio $\varepsilon \equiv \mu/M$ and size of the smaller body, the secondary's motion is simply a geodesic of the larger black hole, a limit that is very well understood.  Important corrections to geodesic motion arise from the smaller body's mass and from its finite size.  Finite mass-ratio effects are known as {\it self forces} \cite{Barack:2009ux,Barack:2018yvs,Pound2021}.  Fundamentally, self forces reflect the fact that the spacetime in which the smaller body moves is not just that of the larger body: the smaller body affects the binary's spacetime, which in turn changes that body's motion.  A well-developed program to compute the self force has been developed over the past several decades \cite{Warburton2012,Osburn2016,Warburton2017,vandeMeent2018,vandeMeent2018_2,Pound2020,Lynch2021,Warburton2021,Upton2021,Mathews2022,Albertini2022_1,Albertini2022_2,vandemeent2023,Blanco2023_1,Wardell2023,Leather2023,Spiers2023}.  For our purposes, it is important to recognize that the self force leads to dissipative corrections (which on average take away energy and other ``conserved'' quantities from the orbit, driving inspiral), and to conservative corrections (which on average leave conserved quantities unchanged, but modify orbit properties such as frequencies versus the geodesic with the same orbital geometry).  Some contributions to the self force are oscillatory, averaging to zero over a single orbit; others accumulate secularly over many orbits.  The leading dissipative self force, for example, accumulates over many orbits.

Finite size effects reflect the fact that real bodies are not zero-size points.  Aspects of a body's finite extent couple to spacetime curvature, and this coupling generates forces relative to a zero-size body's free-fall trajectory.  If the smaller body is itself a black hole, the leading and most important finite size effect is from that body's spin angular momentum \cite{Mathisson2010,Mathisson2010G_2,Papapetrou1951,Dixon1970}.  As it moves through spacetime, a small body's spin precesses, leading to a time-varying spin-curvature force.  Such forces are entirely conservative, changing properties of orbits such as their frequencies.  They have oscillatory aspects, which at leading order average away over an orbit, and secularly accumulating contributions.

The simplest model describing EMRI systems is known as the adiabatic inspiral and waveform.  Adiabatic models are computed by taking the smaller body to follow a geodesic of the background spacetime, and allowing that geodesic to evolve using the leading orbit-averaged dissipative backreaction \cite{Mino2003, Isoyama2019}.  As a matter of principle, it is now possible to compute adiabatic waveforms for essentially any astrophysical extreme-mass ratio system \cite{Drasco2006, Fujita2020, Hughes2021}.  As a matter of practice, fast and efficient adiabatic waveforms can only be computed for a subset of the parameter space \cite{Chua2021, Katz2021}, but work is in progress to expand this space.

Post-adiabatic effects include the conservative self force \cite{Barack:2009ey,Barack:2010ny,Barack:2011ed,vandeMeent:2016hel, Vines:2015efa, Fujita:2016igj,Blanco2023_1} (whose leading orbit-averaged effect is to change orbit frequencies compared to the geodesic), oscillating contributions to the dissipative self force (whose integrated impact on the inspiral is expected to be comparable to the orbit-averaged conservative self force \cite{Hinderer2008}), resonances \cite{Flanagan:2010cd,vandeMeent:2013sza, Berry:2016bit, Gupta:2022fbe} (moments during inspiral when two of the three fundamental orbital frequencies pass through a low-order integer ratio), and the spin-curvature force \cite{Mathisson2010,Mathisson2010G_2,Papapetrou1951,Dixon1970}.  The impact of many of these effects can be computed offline and included in waveform generation in a modular way.  This makes it not too difficult to augment adiabatic waveform generators in order to make waveforms which include important post-adiabatic effects.

The goal of the work that we present here is to show how one can augment adiabatic waveforms to include one particular post-adiabatic effect, the spin-curvature force.  We emphasize strongly that our analysis does {\it not} develop a self-consistent waveform model: we explicitly leave out effects which enter at the same order as the spin-curvature force, but which must be included to have a complete accounting of post-adiabatic effects at this order.  Our goal instead is to show how one can combine data and methods that currently exist in order to make inspirals of spinning bodies into Kerr black holes, and to make the waveforms corresponding to such inspirals.

The particular model we develop in this paper treats inspiral as a sequence of geodesic orbits, evolving from geodesic to geodesic under the combined influence of the spin-curvature force and the orbit-averaged self force.  This allows us to develop an EMRI model that incorporates the most important qualitative dynamics (four distinct orbit and precession frequencies, as well as strong-field backreaction), and to make a waveform that includes these effects.  Other approaches to developing such inspirals would require input that, at present, is not yet ready to be used.  For example, one might imagine treating the inspiral worldline as a sequence of spinning-body orbits (following the prescription laid out in Refs.\ \cite{Drummond2022_1, Drummond2022_2}), then evolving through the sequence by computing the orbit-averaged backreaction at each orbit.  Although we have a good prescription describing such orbits, we do not yet have large data sets which describe backreaction and wave amplitudes from these orbits (although the first calculations describing such data have been performed \cite{Skoupy2023}).  Indeed, it is not yet fully understood how to compute orbit-averaged backreaction on such orbits (see concluding discussion in Ref.\ \cite{Skoupy2023}).

The model we construct and present here is arguably the best that can be done for making spinning body inspiral with tools and data that exist right now.  We propose it as a first tool that can augment existing methods for making adiabatic inspirals and waveforms.  When applied to fast EMRI waveform methods (presently being extended to cover the Kerr parameter space), these waveforms will be useful for science studies assessing the importance of secondary spin for generic spinning-body inspiral.  These waveforms will also serve as a benchmark against which later models can be compared as fast and effective methods for incorporating other post-adiabatic effects become broadly available.

\section{Organization, conventions, and notation of this paper}
\label{sec:org}

We here provide an outline of the paper's organization, as well as a summary of the conventions and notation we use througout.  It is worth emphasizing that much of our analysis is based on bringing together techniques that have been presented at length elsewhere.  As such, several sections of this paper present just a high-level synopsis of these methods.  Several appendices provide detail needed to flesh out the calculations, and summarize material that is presented at length in the references which develop these methods.

Because our analysis is built on bound orbits around Kerr black holes, we briefly review the properties of these orbits in Sec.\ \ref{sec:orbits}.  We begin with the geodesic orbits of non-spinning bodies and their parameterization in \ref{sec:geodesics}, and summarize the properties of spinning body orbits in \ref{sec:spinningorbits}.  In Sec.\ \ref{sec:whichorbits}, we discuss why we choose to anchor our analysis to the properties of geodesic orbits, rather than using spinning-body orbits as our main tool.  We discuss at some length the rationale behind this choice, and why it will be useful as a complementary approach when future data allow us to use spinning-body orbits for broader studies than is possible right now.  Additional details regarding geodesics are given in Appendix \ref{app:geodesicsinkerr}, and regarding spinning-body orbits in Appendix \ref{app:spinningbodymotion}.

In Sec.\ \ref{sec:forcedgeodesics}, we briefly describe the osculating geodesic (abbreviated ``OG'') framework which underlies our inspiral analysis, describing how to map a worldline to a set of geodesics with evolving elements.  We lay out the detailed equations we evolve to generate spinning body inspirals in the Appendix \ref{app:oscelementframework}.  In Sec.\ \ref{sec:spinviaforcedgeod}, we show how to describe spinning-body orbits as forced geodesics, explicitly demonstrating that this approach yields orbits equivalent to those developed using the frequency-domain method of Refs.\ \cite{Drummond2022_1, Drummond2022_2}. We describe how we incorporate the leading adiabatic backreaction in Sec.\ \ref{sec:backreaction}.

In Sec.\ \ref{sec:NIT}, we describe the mathematical scheme underlying the near-identity averaging transformation (abbreviated ``NIT'') in detail.  We outline the notation used in this section in Sec.\ \ref{sec:NITnotation}, then discuss Mino-time and Boyer-Lindquist-time formulations of NITs in Secs.\ \ref{sec:MinoNIT} and \ref{sec:BLNIT} respectively.  We then present the full set of averaged equations of motion for the specific forcing terms studied in this work in Sec.\ \ref{sec:AveragedEoM}. We discuss the details of our NIT implementation in Sec.\ \ref{sec:NITImplement}.  Additional background and details on the NIT are presented in Appendix \ref{app:NITdetails}, and some important details for how we match the OG and NIT calculations in Appendix \ref{app:initialconditions}.

We present results describing spinning body inspirals in Sec.\ \ref{sec:res_inspirals}, and their associated GWs in Sec.\ \ref{sec:res_waveforms}.  We first look at examples of generic (inclined and eccentric) inspirals with aligned secondary spin in Sec.\ \ref{sec:res_inspalign}, and then generalize to arbitrarily oriented spin in Sec.\ \ref{sec:res_inspmisalign}.  We comment that our study of generic inspiral is presently limited by the paucity of data available describing generic strong-field adiabatic radiation reaction.  Though work continues to generate additional such data, we have confined ourselves to the $a = 0.7M$ generic orbit data set that was used in Ref.\ \cite{Hughes2021}.

We begin our discussion of waveforms from spinning body inspirals by briefly reviewing in Sec.\ \ref{sec:Teukolsky} the general principles used to compute waveforms; greater detail can be found in Ref.\ \cite{Hughes2021}.  We then examine in Sec.\ \ref{sec:WFanalysis} the waveforms which correspond to the inspirals presented in Sec.\ \ref{sec:res_inspirals}.  Of particular physical interest is a comparison of waveforms with and without spinning secondary effects, showing the observable imprint that secondary spin has on the waveform.  On a pragmatic level from the standpoint of computations, we also compare waveforms produced with the OG technique versus those using the NIT to generate the trajectory.  We show that these waveforms differ very little, though the NIT produces waveforms significantly more quickly.

Throughout this paper, we work in relativist's units with $G = 1 = c$.  A useful conversion factor in these units is $10^6\,M_\odot = 4.926$ seconds $\simeq 5$ seconds.  We use the (fairly standard) convention that lowercase Greek indices on vectors and tensors denote spacetime coordinate indices.  Latin indices are used on certain quantities to designate elements of a set that holds parameters which describe orbital elements: capital Latin indices are used for seven-element sets, used for the parameters of OGs; lowercase Latin indices are used for two-, three-, and four-element sets, describing the properties of orbits.

\section{Bound orbits of Kerr black holes}
\label{sec:orbits}

In our analysis, we approximate inspiral by a sequence of bound orbits, evolving from orbit to orbit under the influence of orbit-averaged GW backreaction.  We use GW amplitudes computed at each orbit to describe contributions to the waveform from this inspiral.  To set this up, we briefly review the properties of the orbits we use.  All of the details in this section have been presented in depth in other papers, such as Refs.\ \cite{FujitaHikida2009, vandeMeent2019, Hughes2000, Hughes2021, Lynch2021}, so we confine this discussion to a high-level synopsis sufficient to lay out the notation and details we need for this analysis.  Additional important technical details are summarized in Appendices \ref{app:geodesicsinkerr} and \ref{app:spinningbodymotion}.

\subsection{Orbits of non-spinning bodies}
\label{sec:geodesics}

Bound Kerr geodesics can be described using several time parameterizations.  In much of our discussion, we will use the ``Mino time'' variable $\lambda$.  The equations of motion in Boyer-Lindquist coordinates can be written
\begin{align}
\left(\frac{dr}{d\lambda}\right)^2 &= R({r})\;,\qquad\quad\, \left(\frac{d\theta}{d\lambda}\right)^2 = \Theta({\theta})\;,
\nonumber\\
\frac{d\phi}{d\lambda} &= \Phi_r({r}) + \Phi_\theta({\theta})\;,\quad \frac{dt}{d\lambda} = T_r({r}) + T_\theta({\theta})\;.
\label{eq:geods_mino}
\end{align}
Expressions for the functions on the right-hand sides of these equations are presented in Eqs.\ (\ref{eq:geodr})--(\ref{eq:geodt}) of Appendix \ref{app:geodesicsinkerr}.  Mino time $\lambda$ is related to proper time $\tau$ along an orbit by the relation $d\lambda = d\tau/\Sigma$ \cite{Mino2003}, where $\Sigma = r^2 + a^2\cos^2\theta$.  Notice that the factor $\Sigma$ couples the radial and polar motions; when $\lambda$ is the time parameter, the radial motion depends only on $r$, and the polar motion depends only on $\theta$.  This separation means that coordinate-space solutions describing geodesic orbits can be written using simple quadratures; see \cite{Drasco2004, FujitaHikida2009} for further discussion.

The radial and polar motions can both be described using a quasi-Keplerian description, mapping the oscillatory coordinate motion to orbit anomaly angles which increase monotonically with time.  We begin by noting that bound geodesic orbits around a Kerr black hole are contained within a torus that lies in the radius range $r_2 \le {r} \le r_1$ and in the polar angle range $\theta_1 \le {\theta} \le (\pi - \theta_1)$.  It is very useful to remap the radii $r_2$ and $r_1$ using
\begin{equation}
    r_1 = \frac{pM}{1 - e}\;,\qquad r_2 = \frac{pM}{1 + e}\;.
\end{equation}
We have introduced $p$, the orbit's semi-latus rectum, and $e$, its eccentricity.  A geodesic orbit's bounds are then totally set by choosing the parameters $p$, $e$, and $\theta_1$.  Those parameters can be remapped to integrals of the motion $\hat E$ (energy), $\hat L_z$ (axial angular momentum), and $\hat Q$ (Carter constant) which are related to the spacetime's Killing vectors and Killing tensor, and are conserved along any geodesic.  An alternate form of the Carter constant, $\hat K \equiv \hat Q + (\hat L_z - a\hat E)^2$ is also useful.  (The ``hat'' accents indicate that these conserved quantities are defined on geodesics.)  See Refs.\ \cite{Schmidt2002,FujitaHikida2009} for further discussion.

We build the bounds on the radial motion into our parameterization by defining
\begin{align}
    {r} &= \frac{p M}{1 + e\cos\chi_r}\;.
    \label{eq:rdef}
\end{align}
The angle $\chi_r$ is a relativistic analog of the true anomaly angle commonly used to describe orbital dynamics in Newtonian gravity.  We define\footnote{The angle $\chi_r^S$ we use in this analysis is equivalent to $\chi_{r0}$ in Ref.\ \cite{Hughes2021}.  In \cite{Gair2011}, $\psi_0$ is used to denote the initial radial phase, and is equivalent to our $\chi_r^S$, modulo a minus sign.} $\chi_r=\chi_r^F + \chi_r^S$.  The ``$F$'' superscript signifies that $\chi_r^F$ evolves on fast timescales, related to the orbital motion; the ``$S$'' tells us that $\chi_r^S$ evolves on slow timescales, related to the backreaction.  For geodesics (i.e., in the absence of forcing terms), $\chi_r^S$ is a constant, corresponding to the initial radial phase. We later allow $\chi_r^S$ to change with time, accounting for its slow evolution under a perturbing force; see discussion in App.\ \ref{app:oscelementframework}.

The function $R(r)$ defined in Eq.\ (\ref{eq:geods_mino}) and shown in detail in Eq.\ (\ref{eq:geodr}) is a quartic with four roots ordered such that $r_4 \le r_3 \le r_2 \le { r} \le r_1$.  For a bound orbit, the roots $r_1$ and $r_2$ are the physical turning points of the motion, discussed above; the roots $r_3$ and $r_4$ depend in a straightforward way on the orbit parameters $p$, $e$, and $x_I$ (see, e.g., Ref.\ \cite{FujitaHikida2009} for a form that is commonly used).  From the form (\ref{eq:geodr}), we can write 
\begin{equation}
R(r) = (1- \hat E^2)(r_1 - r)(r - r_2)(r - r_3)(r - r_4)\;,
\end{equation}
where $\hat E$ is the orbit's energy introduced above.  It is convenient to introduce parameters $p_3$ and $p_4$ such that
\begin{equation}
    r_3 = \frac{p_3M}{1 - e}\;,\qquad r_4 = \frac{p_4M}{1 + e}\;.
\end{equation}
Using this, we write the radial component of the geodesic equation (\ref{eq:geods_mino}) as a differential equation for $\chi_r$ \cite{Drasco2004}:
\begin{widetext}
\begin{align}
 \frac{d\chi_r}{d\lambda}&=\frac{M\sqrt{1 - \hat E^2}\left[(p - p_3) - e(p + p_3\cos\chi_r)\right]^{1/2}\left[(p - p_4) + e(p - p_4\cos\chi_r)\right]^{1/2}}{1 - e^2} \nonumber \\ 
 &\equiv X_r^F(\chi_r)\;.
 \label{eq:chir}
\end{align}
\end{widetext}
Remapping the oscillatory radial dynamics onto the monotonically evolving angle $\chi_r$ makes the bounded nature of geodesic motion explicit, allowing for straightforward numerical handling of the radial turning points.

Turn now to the polar motion.  Defining ${z} \equiv \cos{\theta}$, we can write the function $\Theta({\theta})$ from Eq.\ (\ref{eq:geods_mino}) (see also Eq.\ (\ref{eq:geodtheta})) in terms of roots $0 \le z_1 \le 1 \le z_2$ \cite{vandeMeent2019}:
\begin{equation}
\Theta({\theta})=\frac{z_1^2 - z^2}{1 - z^2}\left(z^2_2-a^2(1- \hat E^2)z^2\right)\;.
\end{equation}
This form, taken from Ref.\ \cite{vandeMeent2019}, has the advantage that it allows for straightforward evaluation in the $a\rightarrow 0$ limit.  Turning points of the polar motion occur where $z = z_1$, corresponding to when $\theta = \theta_1$ and ${\theta} = \pi - \theta_1$.  The second polar root $z_2$, given by Eq.\ (15) in Ref.\ \cite{vandeMeent2019}, is not actually reached by physical orbits (it generally corresponds to $\cos\theta > 1$).  We define the inclination angle $I$ as 
\begin{equation}
    I = \pi/2 - \mbox{sgn}(\hat L_z)\theta_1\;;
\end{equation}
$I = 0$ corresponds to prograde equatorial orbits, $I = 180^\circ$ to retrograde equatorial, and orbital properties vary smoothly between these extremes.  We put $x_I \equiv\cos I$, from which we see that $z_1 = \sqrt{1 - x_I^2}$.  This allows us to parameterize our polar motion as
\begin{equation}
\cos{\theta} = \sqrt{1 - x_I^2}\cos\chi_\theta = \sin I\cos\chi_\theta\;,
    \label{eq:thdef}
\end{equation}
where $\chi_\theta$ is another relativistic generalization of the ``true anomaly'' angle used in Newtonian orbital dynamics.  As we did for the radial motion, we define\footnote{The angle $\chi_\theta^S$ is equivalent to $\chi_{\theta0}$ used in Ref.\ \cite{Hughes2021}.  In \cite{Gair2011}, $\chi_0$ is used to denote the initial polar phase, and is equivalent to $\chi_\theta^S$ in this analysis, modulo a minus sign.} $\chi_\theta = \chi^F_\theta + \chi_\theta^S$, breaking this anomaly angle into ``fast'' and ``slow'' terms.  In the absence of forcing terms, $\chi_\theta^S$ is a constant, the initial polar phase.  In the osculating element framework (see App.\ \ref{app:oscelementframework}), we promote $\chi_\theta^S$ to a time-varying quantity.  Combining the various reparameterizations with the polar geodesic equation (\ref{eq:geodtheta}) yields an equation governing $\chi_\theta$ \cite{vandeMeent2019,Drasco2004}:
\begin{align}
\frac{d\chi_\theta}{d\lambda} &= \sqrt{z_2^2 - a^2(1 - \hat E^2)(1 - x_I^2)\cos^2\chi_\theta}\nonumber\\
&\equiv X_\theta^F(\chi_\theta)\;.
\label{eq:chitheta}
\end{align}

Bound Kerr geodesics are triperiodic, and can be characterized with frequencies describing the orbit's radial, polar, and axial behavior: the frequencies $\hat\Upsilon_{r,\theta,\phi}$ describe an orbit's radial, polar, and axial frequencies per unit Mino time, and $\hat\Omega_{r,\theta,\phi}$ describe these frequencies per unit Boyer-Lindquist coordinate time.  The Mino-time and coordinate-time frequencies are related by a factor $\hat\Upsilon_t$ that describes\footnote{This factor is labeled $\hat \Gamma$ in many references \cite{Drasco2004, Drasco2006, FujitaHikida2009, Hughes2021}, to reflect the fact that it represents a conversion between two different notions of time, rather than being related to a periodic aspect of orbital motion.  It is however labeled $\hat \Upsilon_t$ in much of the NIT literature, and we follow that convention here.} how much coordinate time accumulates, on average, per unit Mino time along the orbit: $\hat\Omega_{r,\theta,\phi} = \hat\Upsilon_{r,\theta,\phi}/\hat\Upsilon_t$.  The inverse of these frequencies, times $2\pi$, gives the Mino- and coordinate-time periods:
\begin{align}
    \hat\Lambda_{\phi,\theta,r} &= \frac{2\pi}{\hat\Upsilon_{\phi,\theta,r}}\;,
    \\
    \hat T_{\phi,\theta,r} &= \frac{2\pi}{\hat\Omega_{\phi,\theta,r}}\;.
\end{align}
As in our discussion of the constants of motion $\hat E$, $\hat L_z$, and $\hat Q$, the hat accents indicate that these quantities are evaluated on geodesics.  See Ref.\ \cite{FujitaHikida2009} for formulas describing these frequencies, periods, and the factor $\hat\Upsilon_t$.

An action-angle parametrization of geodesic motion is useful for the construction of near-identity transformations in Sec.\ \ref{sec:NIT}.  In this formulation, the Mino-time action angles $q_r$ and $q_z$ are chosen as the orbital phases describing the motion in $r$ and $z$ respectively; explicit formulas connecting these angles to motion in their associated coordinate are given in Refs.\ \cite{FujitaHikida2009, vandeMeent2019}, and are coded into the {\tt KerrGeodesics} package of the Toolkit \cite{Kerrgeodesics}.  We denote by $P_i = \{p,e,x_I \}$ the set of orbital elements.  In this form, the geodesic equations of motion are given by 
\begin{align}
    \frac{dP_j}{d\lambda} &= 0\;, \\
    \frac{dq_{r,z}}{d\lambda} &= \hat{\Upsilon}_{r,z}(\vec{P})\;. \label{eq:AA2}
\end{align}
(Note that $\hat{\Upsilon}_z = \hat{\Upsilon}_\theta$; the period of a complete cycle in $\theta$ is identical that of a complete cycle in $z = \cos\theta$.)  In other words, for geodesics the elements $\vec{P}$ are constants of motion and the right-hand side of Eq.\ (\ref{eq:AA2}) is an orbital frequency determined by $\vec{P}$.  As such, the orbital phases\footnote{Note that the orbital phases $q_{r,z}$ are identical to the ``mean anomaly angles'' $w_{r,\theta}$ used in Refs.\ \cite{Drummond2022_1, Drummond2022_2}.} have solutions $q_{z,r} = \hat\Upsilon_{r,z} \lambda + q_{r,z}^{S}$, where $q_{r,z}^{S}$ is the value of that phase when $\lambda = 0$.  These phases will evolve on the slow timescale when certain post-geodesic forces are introduced.

%

Up to initial conditions, a geodesic orbit can be specified by ``principal orbital elements.'' These are either the constants of motion ($\hat E$, $\hat L_z$, $\hat Q$) or the parameters ($p$, $e$, $x_I$) describing the geometry of the orbit. We can convert between ($\hat E$, $\hat L_z$, $\hat Q$) and ($p$, $e$, $x_I$) using mappings given in Refs.\ \cite{FujitaHikida2009, vandeMeent2019,Hughes2021}. The initial conditions of the orbit are specified by ``positional orbital elements'' which are ($\chi_r^S$, $\chi_\theta^S$, $\phi_0$, $t_0$) in the quasi-Keplerian case and ($q_{r}^S$, $q_{z}^S$, $\phi_0$, $t_0$) in the action-angle case. In order to find the geodesic trajectories for a particular set of orbital elements $\{p, e, x_I, \chi_r^S, \chi_\theta^S, \phi_0, t_0\}$ or $\{p, e, x_I, q_{r}^S, q_{z}^S, \phi_0, t_0\}$, we need only solve differential equations for the radial and polar phases $\chi_r$ and $\chi_\theta$, i.e., Eqs.\ (\ref{eq:chir}) and (\ref{eq:chitheta}); or for $q_r$ and $q_\theta$, i.e., Eqs.\ (\ref{eq:AA2}).

\subsection{Orbits of spinning bodies}
\label{sec:spinningorbits}

The geodesic orbits discussed above describe the motion of a pointlike body freely falling in spacetime.  The equations of motion (\ref{eq:geods_mino}) fundamentally derive from the equation of parallel transport for a freely falling body's 4-momentum:
\begin{equation}
    \frac{Dp^\mu}{d\tau} = 0\;.
    \label{eq:geod_general}
\end{equation}
In this equation, $D/d\tau \equiv u^\alpha\nabla_\alpha$ denotes a covariant derivative with respect to proper time along the trajectory.  The 4-velocity $u^\alpha = dx^\alpha/d\tau$ is the tangent vector to the worldline of this freely falling body.

If the body is not pointlike but has some extended structure, this structure will couple to the spacetime in which it moves, changing its trajectory.  This coupling can be incorporated into the framework describing the body's motion by replacing the right-hand side of (\ref{eq:geod_general}) with a forcing term reflecting how the body's structure couples to spacetime.

The simplest example of such coupling structure is the body's spin angular momentum.  The equation of motion in this case becomes \cite{Mathisson2010G_2, Papapetrou1951, Dixon1970}
\begin{equation}
    \frac{Dp^\mu}{d\tau} = -\frac{1}{2}{R^\mu}_{\nu\lambda\sigma}u^\nu S^{\lambda\sigma}\;.
    \label{eq:spinforce}
\end{equation}
The right-hand side of this equation is the spin-curvature force.  In this equation, ${R^\mu}_{\nu\lambda\sigma}$ is the Riemann tensor of the spacetime through which the spinning body moves, and $S^{\lambda\sigma}$ in a tensor which describes its spin angular momentum.  It is useful to remap this tensor to a vector:
\begin{equation}
S^{\mu} = -\frac{1}{2\mu}{\epsilon^{\mu\nu}}_{\alpha\beta}p_{\nu}S^{\alpha\beta}\;.
\label{eq:spinvec}
\end{equation}
As the body moves through spacetime, its angular momentum precesses according to
\begin{equation}
    \frac{DS^{\mu\nu}}{d\tau} = p^\mu u^\nu - u^\mu p^\nu\;.
    \label{eq:spinprec}
\end{equation}
Note that $p^\mu$ is not parallel to $u^\mu$ in general; the right-hand side of Eq.\ (\ref{eq:spinprec}) is $O(S^2)$.  Equations (\ref{eq:spinforce}) and (\ref{eq:spinprec}) are not sufficient to completely specify the motion of the smaller body, so we augment these equations with a spin supplementarity condition:
\begin{equation}
    p_\mu S^{\mu\nu} = 0\;.
    \label{eq:spinsupp}
\end{equation}
This condition, known as the Tulczejew spin supplementary condition \cite{Tulczyjew1959}, is not unique; other choices could be made.  The physical importance of the spin supplementary condition is to pick out a particular worldline from the many which pass through an extended body.

For extreme mass ratio systems, it makes sense to linearize in the spin of the smaller body: taking the smaller body to be a Kerr black hole, terms linear in spin enter the forcing equations at order $\mu^2$, so terms quadratic in spin enter at order $\mu^4$.  Linearizing, the equations discussed above simplify to
\begin{align}
    \frac{Du^\mu}{d\tau} &= -\frac{1}{2\mu}{R^\mu}_{\nu\lambda\sigma}u^\nu S^{\lambda\sigma}\;,
    \label{eq:spinforcelin}\\
    \frac{DS^\mu}{d\tau} &= 0\;,
    \label{eq:spinpreclin}\\
    u_\mu S^\mu &= 0\;.
    \label{eq:spinsupplin}
\end{align}
Witzany has proven that these linearized equations can be cast as a Hamiltonian system \cite{Witzany2019_1,Witzany2019_2}, and thus that the spin-curvature force is conservative.  A consequence of this is that the linearized equations admit bound orbits.  These orbits can be characterized by energy $E$, axial angular momentum $L_z$, and an analog of either the Carter constant $Q$ or $K \equiv Q + (L_z - aE)^2$, much like geodesic orbits\footnote{It is worth emphasizing that the quantities $E$ and $L_z$ can be defined for motion under the complete set of Papapetrou equations, but analogs of $Q$ and $K$ can be found only when these equations are linearized in spin \cite{Rudiger1981}.  It has recently been shown that analogs of $Q$ and $K$ can be found for the full equations if one includes the next multipole order in the analysis (the secondary's quadrupole moment), though only if that quadrupole moment takes the values appropriate for a Kerr black hole.  See Ref.\ \cite{2023CompereDruart} for further discussion.}, though offset from the geodesic values by an amount that is proportional to the secondary spin $S$.  (Note that we do not write these quantities with hat accents, emphasizing that they are offset from their geodesic analogs.)  Likewise, these orbits have frequencies ($\Omega_r, \Omega_\theta, \Omega_\phi$) describing their coordinate motions which differ from the geodesic values by an amount scaling with $S$.  They also have a ``precession frequency'' $\Omega_s$ which describes the precession of the spin.

References \cite{Drummond2022_1, Drummond2022_2} describe in detail how to construct orbits of spinning bodies using a frequency domain technique to solve the linearized equations (\ref{eq:spinforcelin})--(\ref{eq:spinsupplin}).  For our purposes, a key point is that the resulting motion is similar to geodesic motion, and we can adapt the quasi-Keplerian formulation to describe these orbits.  For example, in the general case, the radial and polar motions can be written
\begin{align}
    r &= \frac{pM}{1 + e\cos\chi_r} + \delta\mathcalligra{r}_S\;,\label{eq:spintrajectoryr}
    \\
    \cos\theta &= \sin I\cos\chi_\theta + \delta\mathcalligra{z}_S\;.\label{eq:spintrajectorytheta}
\end{align}
These expressions resemble the forms used for geodesic motion, with a few key differences. The anomaly angles $\chi_r$ and $\chi_\theta$ used for spinning-body orbits differ from the angles used to describe geodesics:
\begin{equation} 
\chi_r = \chi^{\rm SG}_r + \delta \chi_r^S\;, \; \; \;
\chi_\theta = \chi^{\rm SG}_\theta + \delta \chi_\theta^S\;.
\end{equation}
The quantities $\chi^{\rm SG}_{r,\theta}$ are identical to the anomaly angles used for geodesics, but expanded in a Mino-time Fourier series and with the geodesic frequencies $\hat\Upsilon_{r,\theta}$ shifted to the frequencies $\Upsilon_{r,\theta}$ appropriate for spinning-body orbits (the superscript ``SG'' stands for ``shifted geodesic'').  The terms $\delta\chi^S_{r,\theta}$ are $O(S)$ shifts to the anomaly angles.  See Refs.\ \cite{Drummond2022_1, Drummond2022_2} for details and further discussion.

The libration regions for spinning-body orbits also differ from those of geodesics; this difference is encoded in the functions $\delta\mathcalligra{r}_S$ and $\delta\mathcalligra{z}_S$ introduced in Eqs.\ (\ref{eq:spintrajectoryr}) and (\ref{eq:spintrajectorytheta}).  These functions are both $O(S)$, and are both periodic in harmonics of the spinning body frequencies --- either the set ($\Upsilon_r, \Upsilon_\theta, \Upsilon_s$) or ($\Omega_r, \Omega_\theta, \Omega_s$), depending on which time parameterization is used.

In addition to solutions describing the coordinate-space motion of the smaller body, we need to describe how the orientation of the smaller body's spin evolves over its motion.  We use the closed-form solution describing a parallel-transported vector presented in \cite{vandeMeent2019}.  This solution uses a tetrad, originally developed in Refs.\ \cite{Marck1983, Marck1983_2, Kamran1986}, with legs $\{e_{0\alpha}, e_{1\alpha}, e_{2\alpha}, e_{3\alpha}\}$.  Legs 1 and 2 of this tetrad are related to auxiliary legs $\tilde{e}_{1\alpha}$ and $\tilde{e}_{2\alpha}$ via a precession phase rotation:
\begin{align}
    e_{1\alpha}(\lambda) &= \cos\psi_s(\lambda)\,\tilde{e}_{1\alpha}(\lambda) + \sin\psi_s(\lambda)\,\tilde{e}_{2\alpha}(\lambda)\;,
    \label{eq:tetradleg1}\\
    e_{2\alpha}(\lambda) &= -\sin\psi_s(\lambda)\,\tilde{e}_{1\alpha}(\lambda) + \cos\psi_s(\lambda)\,\tilde{e}_{2\alpha}(\lambda)\;.
    \label{eq:tetradleg2}
\end{align}
Leg 0 is simply the 4-velocity $u_\alpha$ of the orbiting body; expressions for $\tilde{e}_{1\alpha}$, $\tilde{e}_{2\alpha}$, and  $e_{3\alpha}$ can be found in Eqs.\ (48), (50) and (51) of Ref.\ \cite{vandeMeent2019}.  The precession phase\footnote{Note that this phase was written $\psi_p$ in Refs.\ \cite{Drummond2022_1, Drummond2022_2}, with the subscript $p$ standing for ``precession.''  We change notation here to avoid colliding with the use of subscript $p$ to describe how certain forcing terms introduced later in the paper change an orbit's semi-latus rectum.} $\psi_s(\lambda)$ is found by integrating up
\begin{equation}
\frac{d\psi_s}{d\lambda} = \sqrt{K}\left(\frac{(r^2 + a^2)\hat E - a \hat L_z}{\hat K + r^2} + a\frac{\hat L_z- a(1 - z^2)\hat E}{\hat K - a^2 z^2}\right)\;.
\label{eq:precphaseeqn}
\end{equation}
Although an analytic solution to (\ref{eq:precphaseeqn}) exists for geodesic orbits \cite{vandeMeent2019}, we find it useful to explicitly integrate this equation numerically as we evolve through a sequence of orbits to make inspirals.  In this vein, we comment that the terms on the right-hand side of (\ref{eq:precphaseeqn}) depend on the same orbital elements $\{p, e, x_I, q_r^S q_{z}^S, \phi_0, t_0\}$ that we use to characterize geodesics.  We also note that although these functions are written most cleanly as functions of Mino-time $\lambda$, it is straightforward to convert to other time parameterizations.

With the precession phase in hand, the smaller body's spin vector takes the form
\begin{equation}
    S_\alpha = S^0 e_{0\alpha}(\lambda) + S^1 e_{1\alpha}(\lambda) + S^2 e_{2\alpha}(\lambda) + S^3 e_{3\alpha}(\lambda)\;,
    \label{eq:spinvectetrad}
\end{equation}
where $\{S^0, S^1, S^2, S^3\}$ are all constants we select by choosing initial conditions.  Because $e_{0\alpha} = u_\alpha$, the Tulczyjew SSC (\ref{eq:spinsupplin}) requires that $S^0 = 0$.  The constants $S^1$ and $S^2$ denote components of the spin that lie perpendicular to the orbital angular momentum vector, and $S^3$ is the component of the small body's spin aligned with the direction of orbital angular momentum.  This allows us to express $S_\alpha$ in terms of the parallel and perpendicular spin components of the small-body's non-dimensional spin parameter $s$:
 \begin{equation}
    S_\alpha = \mu^2\bigl(s_\perp\cos\phi_s\,e_{1\alpha} + s_\perp\sin\phi_s\,e_{2\alpha} + s_\parallel\,e_{3\alpha}\bigr)\;,
    \label{eq:Smisalign1}
\end{equation}
where $s = \sqrt{s_\perp^2 + s_\parallel^2}$, and $\phi_s$ describes the orientation of the spin vector components. The small body's spin vector will precess only when $S^1$ or $S^2$ are non-vanishing. Refer to Appendix \ref{app:spinningbodymotion} for further discussion about spinning-body orbits.

Note that two dimensionless secondary spin parameters are commonly used in the literature.  The first,
\begin{equation}
    s = \frac{S}{\mu^2} \;,
\end{equation}
satisfies $0 \le s \le 1$.  The other, used for example in \cite{Skoupy2021,Skoupy2022}, is:
\begin{equation}
    \sigma = \frac{S}{\mu M} \;,
\end{equation}
and satisfies $0 \le \sigma \le \mu/M$.  A virtue of this form is that $\sigma$ is of order the mass ratio $\varepsilon$, which can facilitate comparing the magnitude of various terms in our analysis.

\subsection{Which orbits to use?}
\label{sec:whichorbits}

As discussed at length in the Introduction, our goal is to make a model of spinning body inspiral by supplementing a description of orbits which accurately describes motion on short timescales with appropriately averaged radiative backreaction which describes how orbits evolve on long timescales.  In essence, we want to treat inspiral as a sequence of orbits, with backreaction moving us from orbit to orbit in the sequence.

Which notion of orbits should we use? Since our goal is to make a model for an inspiraling spinning body, it is might seem clear that we should begin with orbits of spinning bodies --- use the orbits discussed in Refs.\ \cite{Drummond2022_1, Drummond2022_2}, and evolve from orbit to orbit by computing orbit-averaged GW backreaction on those orbits.  Unfortunately, implementing this scheme is not tenable in the short term.  Studies of backreaction on generic spinning body orbits have only recently been undertaken \cite{Skoupy2023}, and data sets which cover enough parameter space to generate an astrophysically plausible generic inspiral do not yet exist.  In addition, issues of principle remain which mean that, even if such data existed, we do not yet completely understand how to evolve from orbit to orbit using the orbit-averaged backreaction.  In particular, we do not fully understand how to evolve a spinning body's Carter constant due to gravitational radiation reaction (see concluding discussion in Ref.\ \cite{Skoupy2023}).

By contrast, computing backreaction on geodesic orbits is now rather straightforward. Large data sets exist describing backreaction for this case, and more data is being generated and made available in order to extend the ``Fast EMRI Waveform'' (FEW) models \cite{Chua2021, Katz2021,FEW}.  Furthermore, as we describe in more detail in the next section, it is possible to describe spinning body orbits as a sequence of geodesic orbits: we treat the worldline of a spinning body as a sequence of geodesics, with the sequence generated using the forcing terms (\ref{eq:spinforcelin})--(\ref{eq:spinsupplin}).

Because our goal is to make a model describing spinning body inspiral using data and methods available now, the approach we take is to use geodesic orbits forced by a combination of the spin-curvature force and geodesic-averaged GW backreaction.  After confirming that spinning body orbits constructed by forcing geodesics with the spin-curvature forcing terms agree with those constructed using the methods described in Refs.\ \cite{Drummond2022_1, Drummond2022_2}, we make spinning body inspirals by combining the spin-curvature force with orbit-averaged backreaction computed along geodesics.

As we discuss in more detail in our conclusions, it will be worthwhile to compare the results we find using this to results found by directly computing backreaction on spinning body orbits, once large data sets exist which make such calculations practical.  To facilitate this eventual comparison, we release the \textit{Mathematica} code and data which computes the expressions that we use to make the inspirals we develop here as supplementary material for this manuscript.

\section{Forced geodesics}
\label{sec:forcedgeodesics}

In this section, we construct spinning-body inspirals as a sequence of geodesic orbits, using an osculating geodesic (OG) framework to describe the inspiral worldline as a sequence of geodesic orbits.  The OG technique generalizes the venerable method of osculating orbits \cite{moulton1970,Taff1985,Beutler2005} to relativity \cite{Lincoln1990,Pound2008,Gair2011}.  We follow very closely the framework laid out in Ref.\ \cite{Gair2011}, which we summarize in Appendix \ref{app:oscelementframework}.  The key point necessary to understand this calculation is that, as described in Sec.\ \ref{sec:orbits}, both geodesic orbits and the smaller body's precession are entirely characterized by 7 parameters:
\begin{equation}
    \mathcal{E}^A \doteq \{p, e, x_I, q_{r}^S, q_{z}^S, \phi_0, t_0\}\;.
    \label{eq:geodparameters}
\end{equation}
As described at length in Sec.\ \ref{sec:geodesics} and Appendix \ref{app:geodesicsinkerr}, the subset $(p, e, x_I)$ are a geodesic's ``principal orbital elements,'' and fully characterize the coordinate-space torus which a geodesic occupies. The remaining parameters $(q_{r}^S, q_{z}^S, \phi_0, t_0)$ are its ``positional orbital elements,'' and can be regarded as setting the geodesic's initial coordinates on this torus.

The parameters (\ref{eq:geodparameters}) are all constants for geodesic motion.  The OG framework promotes at least some of these parameters to dynamical variables under the influence of some non-geodesic acceleration $a^\mu$.  One can then regard the worldline as a ``geodesic'' whose parameters $\mathcal{E}^A$ evolve under the influence of this acceleration.  See Appendix \ref{app:oscelementframework} for a synopsis of how one develops these evolution equations, and Ref.\ \cite{Gair2011} for a detailed derivation and discussion of the particular frameworks that we use.

We implement two OG schemes: The contravariant quasi-Keplerian formulation discussed in Appendix \ref{sec:contraevol}, and the action-angle formulation discussed in Appendix \ref{sec:AAevol}.  Comparing the results of these two methods is useful for validating our computations.  We also compare to the OG codes used in Refs.\ \cite{Lynch2021, Lynch2023} as an independent check of our implementation.  Because of the relevance of the action-angle formulation for applying the near-identity transformation, we focus on this formulation for the remainder of this analysis.

\subsection{Spinning body orbits as forced geodesics}

\label{sec:spinviaforcedgeod}

We begin by demonstrating the equivalence between spinning-body orbits computed using the frequency-domain approach from Refs.\ \cite{Drummond2022_1,Drummond2022_2} and the forced geodesic approach in this work; see also Appendix \ref{app:spinningbodyparams} for discussion regarding different ways to parameterize spinning-body motion.  First, we compute a spinning-body orbit using the method of Refs.\ \cite{Drummond2022_1, Drummond2022_2}.  We select a $(p, e, x_I)$ triplet that defines a geodesic with radial turning points $r_1 = p/(1 - e)$ and $r_2 = p/(1 + e)$ and polar turning point $z_1 = \sqrt{1 - x_I^2}$.  We then compute the spinning-body trajectory that has the same turning points (on average) as this geodesic \cite{Drummond2022_1,Drummond2022_2}.  Note that the turning points of this spinning-body trajectory differ from the corresponding geodesic due to an $\mathcal{O}(S)$ correction, as discussed in Refs.\ \cite{Drummond2022_1,Drummond2022_2}.

\begin{figure*}
\centerline{\includegraphics[scale=0.6]{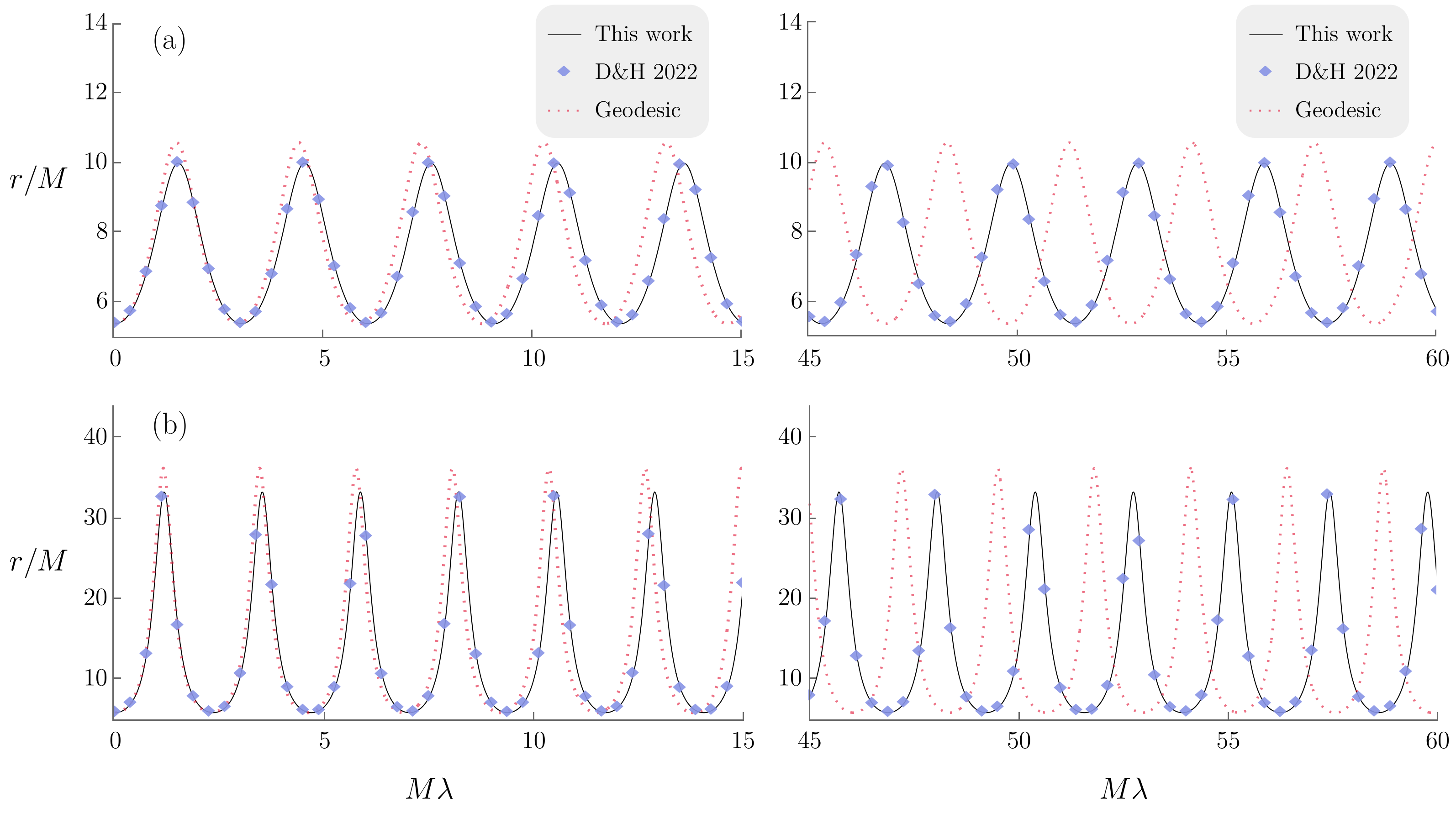}}
    \caption{ Comparison between spinning-body orbits computed using the OG approach in this work (solid black), spinning-body orbits computed using the frequency-domain approach in Refs.\ \cite{Drummond2022_1,Drummond2022_2} (blue diamond markers), and a geodesic, non-spinning orbit with the same parameters (dotted red). The orbit shown in the top panels (a) has initial parameters given by $a=0.7 M$, $p_{IC}=7.138$, $e_{IC}=0.326$, and  $x_{IC}=0.966$ while the orbit shown in the bottom panels has initial parameters given by $a=0.5 M$, $p_{IC}=10.122$, $e_{IC}=0.721$, and  $x_{IC}=0.966$. The ``IC" subscript indicates that these are ``matched initial conditions'' orbital parameters: the geodesic orbit defined by the triplet ($p_{IC}$, $e_{IC}$, $x_{IC}$) (plotted with the red dashed line) has the same initial conditions as the spinning-body orbit (plotted with the black solid line). There is also a ``matched turning point'' description of orbital parameters used in Refs.\ \cite{Drummond2022_1,Drummond2022_2}, where the geodesic defined by ($p_{TP}$, $e_{TP}$, $x_{TP}$) and the corresponding spinning-body orbit have matched turning points.  For completeness, the ``matched turning point" orbital elements for the two spinning-body orbits pictured here are: ($p_{TP}=7$, $e_{TP}=0.3$, $x_{TP}=0.966$) for the top panels and ($p_{TP}=10$, $e_{TP}=0.7$, $x_{TP}=0.966$) for the bottom panels. See Appendix \ref{app:spinningbodyparams} for further details. The small body has mass ratio $\varepsilon =10^{-1}$ and a spin vector with $s = 1$ and $s_\parallel = s$.  Note that this mass ratio is rather far from the EMRI limit; we use this value here to make the effects of spin-curvature coupling more apparent to the eye.}
    \label{fig:diffparams}
\end{figure*}

Next we compute the same spinning-body trajectory with the OG approach used in this work. In order to do this, we find the triplet $(p_{IC},e_{IC},x_{IC})$ which defines a geodesic orbit with the same initial conditions (coordinate positions and four-velocities) as the spinning-body orbit we computed using the method in \cite{Drummond2022_1,Drummond2022_2}; details of the mapping between the two formulations are in Appendix \ref{app:spinningbodyparams}.  We find that OG solutions match for many cycles the corresponding spinning-body orbit computed using the approach of Refs.\ \cite{Drummond2022_1,Drummond2022_2}.  In Fig.\ \ref{fig:diffparams}, we show two example orbits to demonstrate this.  In this figure, solid black curves show the radial motion for a spinning body computed using the OG method.  The blue diamond markers show the same orbit computed using the frequency-domain method of Refs.\ \cite{Drummond2022_1,Drummond2022_2}.  For reference, we show the orbit of a non-spinning body (red dotted curve) with matching parameters.  Figure \ref{fig:diffparams} shows that the three orbits agree in orbital phase at early times (left panels).  At later times (right panels), the geodesic is completely dephased but the two spinning-body orbits remain matched.  

Figure \ref{fig:diffparams} also shows that, after many cycles, a slight difference develops between the solid black curves (spinning-body orbits generated via OG) and the blue diamonds (spinning-body orbits generated using the method of Refs.\ \cite{Drummond2022_1,Drummond2022_2}). The two methods are entirely equivalent up to first-order in secondary spin, but not at $O(S^2)$; the differences we see are quadratic in secondary spin (see Appendix \ref{app:spinningbodyparams} for detailed discussion).  In this vein, note that we used a rather non-extreme mass ratio $\varepsilon = 0.1$, far beyond the EMRI regime, in this figure.  This ``abuse'' of the large-mass ratio limit was done in order to make the effects of spin-curvature coupling more apparent to the eye.  At mass ratios appropriate for EMRI sources, bearing in mind that scaling as $O(S^2)$ means $O(\varepsilon^4)$, we expect differences to be far less apparent.

\subsection{Backreaction and inspiral}
\label{sec:backreaction}

The leading adiabatic backreaction requires only the orbit-averaged dissipative part of the first-order self force.  Flux balance laws allow us to compute this using only knowledge of GW fluxes at the horizon and infinity.  Such flux balance laws have the form
\begin{equation}
   \left( \frac{d\mathcal{C}}{dt}\right)^{\text{orbit}}=- \left( \frac{d\mathcal{C}}{dt}\right)^{\infty}- \left( \frac{d\mathcal{C}}{dt}\right)^{H}\;.
\end{equation}
where $\mathcal{C}$ corresponds to a conserved quantity along the geodesic such as $E$, $L_z$ or $Q$. We can then calculate the transition of the worldline between each OG using rates of change $dE/dt$, $dL_z/dt$, $dQ/dt$ to construct an inspiral.

Note that in this adiabatic construction we omit the conservative first-order self force as well as oscillatory pieces of the dissipative self force; both of these effects are included in Ref.\ \cite{Lynch2021}.  In computing the GW fluxes, we only include the contribution of the ``monopole'' term of the secondary's stress-energy tensor, which arises from the smaller body's mass.  We thus omit the impact of the ``dipole'' term to this stress-energy, which arises from the smaller body's spin, and is included in Refs.\ \cite{Skoupy2021,Skoupy2022}.  Including effects which we neglect are natural points for further development and future work.

The rates of change of energy $dE/dt$ at infinity and at the horizon are given by \cite{Teukolsky1974}
\begin{align}
    \left(\frac{dE}{dt} \right)^\infty&=\sum_{lmkn}\frac{\left|Z^\infty_{lmkn}\right|^2}{4\pi\omega^2_{mkn}}\;,\\ \left(\frac{dE}{dt} \right)^H&=\sum_{lmkn}\frac{\alpha_{lmkn}\left|Z^H_{lmkn}\right|^2}{4\pi\omega^2_{mkn}}\;;
\end{align}
the corresponding rates of change of angular momentum $dL_z/dt$ are \cite{Teukolsky1974}
\begin{align}
     \left(\frac{dL_z}{dt} \right)^\infty&=\sum_{lmkn}\frac{m\left|Z^\infty_{lmkn}\right|^2}{4\pi\omega^3_{mkn}}\;, \\   \left(\frac{dE}{dt} \right)^H&=\sum_{lmkn}\frac{\alpha_{lmkn}m\left|Z^H_{lmkn}\right|^2}{4\pi\omega^3_{mkn}}\;.
\end{align}
The coefficients $Z^{\infty,H}_{lmkn}$ are obtained by integrating homogeneous solutions of the separated radial Teukolsky equation against this equation's source term.  See Sec.\ III, particularly Eq.\ (3.9) of Ref.\ \cite{Hughes2021} for further details of this calculation, and see Eqs.\ (3.30), (3.31), and (3.32) of that paper for the expression for $\alpha_{lmkn}$.  The mode frequency $\omega_{mkn}$ is related to the geodesic frequencies by
\begin{equation}
\omega_{mkn} = m\hat\Omega_\phi + k\hat\Omega_\theta + n\hat\Omega_r\;.
\label{eq:omegamkn}
\end{equation}

Contributions to the rate of change of the Carter constant $Q$ similarly involve contributions from fields at infinity and fields on the horizon:
\begin{align}
    \left(\frac{dQ}{dt} \right)^\infty &= \sum_{lmkn}\left|Z^\infty_{lmkn}\right|^2\frac{\mathcal{L}_{mkn} + k\hat\Upsilon_\theta}{2\pi\omega^3_{mkn}}\;,
    \\
    \left(\frac{dQ}{dt} \right)^H & =\sum_{lmkn}\alpha_{lmkn}\left|Z^H_{lmkn}\right|^2\frac{\mathcal{L}_{mkn} + k\hat \Upsilon_\theta}{2\pi\omega^3_{mkn}}\;. 
\end{align}
where 
\begin{equation}
    \label{eq:Lmkn}
    \mathcal{L}_{mkn} = m\langle \cot^2\theta \rangle \hat L_z -a^2\omega_{mkn}\langle \cos^2\theta \rangle \hat E\;.
\end{equation}
Here, $\langle f(\theta)\rangle$ denotes a particular averaging with respect to the orbital motion of functions of $\theta$, defined in Eq.\ (2.13) of Ref.\ \cite{Hughes2021}.  It is straightforward to convert from rates of change of the constants of motion $(\hat E, \hat L_z, \hat Q)$ to those of the orbital elements $(p,e,x_I)$ which is the form we use in this article.  See Appendix B of Ref.\ \cite{Hughes2021} for the explicit conversion between the two rates of change. 

\section{Near identity transformations}
\label{sec:NIT}

The OG framework described in the previous section is computationally expensive, requiring us to evaluate forcing terms multiple times per orbit cycle.  The computational cost associated with this approach thus grows with the number of orbits, scaling inversely with the system's mass ratio.  Near-identity transformations (NITs) have proven to be powerful tools for modeling EMRI systems \cite{vandeMeent2018_2, Lynch2021, Lynch2022, Lynch2023} by introducing an averaging that makes it possible to include inspiral physics without needing to track the system's cycle-by-cycle orbital-time dynamics, substantially reducing the model's computational cost.  NITs are an established mathematical procedure \cite{Kevorkian1987}, used in celestial mechanics and other domains, that averages a system's short timescale behaviour while preserving the secular evolution on longer timescales.  In this section, we describe how to apply NITs to model the inspiral of spinning bodies, substantially reducing the computational cost of making such models.  In our results (Secs.\ \ref{sec:res_inspirals} and \ref{sec:res_waveforms}), we show that this reduction in computational cost does not involve a loss of modeling accuracy.

\subsection{NIT background: Notation and generalities}
\label{sec:NITnotation}

We begin by introducing important notation and definitions which will be used throughout this section.  Certain sets of related quantities will be organized into ``vectors,'' denoted with an overarrow.  For example, the set of principal orbital elements are organized into a vector $\vec{P} = (p,e,x_I)$, the phases into $\vec{q} = (q_r, q_z)$, and extrinsic quantities $\vec{\mathcal{X}} = (t,\phi)$.  As introduced in Sec.\ \ref{sec:spinningorbits}, we denote spin-precession phase by $\psi_s$.
It is also useful to define a vector containing both orbital and spin phases: $\vec Q =  (q_r,q_z,\psi_s)$.  Finally, it will be useful later, particularly when we begin to construct waveforms, to refer to the complete set of phases including the azimuthal phase.  We denote this set $\vec{\mathcal{Q}} = (q_r,q_z,\phi,\psi_s)$.  (Notice that these ``vectors'' do not have a consistent number of components.)

The NIT of a quantity $A$ will be denoted by $\nit{A}$ and defined by the form
\begin{equation}
\nit{A}=A+\varepsilon \mathcal{A}^{(1)} +\varepsilon^2 \mathcal{A}^{(2)}+\mathcal{O}(\varepsilon^3)\;,
\end{equation}
where the transformation functions $\mathcal{A}^{(n)}$ are required to be smooth, periodic functions of the orbital phases $\vec{q}$.  The transformation functions introduced in this section are: $Y_j^{(n)}$, used to effect the NIT of the vector $\vec P$; $X_i^{(n)}$, used for the phase $\vec q$; $W_s^{(n)}$, used for the spin-precession phase $\psi_s$; and $Z_k^{(n)}$, used for the extrinsic quantities $\vec{\mathcal{X}}$. The superscript $(n)$ indicates the term appears at $n$-th order in the expansion in mass ratio $\varepsilon$.  After undergoing the NIT, these quantities are denoted with two accents, a tilde denoting the NIT, and the overarrow as our vector shorthand for these sets.  For example, $\vec{\nit{P}}$ denotes the set of transformed principal orbit elements $(\nit{p},\nit{e},\nit{x}_I)$.

It will sometimes be useful to decompose functions into a Fourier series.  We use the convention
	\begin{equation} \label{eq:Fourier}
		A(\vec{P},\vec{Q}) = \sum_{\vec{\kappa} \in \mathbb{Z}^{j_{\text{max}}}} A_{\vec{\kappa}}(\vec{P}) e^{i \vec{\kappa} \cdot \vec{Q}}\;,
	\end{equation}
where $j_{\text{max}}$ is the number of phases, and $\vec{\kappa}$ is a vector of integers with $j_{\text{max}}$ components.  Any component of $\vec{\kappa}$ which attaches to the spin phase runs over the set $-1$, $0$, $1$; the other components run formally from $-\infty$ to $\infty$.  The dot product used in the exponent is the usual Euclidean, Cartesian one: $\vec{\kappa}\cdot\vec{Q} = \kappa_i Q_j \delta_{ij}$, where $\delta_{ij}$ is the identity.  Using this Fourier series, we can split $A(\vec{P}, \vec{Q})$ into an averaged piece $\avg{A} (\vec{P})$ given by
	\begin{align}\label{eq:average}
		\avg{A} (\vec{P}) &=  A_{\vec{0}} (\vec{P})
        \nonumber\\
        &= \frac{1}{(2\pi)^{j_{\text{max}}}} \idotsint_{\vec{Q}} A(\vec{P},\vec{Q})\,d q_1 \dots d q_{j_{\text{max}}}\;,
	\end{align}
and an oscillating piece given by
	\begin{equation}\label{eq:oscillating}
		\osc{A}(\vec{P},\vec{Q}) = A(\vec{P},\vec{q}) - \avg{A}(\vec{P})  = \sum_{\vec{\kappa} \neq \vec{0}} A_{\vec{\kappa}}(\vec{P}) e^{i \vec{\kappa} \cdot \vec{Q}}\;. 
	\end{equation}
Note that the Greek subscript with a vector accent (e.g., $A_{\vec \kappa}$) indicates a Fourier index, in contrast to a Latin subscript with no vector accent (e.g., $A_j$), which denotes a component of the vector.

\subsection{Mino-time formulation} 
\label{sec:MinoNIT}

We begin by writing down the form of the equations that we want to average. First observe that the rate of change of the spin phase is given by (\ref{eq:precphaseeqn}). We define the right-hand side of this equation as $f_s^{(0)}$:
\begin{align}
    \frac{d \psi_s}{d \lambda} &= \sqrt{\hat K} \left( \frac{(r^2 + a^2) \hat E - a \hat L_z}{\hat K + r^2} + a \frac{\hat L_z - a (1 - z^2) \hat E}{\hat K - a^2 z^2} \right) \nonumber
    \\ &\equiv f_s^{(0)}\;.
\end{align}
The phase $\psi_s$ has an analytic solution in the form
\begin{equation}
    \psi_s = \Upsilon_s^{(0)} \lambda + \psi_{sr}(q_r) +  \psi_{sz}(q_z)  \;,
\end{equation}
where $\Upsilon_s^{(0)}$ is the Mino-time spin frequency.  (We add the superscript $(0)$ to the various Mino-time geodesic frequencies when they are used in the NIT context, to emphasize that they do not include information about the secondary at $O(\varepsilon)$ or higher.)  Expressions for $\psi_{sr}(q_r)$ and $\psi_{sz}(q_z)$ can be found in Eqs.\ (57) and (58) of Ref.\ \cite{vandeMeent2019} where they are denoted $\psi_{r}(q_r)$ and  $\psi_{z}(q_z)$.

To post-adiabatic order, the equations of motion of the system can be written schematically as
	\begin{subequations}\label{eq:Generic_EMRI_EoM}
		\begin{align}
			\begin{split}
				\frac{d P_j}{d \lambda} &= \mr F_j^{(1)} (\vec{P}, \vec{q},\psi_s) + \mr^2 F_j^{(2)} (\vec{P}, \vec{q},\psi_s)  \;,\label{eq:Pjdot}
			\end{split}\\
			\begin{split}
				\frac{d q_i}{d \lambda} &=  \Upsilon_i^{(0)} (\vec{P}) +  \mr f_i^{(1)} (\vec{P}, \vec{q},\psi_s)\;,\label{eq:qidot}
			\end{split}\\
			\begin{split}
				\frac{d \psi_s}{d \lambda} &=  f_s^{(0)} (\vec{P}, \vec{q})\;, \label{eq:psidot}
			\end{split}\\
			\begin{split}
				\frac{d \mathcal{X}_k}{d \lambda} &= f_k^{(0)}(\vec{P}, \vec{q})\;.
			\end{split}
		\end{align}
	\end{subequations}
Here, the forcing terms are given by
\begin{subequations}
 \begin{equation}
     F_j^{(1)} = F_{j,\text{GSF}}^{(1)} (\vec{P},\vec{q}) + s F_{j,\text{SCF}}^{(1)}(\vec{P},\vec{q}, \psi_s)\;,
 \end{equation}
 \begin{equation}
     f_i^{(1)} = f_{i,\text{GSF}}^{(1)}(\vec{P},\vec{q}) + s f_{i,\text{SCF}}^{(1)}(\vec{P},\vec{q}, \psi_s)\;,
 \end{equation}
 \begin{equation}
     F_j^{(2)} = F_{j,\text{GSF}}^{(2)}(\vec{P},\vec{q})\;,
 \end{equation}
\end{subequations}
where $s$ is the spin of the secondary scaled such that $\|s \| \leq 1$ as discussed in Sec.\ \ref{sec:spinningorbits}.  The terms $F_{i,\text{GSF}}$ and $f_{i,\text{GSF}}$ are due to the gravitational self-force, while $F_{i,\text{SCF}}$ and $f_{i,\text{SCF}}$ are due to the spin-curvature force.  It is worth remarking that although these terms are derived from the gravitational self force and the spin-curvature force, they are not identical to these forces; they are essentially projections of certain components of these forces.

The averaged variables, $\nit{P}_j$, $\nit{q}_i$, $\nit{\psi}_s$, and $\nit{\mathcal{X}}_k$, are related to the OG variables  $P_j$, $q_i$, $\psi_s$, and $\mathcal{X}_k$ via
	\begin{subequations}\label{eq:transformation}
		\begin{align}
			\begin{split}\label{eq:transformation1}
				\nit{P}_j &= P_j + \sp Y_j^{(1)}(\vec{P},\vec{q},\nit{\psi}_s) + \sp^2 Y_j^{(2)}(\vec{P},\vec{q},\nit{\psi}_s) + \HOT{3}\;,
			\end{split}\\
			\begin{split}
				\nit{q}_i &= q_i + \sp X_i^{(1)}(\vec{P},\vec{q},\nit{\psi}_s) +\sp^2 X_i^{(2)}(\vec{P},\vec{q}\;,\nit{\psi}_s) + \HOT{3},
			\end{split}\\
			\begin{split}
				\nit{\psi}_s &= \psi_s + W_s^{(0)}(\vec{P},\vec{q}) + \sp W_s^{(1)}(\vec{P},\vec{q}\;,\nit{\psi}_s) + \HOT{2},
			\end{split}\\
			\begin{split}
				\nit{\mathcal{X}}_k &= \mathcal{X}_k + Z_k^{(0)}(\vec{P},\vec{q}) +\sp Z_k^{(1)}(\vec{P}\;,\vec{q}) + \HOT{2}.
			\end{split}
		\end{align}
	\end{subequations}
	As noted previously, the transformation functions $Y_j^{(n)}$, $X_i^{(n)}$, $W_s^{(n)}$, and $Z_k^{(n)}$ are smooth, periodic functions of the orbital phases $\vec{Q}$. At leading order, Eqs.~\eqref{eq:transformation} are identity transformations for $P_j$ and $q_i$, but not for $\mathcal{X}_k$ and $\psi_s$ due to the presence of zeroth-order transformation terms $Z_k^{(0)}$ and $W_s^{(0)}$ respectively. Details about the derivation of Mino-time quantities are given in Appendix \ref{app:NITMinoderiv} and a summary of relevant Mino-time definitions can be found in Appendix \ref{sec:summaryMinoNIT}. 
 
In summary, the equations of motion for the averaged variables $\nit{P}_j, \nit{q}_i$, $\nit{\psi}_s$, and $\nit{\mathcal{X}}_k$ take the form 
	\begin{subequations}\label{eq:transformed_EoM}
		\begin{align}
			\begin{split}
				\frac{d \nit{P}_j}{d \lambda} &=\varepsilon \nit{F}_j^{(1)}(\vec{\nit{P}}) + \varepsilon^2 \nit{F}_j^{(2)}(\vec{\nit{P}}) +  \mathcal{O}(\varepsilon^3)\;,
			\end{split}\\
			\begin{split}
				\frac{d \nit{q}_i}{d\lambda} &= \Upsilon_i^{(0)}(\vec{\nit{P}}) +\varepsilon \Upsilon_i^{(1)}(\vec{\nit{P}}) + \mathcal{O}(\varepsilon^2)\;,
			\end{split}\\
			\begin{split}
				\frac{d \nit{\psi}_s}{d\lambda} &= \Upsilon_s^{(0)}(\vec{\nit{P}})+ \mathcal{O}(\varepsilon)\;,
			\end{split}\\
			\begin{split}
				\frac{d \nit{\mathcal{X}}_k}{d \lambda} &= \Upsilon_k^{(0)}(\vec{\nit{P}}) + \varepsilon \Upsilon_k^{(1)}(\vec{\nit{P}}) +  \mathcal{O}(\varepsilon^2)\;.
			\end{split}
		\end{align}
	\end{subequations}
The explicit forms for $\nit{F}_j^{(1)}$, $\nit{F}_j^{(2)}$, $\Upsilon_i^{(1)}$, and $\Upsilon_k^{(1)}$ can be found in Appendix \ref{sec:summaryMinoNIT}.
 
Crucially, the NIT equations of motion \ref{eq:transformed_EoM} are independent of the orbital phases $\vec{Q}$, meaning these differential equations are fast to evaluate. Another crucial point is that, in the extreme mass ratio limit $\varepsilon \rightarrow 0$, the solutions to the NIT equations \ref{eq:transformed_EoM} tend to the solutions for OG equations \ref{eq:Generic_EMRI_EoM}. 

\subsection{Boyer-Lindquist-time formulation}
\label{sec:BLNIT}

The above equations of motion \ref{eq:transformed_EoM} are parameterized in terms of Mino time $\lambda$. It is significantly more convenient for waveform generation purposes to have equations of motion parameterized in terms of Boyer-Lindquist time.  Thus, we perform a second averaging transformation as first outlined in Ref.~\cite{Pound2021} and implemented in Refs.~\cite{Lynch2022,Lynch2023}.

We relate the Mino-time averaged variables $\vec{\nit{P}}=(\nit{p},\nit{e}_,\nit{x}_I)$ and $\vec{\nit{\mathcal{Q}}}= ( \nit{q}_r, \nit{q}_z, \nit{\psi}_s,\nit{\phi} )$ to the Boyer-Lindquist-time averaged variables $\vec{\mathcal{P}}=( p_\varphi,e_\varphi,x_\varphi)$ and $\vec{\varphi}=(\varphi_r, \varphi_z, \varphi_s,\varphi_\phi )$ via:
\begin{subequations}\label{eq:t_param_transformation}
	\begin{align}
	\begin{split}\label{eq:transformation2}
		\mathcal{P}_j &= \nit{P}_j + \sp \Pi_j^{(1)}(\vec{\nit{P}},\vec{\nit{q}}) + \sp^2 \Pi_j^{(2)}(\vec{\nit{P}},\vec{\nit{q}}) + \HOT{3}\;,
	\end{split}\\
	\begin{split}
		\varphi_i &= \nit{\mathcal{Q}}_i + \Delta \varphi_i +  \sp \Psi_i^{(1)}(\vec{\nit{P}},\vec{\nit{q}}) + \HOT{2}\;,
	\end{split}
	\end{align}
\end{subequations}
where $\Delta \varphi_i= \Omega_i^{(0)}(\vec{\nit{P}}) \Delta t^{(0)}$ and $\Omega_i^{(0)}$ is the Boyer-Lindquist fundamental frequency of the tangent geodesic.

To obtain the equations of motion for $\vec{\mathcal{P}}$ and $\vec{\varphi}$, we take the time derivative of Eq.~\eqref{eq:t_param_transformation}, substitute the expression for the NIT equations of motion, and then use the inverse transformation of Eq.\ \eqref{eq:t_param_transformation} to ensure that all functions are expressed in terms of $\vec{P}$ and $\vec{\nit{q}}$.  We then expand order by order in $\varepsilon$.  We chose the oscillatory functions $\Delta t$, $\Psi_i^{(1)}$, $\Pi_j^{(1)}$, and  $\Pi_j^{(2)}$ in order to cancel out any oscillatory terms that appear at each order in $\mr$.  This results in averaged equations of motion that take the following form: 
\begin{subequations}\label{eq:t_transformed_EoM}
	\begin{align}
	\begin{split}
		\frac{d \mathcal{P}_j}{dt} &= \varepsilon \Gamma_j^{(1)}(\vec{\mathcal{P}}) + \varepsilon^2 \Gamma_j^{(2)}(\vec{\mathcal{P}}) +  \mathcal{O}(\varepsilon^3)\;,
	\end{split}\\
	\begin{split} \label{eq:Phase_Transformation}
		\frac{d\varphi_\alpha}{dt} &= \Omega^{(0)}_\alpha(\vec{\mathcal{P}}) +\varepsilon \Omega_\alpha^{(1)}(\vec{\mathcal{P}}) + \mathcal{O}(\varepsilon^2)\;.
	\end{split}
	\end{align}
\end{subequations}

These equations of motion are related to the Mino time averaged equations of motion~\eqref{eq:transformed_EoM} with the adiabatic terms given by
\begin{subequations}
	\begin{align}
		\Gamma^{(1)}_j = \frac{\nit{F}_j^{(0)}}{\Upsilon_t^{(0)}}\;, \quad \Omega^{(0)}_\alpha = \frac{\Upsilon_\alpha^{(0)}}{\Upsilon_t^{(0)}}\;,
		\tag{\theequation a-b}
	\end{align}
\end{subequations}
and the post-adiabatic terms given by
\begin{subequations}
	\begin{align}
		\begin{split}
		\Gamma^{(2)}_j & = \frac{1}{\Upsilon_t} \Big( 
		\nit{F}^{(2)}_j + \nit{F}^{(1)} \frac{\partial}{\partial P_j} \avg{\Pi^{(1)}_j} \\
		& - \avg{f_t^{(0)} \Pi_k^{(1)}} \PD{\Gamma_j^{(1)}}{P_k} - \Upsilon_t^{(1)} \Gamma^{(1)}_j \Big)\;,
		\end{split}\\
		\begin{split}
			\Omega^{(1)}_\alpha & = \frac{1}{\Upsilon_t^{(0)}} \Big( \Upsilon^{(1)}_\alpha + \nit{F}^{(1)}_j \avg{\PD {\Delta \varphi_\alpha}{P_j}} \\
			& - \avg{f_t^{(0)} \Pi_k^{(1)}} \PD{\Omega_\alpha^{(0)}}{P_k}  - \Upsilon_t^{(1)} \Omega^{(1)}_i \Big)\;.
		\end{split}
	\end{align}
\end{subequations}
This constrains the oscillating pieces of our transformation to be 
\begin{subequations}
	\begin{align}
	\begin{split} \label{eq:Delta_t}
		\Delta t = &\sum_{\kappa \neq 0} \frac{f_{t,\vec{\kappa}}^{(0)}}{ -i \vec{\kappa} \cdot \vec{\Upsilon}^{(0)}} = -\osc{Z}_t^{(0)}\;,
	\end{split}\\
	\begin{split} \label{eq:leading_order_t_NIT}
		\osc{\Pi}^{(1)}_j = &\sum_{\kappa \neq 0} \frac{f_{t,\vec{\kappa}}^{(0)}}{ -i \vec{\kappa} \cdot \vec{\Upsilon}^{(0)}} \Gamma_j^{(1)} = -  \osc{Z}_t^{(0)} \Gamma_j^{(1)}\;, \text{ and}
	\end{split}\\
	\begin{split}
		\Psi^{(1)}_{\alpha,\vec{\kappa}} = & \frac{i}{\vec{\kappa} \cdot \vec{\Upsilon}^{(0)} }\Biggl( 
		\PD{\Delta \varphi_{\alpha,\vec{\kappa}}}{P_j} \nit{F}_j^{(1)}  - \frac{f^{(0)}_{t,\vec{\kappa}}}{\Upsilon^{(0)}_t} \Upsilon^{(1)}_t \Omega^{(0)}_t
		\\& +\sum _{\vec{\kappa}' \neq \vec{0}} \Biggl[ \left( i \vec{\kappa}' \cdot \vec{X}^{(1)}_{\vec{\kappa} - \vec{\kappa}'} f_{t,\vec{\kappa}'}^{(0)} + Y^{(1)}_{j,\vec{\kappa} - \vec{\kappa}'} \PD{f^{(0)}_{t,\vec{\kappa}'}}{P_j} \right) \Omega^{(0)}_\alpha 
		\\& - \Pi^{(1)}_{j,\vec{\kappa} - \vec{\kappa}'} f_{t, \vec{\kappa}'}^{(0)} \PD{\Omega^{(0)}_\alpha}{P_j} \Biggr] \Biggr)\;.
	\end{split}
	\end{align}
\end{subequations}

We are free to chose the averaged pieces of $\Pi_j^{(1)}$, and we make the simplification that $\avg{\Pi_j^{(1)}} = 0$. 
With this and the identity $\avg{ f_t^{(0)} (\int{f}_t^{(0)} d \vec{q} )} = 0 $, we get the simplification $\avg{f_t^{(0)} \Pi_j^{(1)}} = 0$.  The expressions for $\Gamma^{(2)}_j$ and $\Omega_\alpha^{(1)}$ then simplify to
\begin{subequations}
	\begin{align}
	\begin{split}
		\Gamma^{(2)}_j = \frac{1}{\Upsilon_t^{(0)}} \Big( 
		\nit{F}^{(2)}_j -\Upsilon_t^{(1)} \Gamma^{(1)}_j \Big)\;,
	\end{split}\\
	\begin{split}
			\Omega^{(1)}_\alpha = \frac{1}{\Upsilon_t^{(0)}} \Big( \Upsilon^{(1)}_\alpha -\Upsilon_t^{(1)}\Omega^{(0)}_\alpha \Big)\;.
	\end{split}
	\end{align}
\end{subequations}

A useful aspect of these equations of motion is that their solutions $\vec{\mathcal{P}}(t)$ and $\vec{\varphi}(t)$ are exactly what is required to feed into waveform generating schemes, as shown in Appendix B of \cite{Lynch2022}.  Once these solutions are constructed, it is then straightforward to augment adiabatic waveform construction schemes \cite{Hughes2021, Chua2021, Katz2021} to include the post-adiabatic effects these solutions describe.  It is also worth noting that the additional averaging associated with Boyer-Lindquist time could be circumvented by using closed-form expressions for the geodesic orbits in terms of action angles associated with Boyer-Lindquist frequencies, i.e., $\vec{\varphi}$. This has been achieved already for bound orbits in Schwarzschild spacetime via a small eccentricity expansion \cite{Witzany2022}.

\subsection{Averaged spinning-body equations of motion}
\label{sec:AveragedEoM}

\begin{figure*}
\includegraphics[scale=0.55]{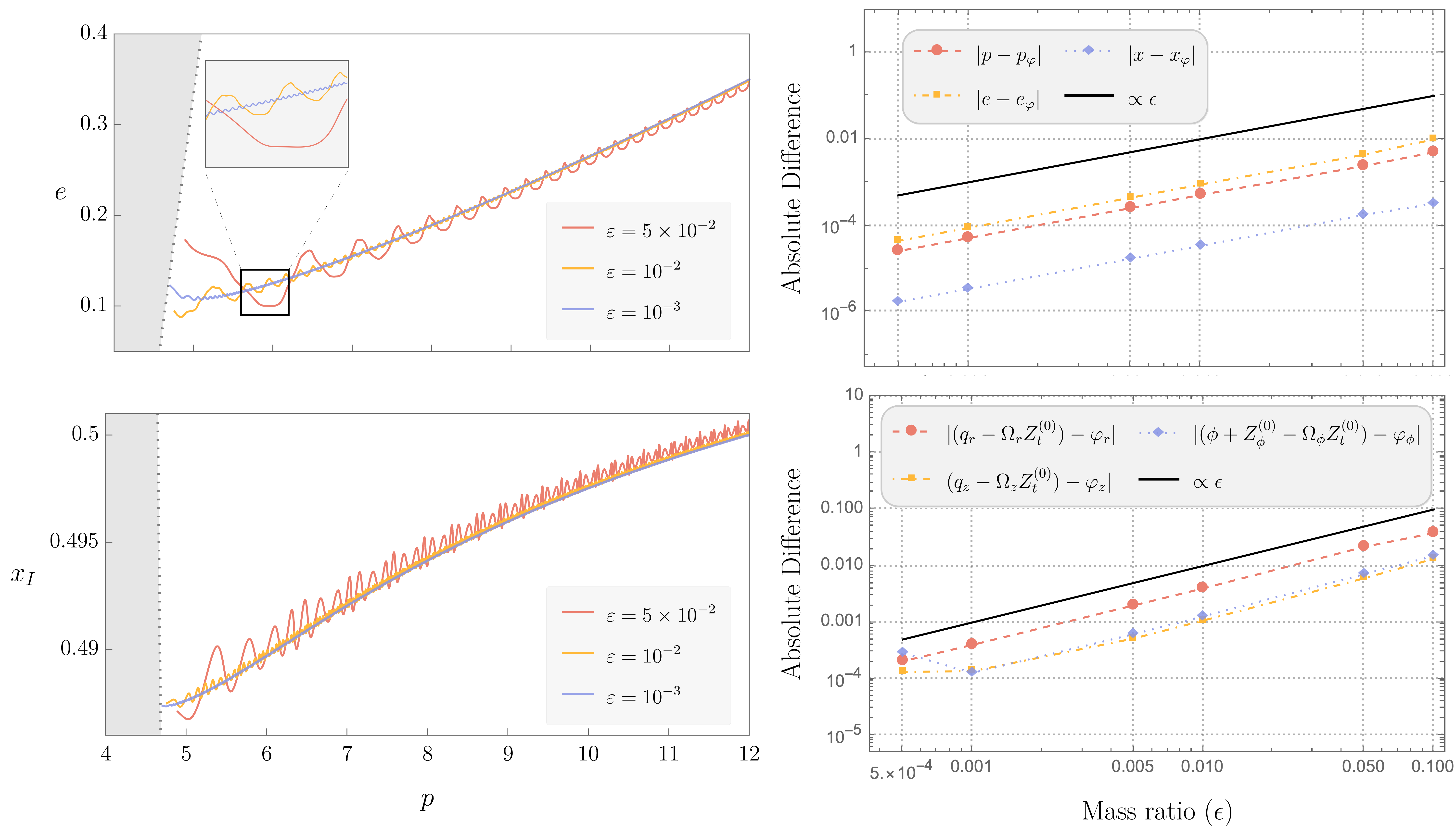}
    \caption{Spinning-body inspiral for different mass ratios.  Left-hand panels show inspirals obtained using the OG equations of motion for mass ratios $\varepsilon = 5\times10^{-2}$ (red; large oscillations), $\varepsilon =10^{-2}$ (yellow; medium oscillations) and $\varepsilon = 10^{-3}$ (blue, small oscillations).  We again note that these mass ratios are larger than those expected for EMRI systems, and are used here in order to amplify the impact of spin-curvature coupling for visual purposes.  The initial parameters used are $p = 12$, $e = 0.35$, $x_I = 0.5$, $q_r = 0$, and  $q_z = 0$.  Right-hand panels show the absolute difference in orbital elements of a spinning-body inspiral comparing the OG and NIT methods; NIT orbital elements are labeled with subscript $\varphi$.  In these right-hand panels, we initially set $e = 0.22$, $x_I = 0.699$, $q_r = 0$, and $q_z = 0$.  Data shown corresponds to the system evolving from $p = 9.45$ to $p = 9$.  As expected, the absolute differences track with the $\varepsilon$ curve (solid, black).  For all data in this figure, the small body orbits a black hole with spin $a = 0.7 M$ and the magnitude and orientation of the small body's spin is specified by $s = 1$, $s_\parallel = s$.}
    \label{fig:diffepsilon}
\end{figure*}

In the previous sections, we derived equations of motion to post-adiabatic order by assuming that the gravitational self-force is known to $O(\varepsilon^2)$.  As of now, it is only feasible to mass produce data describing the leading-order dissipative radiation reaction via flux balance laws (and this has only been done so far for a fairly limited range of parameters).  Although tools exist to compute more of the first-order GSF \cite{vandeMeent2018}, doing so is computationally expensive, and the second-order GSF for generic Kerr remains far off. This means that we set the second-order corrections to zero, $F^{(2)}_{j,\text GSF} = 0$, and we have no conservative contributions from the self-force, $f^{(1)}_{i,\text GSF} = 0$.

The other force driving the evolution is the spin-curvature force which has no dissipative effects.  As such, its orbit average is zero and so the terms which change the principal orbit elements, $F^{(1)}_{j,\text{SCF}}$, vanish on average: $\avg{F^{(1)}_{j,\text{SCF}}} = 0$. The resulting averaged equations of motion parameterized in Mino-time $\lambda$ are given by:
\begin{align}
\begin{split}
    \frac{d \tilde{p}}{ d \lambda} &= \mr  \nit{F}_p^{(1)}(\nit{p},\nit{e}, \nit{x}_I) \;,
\end{split}\\
\begin{split}
    \frac{d \tilde{e}}{ d \lambda} &= \mr  \nit{F}_e^{(1)}(\nit{p},\nit{e}, \nit{x}_I)\;,
\end{split}\\
\begin{split}
    \frac{d \tilde{x}_I}{ d \lambda} &= \mr  \nit{F}_x^{(1)}(\nit{p},\nit{e}, \nit{x}_I)\;,
\end{split}\\
\begin{split}
    \frac{d \tilde{q}_r}{ d \lambda} &= \Upsilon_r^{(0)}( \nit{p},\nit{e}, \nit{x}_I) + \mr s \Upsilon_r^{(1)}(\nit{p},\nit{e}, \nit{x}_I)\;,
\end{split}\\
\begin{split}
    \frac{d \tilde{q}_z}{ d \lambda} &= \Upsilon_z^{(0)}(\nit{p},\nit{e}, \nit{x}_I) + \mr s \Upsilon_z^{(1)}(\nit{p},\nit{e}, \nit{x}_I)\;,
\end{split}\\
\begin{split}
    \frac{d \tilde{\phi}}{ d \lambda} &= \Upsilon_\phi^{(0)}(\nit{p},\nit{e}, \nit{x}_I) + \mr s \Upsilon_\phi^{(1)}(\nit{p},\nit{e}, \nit{x}_I)\;,
\end{split}\\
\begin{split}
    \frac{d \tilde{t}}{ d \lambda} &= \Upsilon_t^{(0)}(\nit{p},\nit{e}, \nit{x}_I) + \mr s \Upsilon_t^{(1)}( \nit{p},\nit{e}, \nit{x}_I)\;,
\end{split}\\
\begin{split}
    \frac{d \tilde{\psi}_s}{ d \lambda} &= \Upsilon_s^{(0)}(\nit{p},\nit{e}, \nit{x}_I) \;.
\end{split}
\end{align}

Many of these terms are simply related to the transformed force terms averaged over a single orbit, which are as follows:
\begin{subequations} \label{eq:Mino_Time_NIT_Terms}
\begin{equation} \label{eq:NIT_Relationship1}
	\nit{F}_p^{(1)} = \avg{F^{(1)}_{p,\text{GSF}}},\quad \nit{F}_{e}^{(1)} = \avg{F^{(1)}_{e,\text{GSF}}}\quad \nit{F}_{x}^{(1)} = \avg{F^{(1)}_{x,\text{GSF}}}, \tag{\theequation a-c}
\end{equation}
\begin{equation} \label{eq:NIT_Relationship2}
	\Upsilon_r^{(1)}= \avg{f^{(1)}_{r,\text{SCF}}}, \quad \Upsilon_z^{(1)}= \avg{f^{(1)}_{z,\text{SCF}}},  \tag{\theequation d-e}
\end{equation}
\begin{equation}\label{eq:NIT_Relationship3}
	\Upsilon_s^{(0)} =  \avg{f_{s}^{(0)}}\;, \quad \Upsilon_\phi^{(0)} =  \avg{f_{\phi}^{(0)}}\;,  \quad \Upsilon_t^{(0)} =  \avg{f_{t}^{(0)}} \tag{\theequation f-h},
\end{equation}
\end{subequations}
where $\Upsilon_s^{(0)}$, $\Upsilon_\phi^{(0)}$, and $\Upsilon_t^{(0)}$ are the Mino-time precession, azimuthal, and time  fundamental frequencies respectively which are known analytically \cite{FujitaHikida2009,vandeMeent2019}.
The remaining terms are more complicated and are given in terms of an operator $\mathcal{N}$ which we define in Appendix~\ref{sec:NITOperator}.  These remaining terms are given by:
\begin{equation}\label{eq:NIT_Relationship4}
	\begin{split}
	  \Upsilon_s^{(1)} =  \mathcal{N}(f_s^{(0)})\;, \quad \Upsilon_\phi^{(1)} =  \mathcal{N}(f_\phi^{(0)})\;, \quad \Upsilon_t^{(1)} =  \mathcal{N}(f_t^{(0)})\;.
	\end{split}
	\tag{\theequation i-k}
\end{equation}

The leading order near-identity transformation for the orbital elements needed for the initial conditions is given by:
\begin{align}\label{eq:NIT_Y_Generic_Spin}
    \begin{split}
		& \osc{Y}_j^{(1)}  \equiv \sum_{(\kappa_r,\kappa_z)\neq (0,0)} \frac{i F_{j,\text{GSF},\kappa_r,\kappa_z}^{(1)}}{\kappa_r \Upsilon^{(0)}_r + \kappa_z \Upsilon^{(0)}_z}  e^{i (\kappa_r q_r + \kappa_z q_z)} \\
   & + \sum_{(\kappa_r, \kappa_z,\kappa_s) \neq (0,0,0)} \frac{i s F_{j,\text{SCF},\kappa_r,\kappa_z, \kappa_s}^{(1)}}{\kappa_r \Upsilon^{(0)}_r + \kappa_z \Upsilon^{(0)}_z + \kappa_s \Upsilon^{(0)}_s} \\ & \times e^{i (\kappa_r q_r + \kappa_z q_z + \kappa_s \psi_s)}\;.
  \end{split}
\end{align}

With this all in hand, we can now derive the averaged equations of motion parameterized by Boyer-Lindquist time $t$ for the phases $\vec{\varphi} = \{\varphi_r, \varphi_z,\varphi_\phi\, \varphi_s \}$ and orbital elements $\vec{\mathcal{P}} = \{ p_\varphi,e_\varphi, x_\varphi\}$ in form
\begin{subequations}
	\begin{align}
		\begin{split}
			\frac{d p_\varphi}{dt} &= \mr \Gamma_p^{(1)}(p_\varphi,e_\varphi,x_\varphi) + \mr^2  \Gamma_p^{(2)}(p_\varphi,e_\varphi,x_\varphi)\;,
		\end{split}\\
            \begin{split}
			\frac{d e_\varphi}{dt} &= \mr \Gamma_e^{(1)}(p_\varphi,e_\varphi, x_\varphi) + \mr^2  \Gamma_e^{(2)}(p_\varphi,e_\varphi,x_\varphi)\;,
		\end{split}\\
		\begin{split}
			\frac{d x_{\varphi}}{dt} &= \mr \Gamma_x^{(1)}(p_\varphi,e_\varphi, x_{I,\varphi}) + \mr^2 \Gamma_x^{(2)}(p_\varphi,e_\varphi,x_\varphi)\;,
		\end{split}\\
            \begin{split}
			\frac{d\varphi_r}{dt} &= \Omega^{(0)}_r(p_\varphi,e_\varphi,x_\varphi) +\mr s \Omega_r^{(1)}(p_\varphi,e_\varphi,x_\varphi)\;,
		\end{split}\\
		\begin{split}
			\frac{d\varphi_z}{dt} &= \Omega^{(0)}_z(p_\varphi,e_\varphi,x_\varphi) +\mr s \Omega_z^{(1)}(p_\varphi,e_\varphi,x_\varphi)\;,
		\end{split}\\
		\begin{split}
			\frac{d\varphi_\phi}{dt} &= \Omega^{(0)}_\phi(p_\varphi,e_\varphi,x_\varphi) +\mr s \Omega_\phi^{(1)}(p_\varphi,e_\varphi,x_\varphi)\;,
		\end{split}\\
        \begin{split}
			\frac{d\varphi_s}{dt} &= \Omega^{(0)}_s(p_\varphi,e_\varphi,x_\varphi)\;.
		\end{split}
	\end{align}
\end{subequations}
The leading order terms in these equations are given by
\begin{subequations}\label{eq:t_NIT_Terms1}
\begin{equation} \label{eq:t_NIT_Relationship1}
	\nit{\Gamma}_p^{(1)} = \nit{F}_p^{(1)}/\Upsilon_t^{(0)}, \quad  \nit{\Gamma}_e^{(1)} = \nit{F}_p^{(1)}/\Upsilon_t^{(0)}, \quad  \nit{F}_x^{(1)} = \nit{F}_x^{(1)}/\Upsilon_t^{(0)}\;,\tag{\theequation a-c}
\end{equation}
\begin{equation}\label{eq:t_NIT_Relationship2}
	\nit{\Omega}_r^{(0)} =  \Upsilon_r^{(0)} /\Upsilon_t^{(0)},\quad \nit{\Omega}_z^{(0)} =  \Upsilon_z^{(0)} /\Upsilon_t^{(0)}, \tag{\theequation d-e}
\end{equation}
\begin{equation}\label{eq:t_NIT_Relationship3}
	\nit{\Omega}_\phi^{(0)} =  \Upsilon_\phi^{(0)} /\Upsilon_t^{(0)},\quad \nit{\Omega}_s^{(0)} =  \Upsilon_s^{(0)} / \Upsilon_t^{(0)}\;. \tag{\theequation f-g}
\end{equation}
\end{subequations}
The sub-leading terms are given by 
\begin{subequations} \label{eq:t_NIT_Terms2}
	\begin{align}
		\begin{split}
			\Gamma^{(2)}_p = -\Upsilon_t^{(1)} \Gamma^{(1)}_p / \Upsilon_t^{(0)}\;,
		\end{split}\\
            \begin{split}
			\Gamma^{(2)}_e =  -\Upsilon_t^{(1)} \Gamma^{(1)}_e / \Upsilon_t^{(0)}\;,
		\end{split}\\
		\begin{split}
			\Gamma^{(2)}_x = -\Upsilon_t^{(1)} \Gamma^{(1)}_x / \Upsilon_t^{(0)}\;,
		\end{split}\\
            \begin{split}
			\Omega^{(1)}_r = \frac{1}{\Upsilon_t^{(0)}} \Big( \Upsilon^{(1)}_r -\Upsilon_t^{(1)}\Omega^{(0)}_r \Big)\;, \label{OmegarNIT}
		\end{split}\\
		\begin{split}
			\Omega^{(1)}_z = \frac{1}{\Upsilon_t^{(0)}} \Big( \Upsilon^{(1)}_z -\Upsilon_t^{(1)}\Omega^{(0)}_z \Big)\;, \label{OmegazNIT}
		\end{split}\\
		\begin{split}
			\Omega^{(1)}_\phi = \frac{1}{\Upsilon_t^{(0)}} \Big( \Upsilon^{(1)}_\phi -\Upsilon_t^{(1)}\Omega^{(0)}_\phi \Big)\;. \label{OmegaphiNIT}
		\end{split}
	\end{align}
\end{subequations}

The aligned spin case has equations in the same form as in the arbitrarily oriented case. The main difference is that we no longer have to evolve the precession phase $\psi_s$ or $\varphi_s$.  The other consequence is that the leading order NIT for the orbital elements reduces to 
\begin{align}
\label{eq:NIT_Y_Aligned_Spin}
    \begin{split}
		 \osc{Y}_j^{(1)}  \equiv & \sum_{(\kappa_r,\kappa_z)\neq (0,0)} \frac{i \left( F_{j,\text{GSF},\kappa_r,\kappa_z}^{(1)} + s F_{j,\text{SCF},\kappa_r,\kappa_z}^{(1)} \right)}{\kappa_r \Upsilon^{(0)}_r + \kappa_z \Upsilon^{(0)}_z} \\ & \times  e^{i (\kappa_r q_r + \kappa_z q_z)}\;.
  \end{split}
\end{align} 

The difference between the OG and averaged quantities scales linearly with the mass ratio as can be seen in Fig.\  \ref{fig:diffepsilon}. See Appendix \ref{app:initialconditions} for a discussion of the choice of initial conditions in the context of OG and NIT inspirals.

\subsection{Implementation}
\label{sec:NITImplement}

To implement the NIT procedure in practice, we must perform a series of offline steps.  We first generate a grid to cover a section of the 4-dimensional Kerr parameter space that we wish to examine.  We fix $a/M = 0.7$, and choose our principal elements $P_j = (p, e, x_I)$ in the range from $P_{j,\text{min}}$ to $P_{j, \text{max}}$ in steps of $P_{j, \text{step}}$.  For all the analyses we present in this paper, we use $e_\text{min} = 0.05$, $e_\text{max} = 0.22$, $e_\text{step} = 0.005$, and $x_{I, \text{min}} = 0.69$, $x_{I, \text{max}} = 0.701$, $x_{I, \text{step}} = 0.001$.  The resolution we use in $p$ varies depending on our goal.  For the convergence study in Fig.\ \ref{fig:diffepsilon}, we use $p_\text{min} = 9$, $p_\text{max} = 9.5$, $p_\text{step} = 0.002$; for calculating the full trajectory, we use a coarser grid that covers a wider range of parameter space: $p_\text{min} = 3.2$, $p_\text{max} = 10$, $p_\text{step} = 0.02$.  We select this region in order to avoid low order transient resonances\footnote{Note that transient self-forced resonances are not a concern in this work because we do not include self-force terms that would produce them in this analysis.  Such terms are likely to be incorporated in the future.} where our NIT procedure breaks down, though methods for dealing with resonances have been developed elsewhere \cite{Lynch2022}.

At each point in this grid, we use a fast Fourier transform to numerically decompose the OG functions into Fourier modes, and then sum them together in accordance with Eqs.\ \eqref{eq:Mino_Time_NIT_Terms}, \eqref{eq:NIT_Y_Generic_Spin}, \eqref{eq:t_NIT_Terms1}, and \eqref{eq:t_NIT_Terms1} to produce the averaged terms needed in our NIT equations of motion and the modes of the leading order transformation terms needed to set the initial conditions.  These data are then interpolated using Hermite polynomials with \textit{Mathematica}'s \texttt{Interpolate} function. Overall, these offline steps take about $3$ hours running in parallel (10 cores) using $3$-GHz-class Apple M1 processors.

By contrast, the online steps are computationally cheap.  One loads the interpolants produced by the offline analysis, sets initial conditions using Eqs.~\eqref{eq:NIT_ICs}, \eqref{eq:Phase_ICs}, and \eqref{eq:Orbital_ICs}, and then numerically solves the equations using \textit{Mathematica}'s \texttt{NDSolve}.  The resulting equations of motion can then be solved in less than a second, regardless of mass ratio.  This is compared with the minutes to multiple hours (depending on mass ratio) required by the OG method.  In the supplementary material, we provide the interpolants, radiation-reaction data, and a \textit{Mathematica} notebook to rapidly compute this trajectory.

\section{Results I: Inspirals}
\label{sec:res_inspirals}

We present our results in two parts: the inspirals we find combining spin-curvature coupling with radiation reaction (this section), and the waveforms produced by those inspirals (following section).

\subsection{Aligned spin}
\label{sec:res_inspalign}

\begin{figure}
\centerline{\includegraphics[scale=0.64]{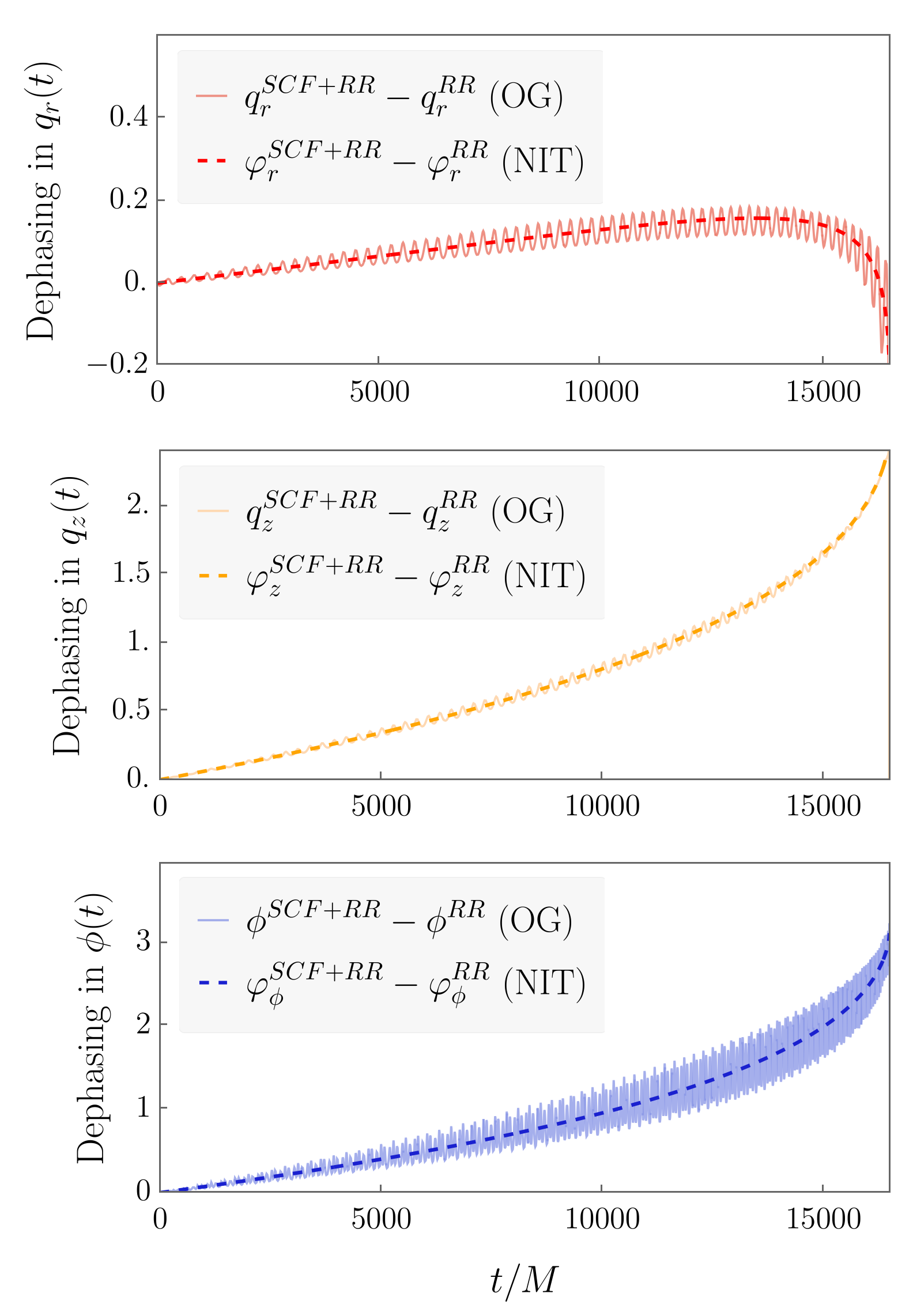}}
    \caption{Dephasing in $q_r(t)$, $q_z(t)$, and  $\phi(t)$ for a spinning body relative to a non-spinning body with mass ratio $\varepsilon = 10^{-2}$ orbiting a black hole with spin $a = 0.7 M$. The magnitude and orientation of the small body's spin is specified by $s=1$, $s_\parallel=s$.  Dashed lines show the dephasing computed using the NIT; solid lines show the dephasing given by the OG equations.  Top panel shows the radial dephasing $q_r^{SCF+RR} - q_r^{RR}$ (red), middle shows dephasing in the polar angle $q_z^{SCF+RR} - q_z^{RR}$ (yellow), and bottom shows dephasing in axial angle $\phi^{SCF+RR} - \phi^{RR}$ (blue).  In all panels, solid lines show the OG computation, dashed shows the NIT results.  The inspiral used for all panels has the initial conditions $p = 10$, $e = 0.2$, $x_I = 0.7$, $q_r = 0$, $q_z = 0$, and $\phi = 0$.}
    \label{fig:OGNITphases}
\end{figure}

\begin{figure*}
\centerline{\includegraphics[scale=0.55]{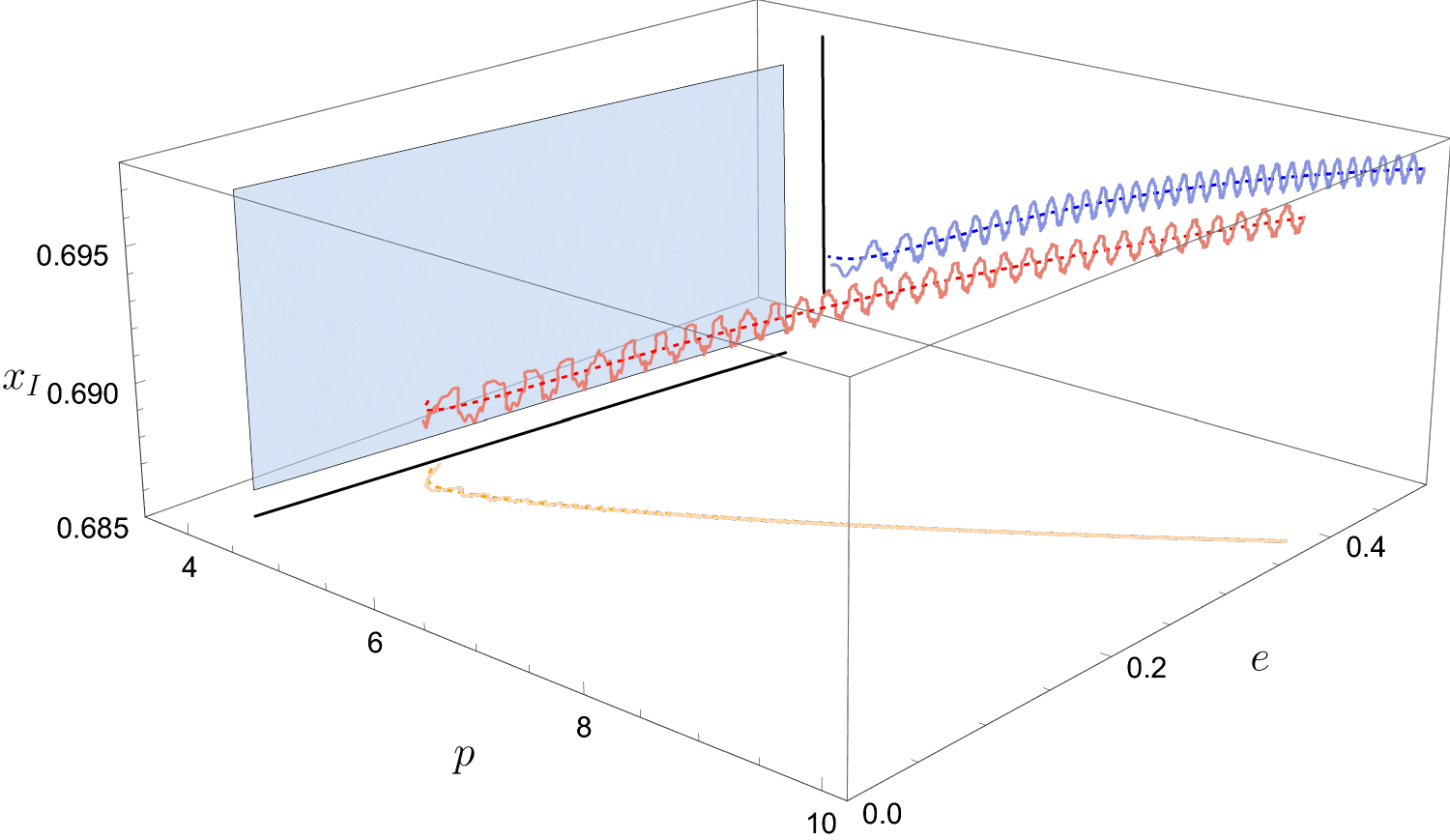}}
    \caption{The trajectory in $p$-$e$-$x_I$ space for an example generic inspiral. This inspiral (red curve) begins at $(p, e, x_I) = (10, 0.38, 0.6967)$ and ends at the LSO (the light blue plane).  The dashed curves show a non-spinning body's inspiral; solid curves are for the inspiral of a spinning small body.  The orange curves show the projection of the inspiral onto the $p$-$e$ plane; the solid black line in this plane is the projection of the last stable orbit at the final value of $x_I$.  (This projection is the same as the top panel of Fig.\ \ref{fig:2Dinspiral}.)  The blue curves show the projection of the inspiral onto the $p$-$x_I$ plane; the solid black curve in this plane is the projection of the LSO at the final value of $e$.  (This projection is the same as the middle panel of Fig.\ \ref{fig:2Dinspiral}.)  We use mass-ratio $\varepsilon = 0.005$ and small-body spin $s = 1$, with $s_\parallel = 0.9$ and $\phi_s = \pi/2$.  See Fig.\  15 in Ref.\ \cite{Hughes2021} for comparison.}
    \label{fig:3Dinspiral}
\end{figure*}

\begin{figure*}
\centerline{\includegraphics[scale=0.47]{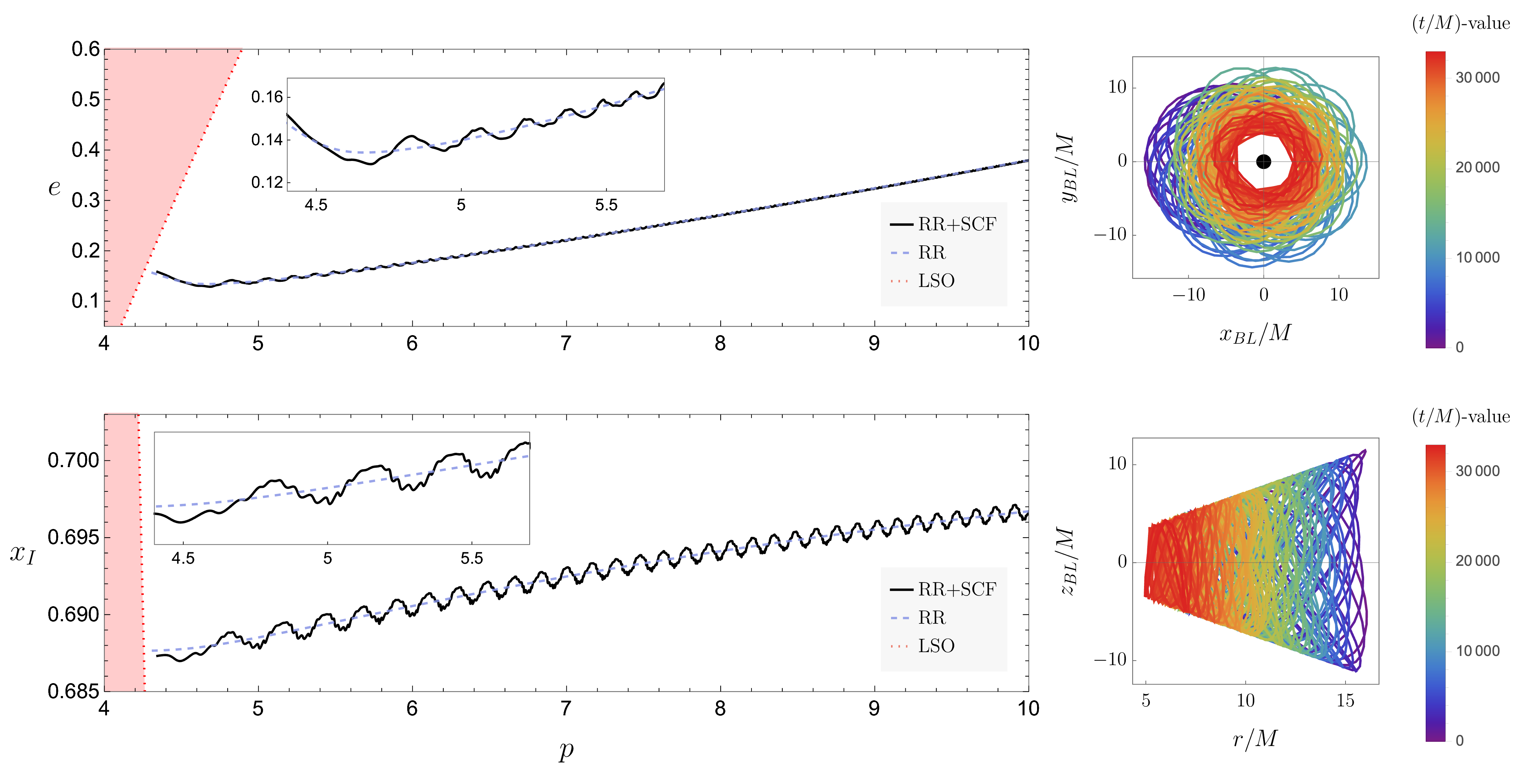}}
    \caption{Evolution of $p$ versus $e$ (top left) and evolution of $p$ versus $x_I$ (bottom left) for the inspiral shown in Fig.\ \ref{fig:3Dinspiral}.  Solid black curves show spinning body inspiral; blue dashed curves show non-spinning body inspiral.  In both plots, the last stable orbit (LSO) is shown by the red dotted curve.  The insets show close-ups of inspiral near the LSO.  Right-hand panels show projections of the worldline onto the $x_{BL}$-$y_{BL}$ and $r$-$z_{BL}$ planes (where $x_{BL}$, $y_{BL}$, $z_{BL}$ are Cartesian-like representations of Boyer-Lindquist coordinates: $x_{BL} = r\sin\theta\cos\phi$, etc.), with color encoding the time evolution (early times in purple and late times in red). Parameters are identical to those used in Fig.\ \ref{fig:3Dinspiral}.}
    \label{fig:2Dinspiral}
\end{figure*}

\begin{figure}[!]
\centerline{\includegraphics[scale=0.55]{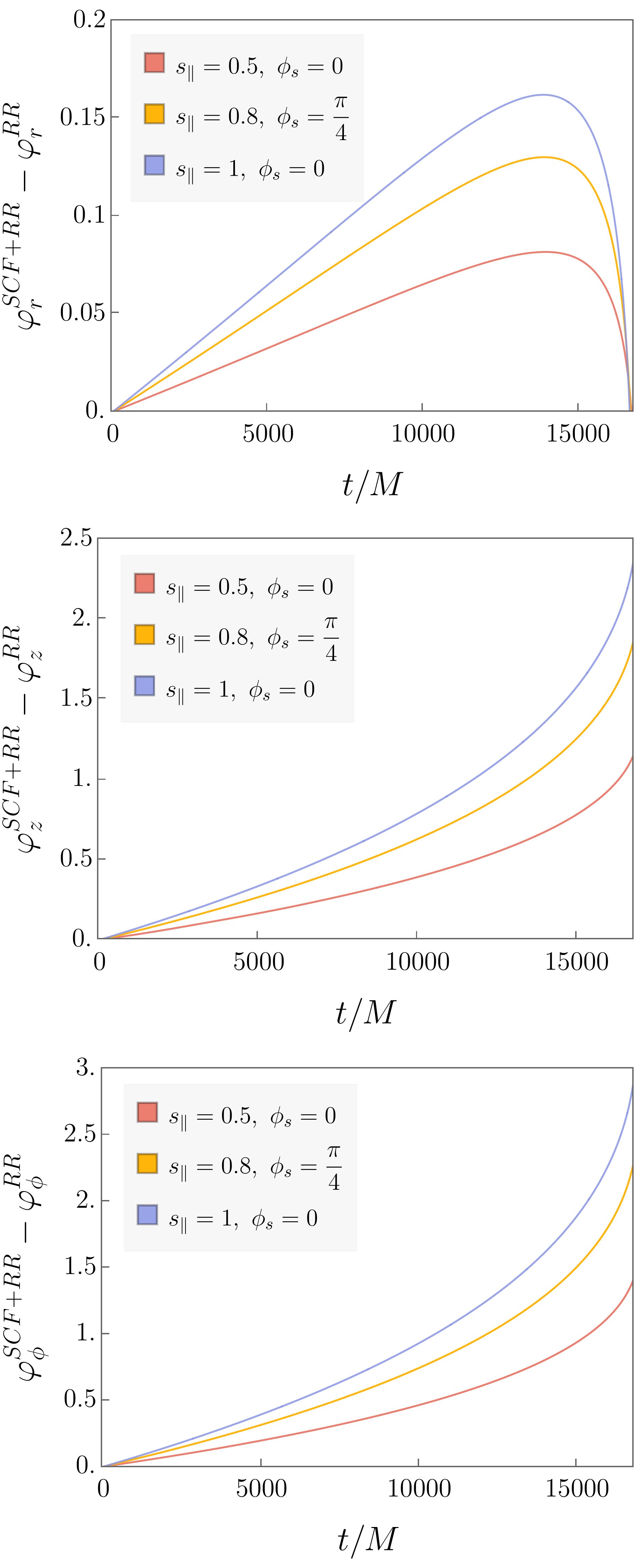}}
\caption{The averaged dephasing of $\varphi_r(t)$, $\varphi_z(t)$, and  $\varphi_\phi(t)$ for a small body with a misaligned spin vector relative to a non-spinning body for three different values of spin alignment: $s_\parallel=1$ (blue), $s_\parallel=0.8$ (orange) and $s_\parallel=0.5$ (red). The magnitude of the small body's spin is $s=1$; $\phi_s$ is zero except for the orange curve which has $\phi_s=\pi/4$. The small body has mass ratio $\varepsilon=10^{-2}$ and is orbiting a black hole with spin $a=0.7 M$. For all panels, $p=10$, $e=0.2$, $x_I=0.7$, $q_r^S=0$, $q_z^S=0$, and  $\phi=0$ initially.  In all three cases, the dephasing is simply proportional to $s_\parallel$: $s_\parallel = 1$ shows the largest effect; the curves with $s_\parallel = 0.8$ and $s_\parallel = 0.5$ track that curve, but with magnitudes smaller by factors of $0.8$ and $0.5$, respectively.}
   \label{fig:precessplot1}
\end{figure}

We begin by examining a set of generic (inclined and eccentric) inspirals with aligned secondary spin and mass ratios $\varepsilon = 5 \times 10^{-2}$, $10^{-2}$, and  $10^{-3}$ (left panel of Fig.\ \ref{fig:diffepsilon}).  As we have emphasized elsewhere, we expect astrophysical EMRI systems to have mass ratios of $10^{-4}$ or smaller; we use a larger mass ratio here to augment and clearly show spinning body effects.  Each example we consider begins at $p = 12$, $e = 0.35$, $x_I = 0.5$.  We look at inspiral into black holes with $a/M = 0.7$.  The left-hand panel of Fig.\ \ref{fig:diffepsilon} shows these inspirals in the $(p, e)$ plane (top) and the $(p, x_I)$ plane (bottom).  In all cases, $p$ decreases due to radiation reaction until the system reaches the LSO (shown as a dotted line); $e$ decreases for much of the inspiral, showing an uptick near the LSO (a well-known strong-field characteristic of GW driven inspiral \cite{Loutrel2019}). The inspiral increases in inclination (corresponding to a decrease in $x_I$) all the way to the LSO, with no deep strong-field reversal of sign unlike the $p$-$e$ trajectory.

In the left panel of Fig.\ \ref{fig:diffepsilon}, we see that the amplitude of the oscillations increases with increasing mass ratio $\varepsilon$, while the number of oscillations increases inversely with mass ratio.  This is because the duration of inspiral scales inversely with $\varepsilon$, changing the number of orbital cycles the inspiral passes through before reaching the LSO. The difference between OG and averaged quantities also decreases with decreasing $\varepsilon$ (right panel of Fig.\  \ref{fig:diffepsilon}); this is a useful validation of the NIT procedure. In the bottom right panel of Fig.\ \ref{fig:diffepsilon}), there is an uptick in the value of $|(\phi +Z_\phi^{(0)}-\Omega_\phi Z_t^{(0)})-\varphi_\phi|$ for mass ratio $\varepsilon=10^{-4}$; this is due to numerical error floor in the OG solver as well as interpolation error in the NIT solution.  We expect this error could be reduced with a more computationally expensive online (higher precision numerical solver) or offline (higher precision interpolation) step.

The curves in Fig.\ \ref{fig:OGNITphases} show the dephasing of a generic inspiral due to spin-curvature force.  We show the difference between various phases computed using only adiabatic radiation reaction (denoted by ``RR"), and radiation reaction plus the spin-curvature force (denote by ``SCF + RR'').  The dashed lines in all panels show the averaged (NIT) dephasing $\varphi_y^{SCF+RR} - \varphi_y^{RR}$; $y = r$ is shown in the top panel, $y = z$ in the middle, and $y = \phi$ in the bottom.  (We remind the reader that $\varphi_\alpha$ represents the averaged phases parameterized in Boyer-Lindquist time.)  The solid curves in the three panels show these dephasings computed using the OG equations.

The inclusion of the spin-curvature force, which is conservative \cite{Witzany2019_1,Blanco2023}, will lead to secular changes to the evolution of the phases. In Fig.\ \ref{fig:OGNITphases}, we see secular corrections to the phases accumulate when post-adiabatic effects are included. The evolution of the radial dephasing $\varphi_r^{SCF+RR}-\varphi_r^{RR}$ is not monotonic, increasing to a maximum value and subsequently decreasing to less than zero.  The secular dephasing of both $\varphi_z^{SCF+RR}-\varphi_z^{RR}$ and $\varphi_\phi^{SCF+RR}-\varphi_\phi^{RR}$ by contrast is monotonic. 

As discussed in previous sections of this paper, short timescale oscillations in solutions to the OG equations of motion are removed by the NIT averaging procedure, isolating the longer timescale, secular evolution (compare the solid and dashed curves in Fig.\ \ref{fig:OGNITphases}).  For more extreme mass ratios, the difference in time scales is significant, and it greatly reduces computational cost to compute on only the longer secular timescale.  The oscillations in the solution to the OG equations contain harmonics of multiple frequencies; this complexity in harmonic structure is especially clear in the bottom panel of Fig.\ \ref{fig:OGNITphases} which displays $\phi^{SCF + RR} - \phi^{RR}$.  In this spin-aligned case, harmonics of $\Omega_r$ and $\Omega_z$ (or equivalently $\Omega_\theta$) contribute to the structure.  In the spin-misaligned case we examine in the next section, harmonics of $\Omega_s$ are also present. 

\subsection{Misaligned spin}
\label{sec:res_inspmisalign}

We now look at an example of generic spinning body inspiral with misaligned small-body spin.  The red curves in Fig.\ \ref{fig:3Dinspiral} show a generic inspiral, both with (solid line) and without (dashed line) the spin-curvature force.  The orange curve shows the projection of the inspiral onto the $p$-$e$ plane; the blue curve shows the projection onto the $p$-$x_I$ plane.  Just as in the aligned case, the projection onto the $p$-$e$ plane shows a decrease in eccentricity throughout most of the inspiral, and then ticks up shortly before reaching the LSO (depicted by a black line).  The inspiral increases in inclination (corresponding to a decrease in $x_I$) all the way to the LSO, with no deep strong-field reversal of sign unlike the $p$-$e$ trajectory.

Figure \ref{fig:2Dinspiral} shows a more detailed depiction of the projections of the inspiral onto the $p$-$e$ and $p$-$x_I$ planes (leftmost panels of the first two rows).  Each panel includes an inset which zooms in on the inspiral close to the LSO. The secular evolution of the principal orbital elements $p$, $e$, and  $x_I$ is unaffected by the presence of the spin-curvature force, but this force drives oscillations about the secular trajectory.  Notice that the generic inspiral has harmonic structure at multiple timescales --- the oscillations have a more complicated structure than we saw in the case of aligned inspirals.  This more intricate harmonic structure is because there are terms in the equations of motion which are periodic with the four frequencies $\Omega_r$, $\Omega_\theta$, $\Omega_\phi$, and  $\Omega_s$.  Harmonics at frequency $\Omega_s$ are due to the precession of the small-body's spin vector.  Oscillations in the $x_I$-$p$ trajectory are particularly complex, involving beats between all four frequencies.

The right-hand panels of Fig.\ \ref{fig:2Dinspiral} show the inspiral trajectory in a Cartesian representation of the Boyer-Lindquist coordinates: we define $x_{BL} = r\sin\theta\cos\phi$, $y_{BL} = r\sin\theta\sin\phi$, $z_{BL} = r\cos\theta$, with $r$, $\theta$, and $\phi$ the Boyer-Lindquist coordinates along the inspiral.  In the $r$-$z_{BL}$ inspiral projection, we see that the maximum $|z_{BL}|$ decreases as inspiral progresses.  Although the inclination angle $I$ increases during inspiral, the effect is quite small.  The shrinking of $r$ due to radiative backreaction is much more significant, so $|z_{BL}| = |r\cos\theta|$ decreases overall. 

Figure \ref{fig:precessplot1} shows how the misalignment of the small-body spin modifies the inspiral.  From top to bottom, the three panels show the dephasing of the spinning-body phases ($\varphi_r^{SCF+RR}$, $\varphi_z^{SCF+RR}$, and  $\varphi_\phi^{SCF+RR}$) relative to those of a non-spinning body ($\varphi_r^{RR}$, $\varphi_z^{RR}$, and  $\varphi_\phi^{RR}$).  We see that the value of $\varphi_y^{SCF+RR}-\varphi_y^{RR}$,  $y\in\{r,z,\phi\}$, is proportional to $s_\parallel$, as expected from previous analyses \cite{Drummond2022_2,Witzany2019_2}.  In all three panels, the blue curve (corresponding to aligned spins, $s_\parallel=1$), shows the largest dephasing.  The maxima of the other two curves, $s_\parallel=0.8$ (orange) and $s_\parallel=0.5$ (red), are exactly 0.8 and 0.5 times the maximum of the $s_\parallel=1$ curve, as expected.  The component of the small body spin misaligned from the orbit does not play any role in this dephasing.  See Appendix \ref{sec:varyinitialconditions} for a discussion about the selection of initial conditions in the case of inspirals with spin precession.

\section{Results II: Waveforms}
\label{sec:res_waveforms}

We wrap up our discussion of spinning-body inspirals by examining the waveforms these inspirals generate.

\subsection{Waveform generation}
\label{sec:Teukolsky}

\begin{figure*}
\centerline{\includegraphics[scale=0.60]{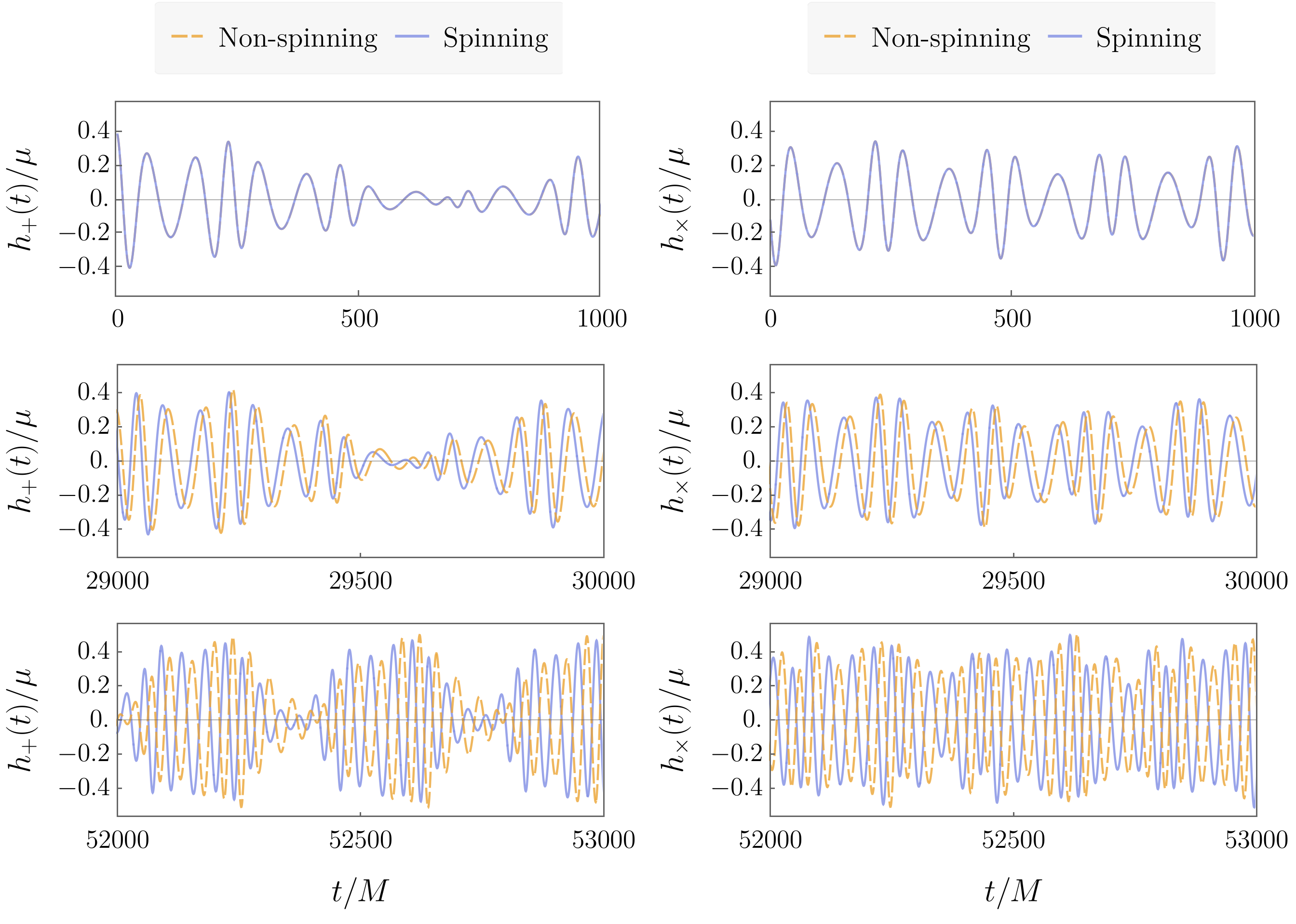}}
    \caption{Evolution of $h_+$ and $h_\times$ for a generic inspiral with mass-ratio $10^{-3}$. Top panel shows the part of the waveform corresponding to an early part of the inspiral, the middle panel shows an intermediate stage and the bottom panel shows the end of the inspiral. The blue (solid) and orange (dashed) curves correspond to spinning and non-spinning small bodies respectively. The mismatch between the two waveforms is $0.2067$. Initial orbital parameters are: $p=7.95$, $e=0.22$, $x=0.699$, $q_r=0$, and  $q_z=0$. The small body orbits a black hole with spin $a=0.7 M$ and the magnitude and orientation of the small body's spin is specified by: $ s=1$ and $s_\parallel=s$. We use the code \texttt{GremlinInsp} to generate these waveforms, with parameters $l_{max}=2$, $k_{max}=4$, and  $n_{max}=10$. }
    \label{fig:generichtotalplot}
\end{figure*}

We write the GW strain in the ``multivoice'' form \cite{Hughes2021}
\begin{align}
\label{eq:heq}
h(t) &\equiv h_+(t) - ih_\times(t) \equiv \frac{1}{r}\sum_{lmkn}h_{lmkn}(t)
\nonumber\\
&= \frac{1}{r}\sum_{lmkn}H_{lmkn}(t) e^{i[m\varphi_{\rm S} - \Phi_{mkn}(t)]}\;.
\end{align}
This form is found by promoting ``snapshot'' waveforms from a geodesic orbit into a sequence of snapshots in which the waveform's properties evolve as inspiral proceeds.  The amplitude of each waveform voice is given by
\begin{equation}
\label{eq:Hlmkn}
    H_{lmkn}(t) = A_{lmkn}(t) S_{lm}\left[\vartheta_{\rm S}; a\omega_{mkn}(t)\right]\;,
\end{equation}
where 
\begin{equation}
\label{eq:Almkn}
    A_{lmkn}(t) = -\frac{2Z^\infty_{lmkn}(t)}{\omega_{mkn}(t)^2}\;.
\end{equation}
For adiabatic inspirals, the phase of each voice is
\begin{align}
\label{eq:Philmkn_adiabatic}
    \Phi_{mkn}(t) &= \int_{t_0}^{t} \left[m\Omega_\phi(t') + k\Omega_\theta(t') + n\Omega_r(t')\right]dt'
    \nonumber\\
    &\equiv \int_{t_0}^t \omega_{mkn}(t')\;,dt'\;.
\end{align}
The waveform $h$ is measured at ($t$, $r$, $\vartheta_{\rm S}$, $\varphi_{\rm S}$); the ``S'' on these angles denotes position on the sky, and distinguishes them from orbit coordinates $(\theta, \phi)$, as well as from the Boyer-Lindquist NIT phases $\varphi_{r,z,\phi}$.  The function $S_{lm}(\vartheta_{\rm S}; a\omega_{mkn})$ is a spheroidal harmonic of spin-weight $-2$.  The strain $h$ is decomposed onto a basis of spheroidal harmonics with indices $lm$, as well as into a discrete frequency spectrum labeled with indices $mkn$.

The dependence on time of the various quantities introduced in the waveform above are inherited from the dynamics of the binary's inspiral.  For example, the complex amplitudes $Z^\infty_{lmkn}(\vec P)$ are pre-evaluated by solving the radial Teukolsky equation on a grid of principal orbit elements, and are then interpolated to generate the waveform at arbitrary points within the grid domain.  As the orbit underlying an EMRI evolves, the orbital elements $\vec P$ likewise evolve.  We denote these evolving elements by $\vec P(t)$, where $t$ parameterizes evolution along the inspiral as seen by a distant observer.  The amplitude $Z^\infty_{lmkn}(t)$ is thus shorthand for $Z^\infty_{lmkn}[\vec{P}(t)]$, and likewise for other quantities which enter the waveform.

In Sec.\ \ref{sec:AveragedEoM}, we wrote down expressions for the Boyer-Lindquist averaged equations of motion for the orbital phases (\ref{eq:Phase_Transformation}). In integral form, the expression for these phases is:
\begin{align}
\varphi_\alpha (t)&=\int_{t_0}^{t^{\rm i}} \left( \Omega_\alpha^{(0)}(t') +\varepsilon \Omega _\alpha^{(1)}(t')+\mathcal{O}(\varepsilon^2) \right)dt' \nonumber \\ &= \int_{t_0}^{t} \left( \Omega_\alpha(t') +\mathcal{O}(\varepsilon^2) \right)dt'\;.
\end{align}
These phases contribute to the waveform voices via
\begin{equation}
\label{eq:Philmkn_postadiabatic}
\Phi_{mkn}(t) = m\varphi_\phi(t) + k\varphi_\theta(t) + n\varphi_r(t) + \mathcal{O}(\varepsilon)\;.
\end{equation}
The Boyer-Lindquist time averaged phases $\varphi_\alpha(t)$ are thus exactly equivalent to the input required for generating multi-voice Teukolsky waveforms \cite{Lynch2023}.  Replacing the adiabatic phase (\ref{eq:Philmkn_adiabatic}) used in the waveform (\ref{eq:heq}) with the phase (\ref{eq:Philmkn_postadiabatic}) is thus a simple and computationally effective way to incorporate spin-curvature physics into inspiral waveforms.  A generalization of this to include other post-geodesic forcing terms should likewise enable simple incorporation of other important post-adiabatic effects into EMRI waveforms.

We compute relativistic waveforms using \texttt{GremlinInsp}\footnote{{\tt GremlinInsp} is a subset of the {\tt Gremlin} package, a C++ code developed by author Hughes to solve the frequency-domain Teukolsky equation for generic bound Kerr orbits.  It is not yet in the public domain due to licensing issues, but an open-source version is under development.  In the meantime, interested parties should contact Hughes regarding this code.}, which accepts as input a worldline (an HDF5 file with datasets $\{t,$ $p(t),$ $e(t),$ $x_I(t),$ $\Phi_r(t),$ $\Phi_\theta(t),$ $\Phi_\phi(t)\}$) and maximum values $l_{max}$, $k_{max}$ and $n_{max}$. The waveform is assembled by performing the sum (\ref{eq:heq}), where the amplitudes $Z^{\infty}_{lmkn}$ have been obtained by solving the Teukolsky equation with a point-particle source \cite{Hughes2021}.

Note that the \texttt{FastEMRIWaveforms} waveform module takes the same inputs from the orbital dynamics \cite{Katz2021}.  As such, replacing the adiabatic equations of motion currently in place with the averaged equations of motion we have developed here, along with setting the initial conditions outlined in Appendix \ref{app:initialconditions}, will provide a very convenient way to incorporate the conservative effects of an arbitrary secondary spin into EMRI waveforms efficient enough for LISA data analysis.  At present, \texttt{FastEMRIWaveforms} can only produce fully relativistic waveforms for eccentric Schwarzschild inspirals.  Work is in progress to extend this package to cover inspirals into Kerr black holes; once that it is done, it should not be difficult to adapt this package further to include the post-adiabatic effect of spin-curvature coupling.

\begin{figure*}
\centerline{\includegraphics[scale=0.65]{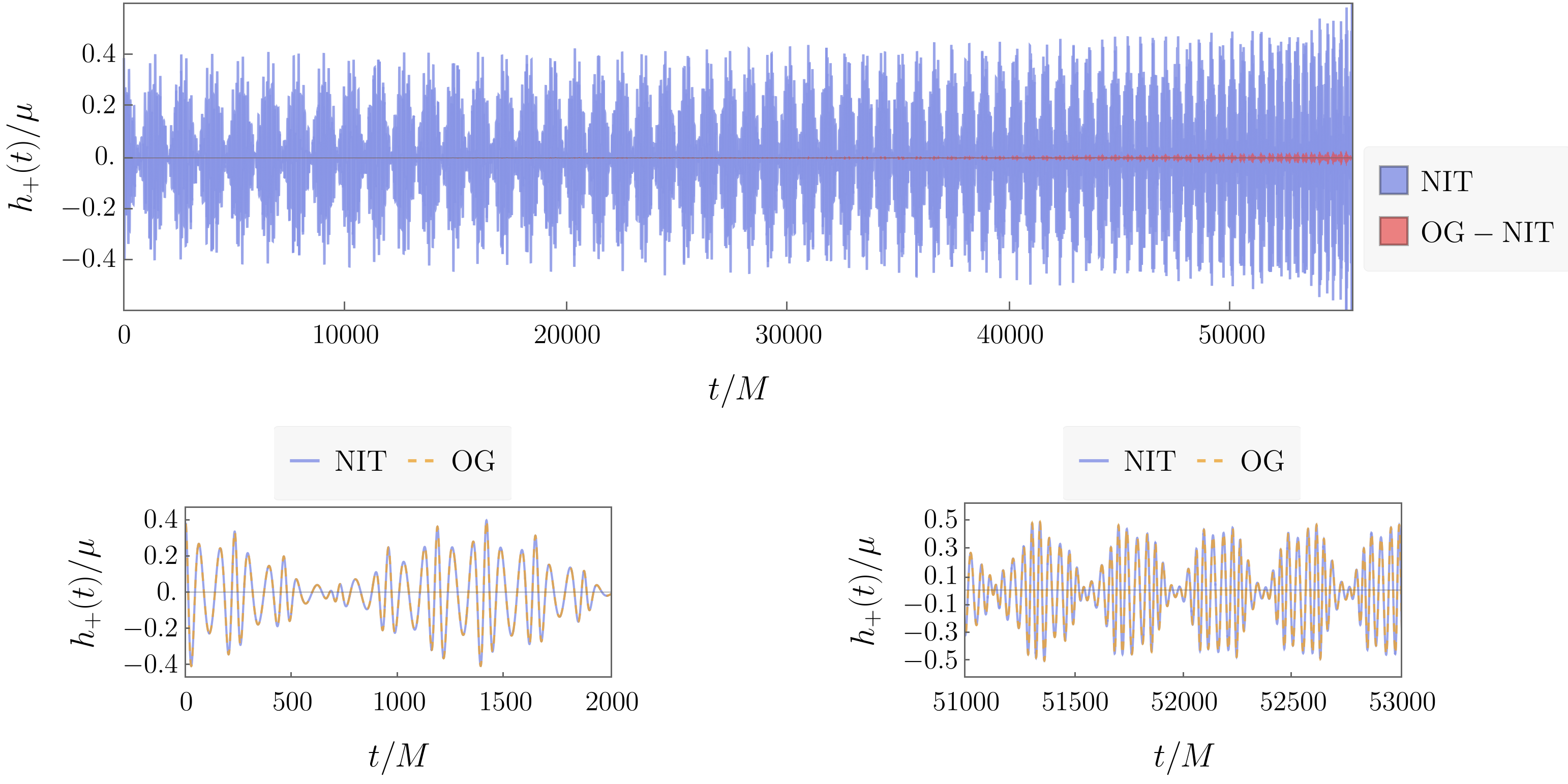}}
    \caption{Comparison of waveforms computed from OG inspiral and NIT inspiral.  Top panel shows $h_+$ for parameters identical to those used in Fig.\ \ref{fig:generichtotalplot}.  Blue curve shows waveform from a NIT inspiral for the entire domain we computed; red shows the difference between the OG and the NIT waveforms.  Bottom left panel shows the early part of the inspiral; bottom right shows the end of inspiral.  The blue solid and orange dashed curves corresponds to NIT and OG inspirals respectively.  The mismatch between these two waveforms, computed using Eq.\ (\ref{eq:mismatchcriteria}), is $\mathcal{M} = 3.462\times10^{-4}$.}
    \label{fig:eqhplot}
\end{figure*}

\subsection{Waveform analysis}
\label{sec:WFanalysis}

We conclude our analysis of waveforms by quantitatively comparing the different physical effects and modeling methods that we have used.  To do this, we use a noise-weighted inner product of two waveforms $h_1$ and $h_2$ given by
\begin{equation}
\langle h_1|h_2\rangle = 2\int^\infty_0 \frac{\tilde h^*_1(f) \tilde h_2(f)+\tilde h_1(f) \tilde h_2^*(f)}{S_n(f)} df\;,
\end{equation}
where $\tilde h(f)$ is the Fourier transform of the time-domain waveform $h(t)$, $\tilde h^*(f)$ is the complex conjugate of $\tilde h(f)$, and $S_n(f)$ is the one-sided power spectral density (PSD) of detector noise.  We use a white noise power spectrum here (i.e., noise independent of frequency); an analysis focusing on astrophysical waveform characteristics (as opposed to assessing more general aspects of waveform modeling) would use noise from a particular detector, such as that projected for the LISA mission \cite{Robson2019}. The fractional waveform overlap $\mathcal{O}$ is defined by
\begin{equation}
\mathcal{O}=\frac{\langle h_1|h_2\rangle}{\sqrt{\langle h_1|h_1\rangle\langle h_2|h_2\rangle}}\;.
\end{equation}
This measure equals equals 1 when $h_1 = h_2$; $\mathcal{O} = 0$ defines ``orthogonal'' waveforms.  Note that, for white noise, $\mathcal{O}$ is independent of the noise amplitude; we thus set $S_n(f) = 1$ for these comparisons.  A closely related notion is the fractional waveform mismatch, $\mathcal{M} = 1 - \mathcal{O}$. We use the \texttt{WaveformMatch} function from the SimulationTools package \cite{SimulationTools} to calculate waveform overlaps.

Using these tools to compare waveforms, we now consider how high the overlap should be for waveforms to be distinguishable in the context of LISA data science.  Following the criteria defined in Ref.~\cite{Lindblom:2008cm}, two waveforms $h_1$ and $h_2$ are defined to be indistinguishable if they satisfy $\langle \delta h |\delta h\rangle < 1$, where $\delta h = h_1 - h_2$. The signal-to-noise ratio (SNR) $\rho$ is defined by $\rho^2\equiv \langle h|h\rangle$.  Combining these definitions and going to the limit $\rho_1 \simeq \rho_2 \equiv \rho$ yields the benchmark that two waveforms with mismatch $\mathcal{M}$ will be indistinguishable if their SNR satisfies
\begin{equation}
\rho\leq \frac{1}{\sqrt{2\mathcal{M}}}\;. \label{eq:mismatchcriteria}
\end{equation}
Two signals being distinguishable according to the criterion (\ref{eq:mismatchcriteria}) is a necessary but not sufficient condition for detectability of a particular effect.  A more concrete measure of whether some effect related to the source physics is detectable should be assessed using a Bayesian maximum-likelihood estimation framework. 

Figures \ref{fig:generichtotalplot} and \ref{fig:eqhplot} display snapshots of gravitational waveforms. Figure \ref{fig:generichtotalplot} shows the plus and cross polarizations for a generic inspiral with mass-ratio $10^{-3}$; the blue curve shows the waveform of a spinning body, while the orange curve shows the waveform of a non-spinning body. The top, middle and bottom panels display the early, intermediate and late stages of the inspiral. If the non-spinning and spinning-body inspirals are initially in phase at the beginning of the inspiral, the dephasing accumulates as the inspiral progresses. This dephasing accumulates a rather large mismatch of $\mathcal{M} = 0.2067$ between spinning and non-spinning waveforms.  Using Eq.\ \ref{eq:mismatchcriteria}, these waveforms would be distinguishable for EMRI signals with SNR $\rho \gtrsim 1.5$.  In other words, if these were real signals, they would be easily distinguishable.

Figure \ref{fig:eqhplot} compares OG and NIT models of $h_+$ for the spinning body generic inspiral shown in Fig.\ \ref{fig:generichtotalplot}. The top panel shows the waveform of the entire inspiral, left bottom shows early in the inspiral, and right bottom shows late times.  The solid blue curve is the waveform computed using the NIT inspiral, while dashed orange corresponds to the waveform computed with the OG inspiral. In the bottom two panels of Fig.\ \ref{fig:generichtotalplot}, we see that the NIT and OG curves lie almost exactly on top of each other, even late in the inspiral.  The difference between the OG and NIT waveforms is shown by the red curve of the top panel of Fig.\ \ref{fig:eqhplot}; a small mismatch, $\mathcal{M} \simeq 0.00035$, accumulates over the inspiral.  According to the criterion (\ref{eq:mismatchcriteria}), the OG and NIT waveforms would be distinguishable as EMRI signals with SNR greater than about 38.  It's worth bearing in mind that this result is for mass ratio $\varepsilon = 10^{-3}$.  The mismatch would be lower, and the SNR needed for signals to be distinguishable would be greater, for EMRI mass ratios $\varepsilon \lesssim 10^{-4}$.  Waveforms computed using the OG and NIT techniques differ only slightly, despite their vastly different computational costs.

\section{Conclusions}
\label{sec:conclude}

We have presented a framework to combine orbit-averaged point-particle GW backreaction with the orbital dynamics of spinning bodies to make inspiral worldlines and gravitational waveforms for spinning bodies bound to Kerr black holes in the extreme mass ratio limit.  The inspirals and GWs produced by this framework are demonstrably incomplete (we discuss below aspects of this model which are ripe for improvement and additional work), but nonetheless make it possible to augment existing models of strong-field inspiral and waveform generation using data and methods available today.

As tools for efficiently computing EMRI waveforms \cite{Chua2021, Katz2021} expand to cover more of the astrophysical parameter space, it should not be difficult using the methods and techniques we have presented to further augment these tools to include the influence of secondary spin.  As show in Sec.\ \ref{sec:res_waveforms}, the leading impact on the waveforms' phase evolution can be found by ``upgrading'' the adiabatic inspiral phase, our Eq.\ (\ref{eq:Philmkn_adiabatic}), to a version that includes the post-adiabatic influence of secondary spin.  This may be particularly useful in the short term for assessing the importance of spin effects for EMRI science.  For example, previous work based on much simpler orbit geometries concluded that secondary spin is likely to have negligible impact on EMRI measurements \cite{Piovano2020, Piovano2021}; re-examining this question for generic orbits and spin orientations may change this conclusion.  A further generalization of this problem may even be useful for examining the impact of secondary structure beyond spin (looking at, for example, the findings of Ref.\ \cite{Rahman2023} to a broader class of orbits).  It should not be too challenging to generalize further to include the post-adiabatic influence of other important post-geodesic effects.

As discussed in Sec.\ \ref{sec:orbits}, another way to approach this problem is to consider orbit-averaged backreaction directly applied to spinning body orbits, following the kind of calculations laid out in Ref.\ \cite{Skoupy2023}.  Indeed, given that spinning body orbits describe the behavior of these inspirals on timescales too short for radiation reaction to be apparent, one might regard this as a more natural approach to this problem.  Performing such a calculation will require large data sets describing backreaction onto spinning body orbits, as well as a better understanding of how to evolve the generalized Carter constant of a spinning body.  In addition, the GWs produced by a spinning body are more complicated than those from a point body: an additional term, linear in the small body's spin tensor, enters the source term of the wave equation.  This changes the instantaneous wave amplitude, and thus changes the rate at which GWs backreact on the system.  The calculation we present here will be a useful tool for assessing the importance of different terms which enter the dynamics of backreaction for spinning-body orbits.  By incorporating the linear-in-secondary-spin flux corrections to our calculation, it would be equivalent (to 1PA order) to using a spinning-body orbit formulation as the basis for the calculation from the outset. Comparing the two approaches would then be a useful validation for both formulations.  We include {\it Mathematica} code and access to the data used to describe backreaction with this paper in order to facilitate making such comparisons.

Secondary spin is one example of an important post-adiabatic effect.  Other effects, especially those related to the gravitational self force \cite{vandeMeent2018, Wardell2023} are also critically important, and must also be included in order to develop accurate EMRI waveform models.  As long as these terms can be considered independently, with each term contributing in a ``modular'' fashion, a framework based on osculating orbits may be particularly suitable to combining the impact of different post-adiabatic effects in a single unified model; by using osculating geodesics as the basis for the calculation, all the post-adiabatic effects will be parameterized in the same way and can be directly combined.  Such a model will be needed before too long in order to accurately assess the importance of various contributors to inspiral and EMRI waveforms.

\section*{Acknowledgements}

LVD, DRB and SAH were supported by NASA ATP Grant 80NSSC18K1091 and NSF Grant PHY-2110384; AGH was supported by ATP Grant 80NSSC18K1091 while at MIT. PL acknowledges support from the Irish Research Council under grant GOIPG/2018/1978.  We thank the anonymous referee of a previous manuscript for critical feedback which we incorporated into this paper. We also thank Ian Hinder and Barry Wardell for the SimulationTools analysis package \cite{SimulationTools}. This work makes use of the Black Hole Perturbation Toolkit \cite{BHPToolkit}, in particular the \texttt{KerrGeodesics} package \cite{Kerrgeodesics}.

\appendix

\section{Geodesics in Kerr spacetime}
\label{app:geodesicsinkerr}

In this appendix, we list formulas and definitions used to describe geodesic orbits of Kerr black holes, which, for brevity, are left out of the main body of this paper.  In Boyer-Lindquist coordinates, the metric for a Kerr black hole with mass $M$ and spin angular momentum $S = aM$ is written \cite{Kerr1963,Boyer1967} 
\begin{align}
ds^2 & = -\left(1 - \frac{2r}{\Sigma}\right)\,dt^2 + \frac{\Sigma}{\Delta}\,dr^2 - \frac{4Mar\sin^2\theta}{\Sigma}dt\,d\phi\nonumber\\
 & + \Sigma\,d\theta^2 + \frac{\left(r^2 + a^2\right)^2 - a^2\Delta\sin^2\theta}{\Sigma}\sin^2\theta\,d\phi^2,\label{eq:kerrmetric}
\end{align}
where 
\begin{equation}
\Delta = r^2 - 2Mr + a^2\;,\qquad \Sigma = r^2 + a^2\cos^2\theta\;.
\end{equation}
The polar angle $\theta$ is measured from the black hole's spin axis (i.e., $\theta = 0$ is the ``North pole'' of the spinning black hole).  This metric has no dependence on coordinates $t$ and $\phi$, and so admits a timelike Killing vector $\xi_{t}^{\alpha}$ and an axial Killing vector $\xi_{\phi}^{\alpha}$. A body freely falling in this spacetime therefore has two constants of motion related to these Killing vectors, the energy per unit mass $\hat E$ and axial angular momentum per unit mass $\hat L_z$:
\begin{align}
\hat E & =-\xi_{t}^{\alpha} u_{\alpha}= -u_{t}\;,\\
\hat L_z & =\xi_{\phi}^{\alpha} u_{\alpha} =  u_{\phi}\;,
\end{align}
where $u^\alpha$ is the 4-velocity of the free falling body.  (The hat accent on these quantities indicates that they are defined on geodesics; very similar constants of the motion can be found for certain non-geodesic orbits, such as spinning-body orbits.)  The Kerr metric also possesses a Killing-Yano tensor $\mathcal{F}_{\mu\nu}$ \cite{Penrose1973}, which has the defining property
\begin{equation}
\nabla_{\gamma}\mathcal{F}_{\alpha\beta}+\nabla_{\beta}\mathcal{F}_{\alpha\gamma}=0\;.
\label{eq:KillingYanoDerivs}
\end{equation}
Carter showed that the Killing tensor $K_{\mu\nu}$, defined as the ``square'' of the Killing-Yano tensor via
\begin{equation}
K_{\mu\nu}=\mathcal{F}_{\mu\alpha}{\mathcal{F}_\nu}^{\alpha}\;,
\end{equation}
yields another constant of motion,
\begin{equation}
\hat K = K_{\alpha\beta} u^{\alpha} u^{\beta}\;,
\end{equation}
known as the ``Carter constant'' \cite{Carter1968}.  When $a = 0$, $\hat K$ is the square of a body's total angular momentum per unit mass.  It is convenient to define a related conserved quantity $\hat Q$, usually also called the Carter constant, by
\begin{align}
  \hat Q &= \hat K - \left(\hat L_z - a\hat E\right)^2\;.
\label{eq:Qdef}
\end{align}
When $a = 0$, $\hat Q$ is the square of a body's total angular momentum per unit mass projected into the $\theta = \pi/2$ plane.  The three constants of motion $(\hat E, \hat L_z, \hat Q)$ are one set of ``principal orbital elements'' (as discussed in Sec.\ \ref{sec:geodesics}) we can use to denote a particular geodesic in the osculating element framework.

The fact that the Kerr spacetime possesses these conserved quantities allows the geodesic equations to be separated as follows \cite{Carter1968}
\begin{align}
\Sigma^2\left(\frac{d{r}}{d\tau}\right)^2 &= [\hat E(r^2 + a^2) - a \hat L_z]^2\nonumber\\
 & \qquad-\Delta[r^2 + (\hat L_z - a\hat E)^2 + \hat Q]\nonumber\\
 & \equiv R({r})\;,\label{eq:geodr}\\
\Sigma^2\left(\frac{d{\theta}}{d\tau}\right)^2 &= \hat Q-\cot^2{\theta}\hat L_z^2 - a^2\cos^2{\theta}(1 - \hat E^2)\nonumber\\
 & \equiv\Theta({\theta})\;,\label{eq:geodtheta}\\
\Sigma\frac{d  {\phi}}{d\tau} &= a\hat E\left(\frac{r^2 + a^2}{\Delta} - 1\right) - \frac{a^2\hat L_z}{\Delta} + \csc^2{\theta} \hat L_z \nonumber\\
 & \equiv\Phi_r({r}) + \Phi_\theta({\theta})\;,\label{eq:geodphi}\\
\Sigma\frac{d  {t}}{d\tau} &= \hat E \frac{(r^2 + a^2)^2}{\Delta} + a \hat L_z\left(1 - \frac{r^2 + a^2}{\Delta}\right)\nonumber\\
&\qquad - \hat E a^2\sin^2{\theta} \nonumber\\
 & \equiv T_r({r})+T_\theta({\theta})\;.\label{eq:geodt}
\end{align}
When the motion is parameterized using proper time $\tau$ as above, equations (\ref{eq:geodr}) -- (\ref{eq:geodt}) do not entirely separate because the quantity $\Sigma(r,\theta)$ couples the radial and polar kinematics.  Mino time $\lambda$, defined by $d\lambda = d\tau/\Sigma$, allows us to separate these equations \cite{Mino2003, Drasco2004}.  It is straightforward to convert from $\lambda$ to Boyer-Lindquist time $t$, which describes quantities as measured by a distant observer, by using $d t/d\lambda$.

Any function of $r$ and $\theta$ evaluated along a geodesic can be expressed in a Fourier series as harmonics of the radial and polar frequencies.  A particularly useful form for many of our purposes uses the coordinate-time frequencies, since those correspond to frequencies as seen by distant observers.  As discussed in Ref.\ \cite{Drasco2004}, a function $f(r,\theta)$ evaluated along a geodesic can be written
\begin{equation}
    f[r(t),\theta(t)] = \sum_{k,n} f_{kn} e^{-i(k\hat\Omega_\theta + n\hat\Omega_r)t}\;.
\end{equation}
The sums over $k$ and $n$ are formally taken from $-\infty$ to $\infty$; for most numerical applications, the sums converge to an acceptable level of numerical error at maximum values that are not too large (several tens for fractional errors of $10^{-7}$ or smaller in most cases, though going up to hundreds for $n$ when studying highly eccentric strong-field orbits).  The Fourier amplitudes are found by integrating the functions over their Mino-time periods, with a factor of the geodesic function $dt/d\lambda$ from Eq.\ (\ref{eq:geods_mino}) \cite{Drasco2004}:
\begin{align}
    f_{kn} = \frac{(2\pi)^2}{\hat\Upsilon_t\hat\Lambda_r\hat\Lambda_\theta}\int_0^{\Lambda_r}\int_{0}^{\Lambda_\theta}f\left[r(\lambda_r), \theta(\lambda_\theta)\right]\frac{dt}{d\lambda}\,d\lambda_r\,d\lambda_\theta\;.
\end{align}
This calculation takes advantage of the fact that the Mino-time parameterization completely separates the radial and polar equations of motion, and treats the two degrees of freedom separately in performing the integral.

We also need expressions for the coordinate time $t$ and axial angle $\phi$ as functions of $\lambda$:
\begin{align}
t(\lambda) &= t_0 + \hat \Upsilon_t\lambda + \Delta t_r [r(\lambda)] + \Delta t_\theta [\theta(\lambda)]\;, \\
\phi(\lambda) &= \phi_0 + \hat \Upsilon_\phi \lambda+\Delta \phi_r [r(\lambda)]+\Delta \phi_\theta [\theta(\lambda)]\;.
\end{align}
The quantities $t_0$ and $\phi_0$ introduced above denote initial conditions.

We define
\begin{align}
\hat \Upsilon_t &= \langle T_r(r) \rangle +\langle T_\theta(\theta) \rangle \label{eq:Gamma}\;,\\
\hat \Upsilon_\phi &= \langle \Phi_r(r) \rangle +\langle \Phi_\theta(\theta) \rangle \label{eq:Upsphi}\;.
\end{align}
The angle brackets denote an averaging of the function with respect to either the radial or the angular motion of an orbiting body, and are defined precisely in Eqs.\ (2.12) and (2.13) of Ref.\ \cite{Hughes2021}.  The quantity $\hat \Upsilon_t$ is, in an orbit-averaged sense, the rate at which coordinate time $t$ ``ticks'' per unit Mino time $\lambda$; $\hat\Upsilon_\phi$ is a similarly averaged rate at which the axial coordinate advances per unit $\lambda$.  (As mentioned in Sec.\ \ref{sec:geodesics}, $\hat\Upsilon_t$ would be labeled $\hat\Gamma$ following the conventions of much of the literature.)  This means that $\hat\Upsilon_\phi$ is the axial orbit frequency conjugate to Mino time $\lambda$.  We also define 
\begin{align}
\Delta t_r [r(\lambda)] &= \int_0^\lambda \left\{T_r[r(\lambda')] - \langle T_r(r)\rangle\right\}d\lambda'\;,
\\
\Delta t_\theta [\theta(\lambda)] &= \int_0^\lambda \left\{T_\theta[\theta(\lambda')] - \langle T_\theta(\theta)\rangle\right\}d\lambda'\;,
\\
\Delta \phi_r [r(\lambda)] &= \int_0^\lambda \left\{\Phi_r[r(\lambda')] - \langle \Phi_r(r)\rangle\right\}d\lambda'\;,
\\
\Delta \phi_\theta [\theta(\lambda)] &= \int_0^\lambda \left\{\Phi_\theta[\theta(\lambda')] - \langle \Phi_\theta(\theta)\rangle\right\}d\lambda' \;.
\end{align}
We note that Eqs.\ (2.10) and (2.11) of Ref.\ \cite{Hughes2021}, which were intended to be equivalent to the equations above, left out the integrations, incorrectly presenting only the integrands on the right-hand sides of those equations.

\section{Motion of a spinning body}
\label{app:spinningbodymotion}

The motion of a spinning body in curved spacetime obeys the Mathisson-Papapetrou-Dixon (MPD) equations \cite{Papapetrou1951, Mathisson2010,Mathisson2010G_2,Dixon1970} which we introduced in Sec.\ \ref{sec:spinningorbits}. In this Appendix, we provide more detail about these equations and illustrate with examples of spinning-body motion.

\subsection{Constants of motion}
\label{sec:spinningbodyconstants}

The spinning body's worldline admits a constant of motion for each spacetime Killing vector $\xi^{\alpha}$, given by
\begin{equation}
\mathcal{C}=p_{\alpha}\xi^{\alpha}-\frac{1}{2}S^{\alpha\beta}\nabla_{\beta}\xi_{\alpha}\;.
\end{equation}
For spinning body orbits in Kerr, this allows us to generalize notions of energy and axial angular momentum:
\begin{align}
\mathcal{E}^S & = -p_t+\frac{1}{2}\partial_{\beta}g_{t\alpha}S^{\alpha\beta} \label{eq:Espingen}\;,\\ 
\mathcal{L}_z^S & = p_{\phi}-\frac{1}{2}\partial_{\beta}g_{\phi\alpha}S^{\alpha\beta}\;.\label{eq:Lspingen}
\end{align}
No Carter-type integral of the motion exists for spinning bodies in general, although an analogue of this constant exists at linear order in the small body's spin \cite{Rudiger1981}.  It has recently been shown that a Carter-like integral exists up to second-order in the small body's spin for a test body possessing exactly the spin-induced quadrupole moment expected for a Kerr black hole  \cite{Compere2022,2023CompereDruart}.

We define a spin vector from the spin tensor by
\begin{equation}
S^{\mu}=-\frac{1}{2\mu}{\varepsilon^{\mu\nu}}_{\alpha\beta}p_{\nu}S^{\alpha\beta}\;,
\label{eq:spinvec2}
\end{equation}
where
\begin{equation}
\epsilon_{\alpha\beta\gamma\delta}=\sqrt{-g}[\alpha\beta\gamma\delta]\;
\end{equation}
and $[\alpha\beta\gamma\delta]$ is the totally antisymmetric symbol. The magnitude of the spin vector $S$ is defined by
\begin{equation}
S^2=S^{\alpha}S_{\alpha}=\frac{1}{2}S_{\alpha\beta}S^{\alpha\beta}\;,\label{eq:smag}
\end{equation}
and is conserved along the spinning body's wordline. 

At linear order in the small body's spin, Eqs.\ (\ref{eq:Espingen}) and \ref{eq:Lspingen}) simplify, allowing us to define the energy and axial angular momentum per unit mass introduced in Sec.\ \ref{sec:orbits}:
\begin{align}
E^S & = -u_t+\frac{1}{2\mu}\partial_{\beta}g_{t\alpha}S^{\alpha\beta}\;,\\ 
L_z^S & = u_{\phi}-\frac{1}{2\mu}\partial_{\beta}g_{\phi\alpha}S^{\alpha\beta}\;.
\end{align}
At this order, a generalization of the Carter constant is also an integral of the motion \cite{Rudiger1981}:
\begin{equation}
K^S=K_{\alpha\beta}u^\alpha u^\beta+\delta\mathcal{C}^S\;,
\label{eq:Kspin}
\end{equation}
where
\begin{equation}
\delta\mathcal{C}^S = -\frac{2}{\mu^2} p^{\mu}S^{\rho\sigma}\left( {\mathcal{F}^\nu}_{\sigma}\nabla_{\nu}\mathcal{F}_{\mu \rho } - {\mathcal{F}_\mu}^\nu\nabla_{\nu}\mathcal{F}_{\rho\sigma}\right)\;.
\label{eq:Cspin}
\end{equation} 

\subsection{Spinning-body orbits}

We now briefly survey some of the key differences between spinning-body and geodesic orbits; Refs.\ \cite{Witzany2019_2, Drummond2022_1, Drummond2022_2} provide more details.  Spinning-body orbits are qualitatively different from geodesic ones.  If the body's spin is misaligned from the orbit, then its orientation precesses, with a Mino-time frequency $\Upsilon_s$ characterizing this precession; the body's orbital plane likewise precesses at this frequency.  This precession appears in the equations of motion as a variation in the bounds of both the polar and radial libration regions.  Indeed, one finds that the radial and polar motions for a spinning body do not separate when parameterized in Mino time as they do for geodesics \cite{Witzany2019_2, Drummond2022_1, Drummond2022_2}.  Finally, a body's spin also shifts the orbital frequencies relative to the orbital frequencies associated with geodesic orbits.  The well-understood frequencies $\Omega_{r,\theta,\phi}$ which characterize geodesic orbits are each shifted by an amount $\propto s_{\parallel}$, the component of the smaller body's spin parallel to its angular momentum.

\begin{figure*}
\centerline{\includegraphics[scale=0.43]{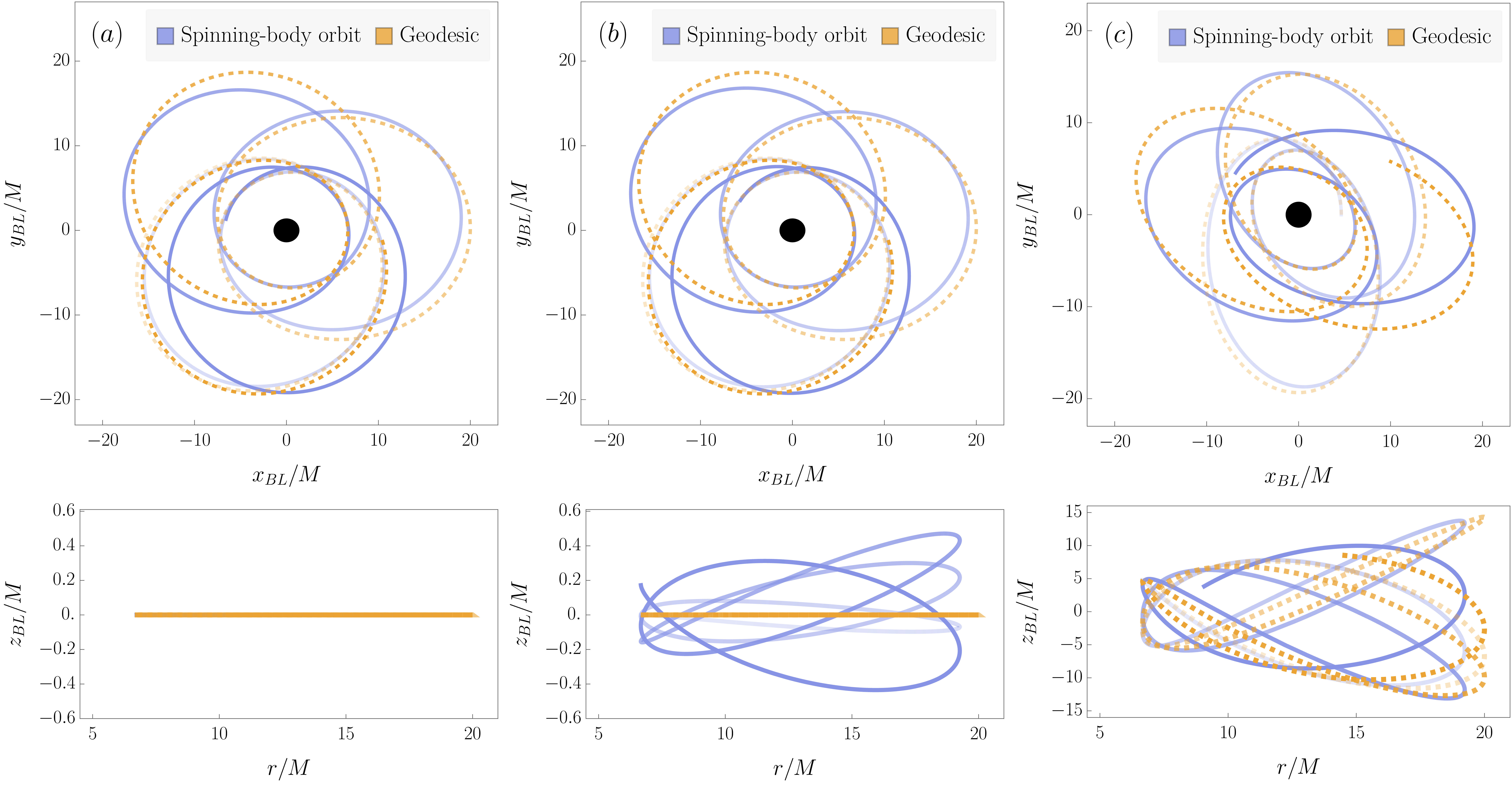}}
    \caption{Comparison of spinning-body (blue) and geodesic (orange) orbit trajectories.  Within each column, the trajectories shown have the same initial conditions.  The top row shows $x_{BL}$-$y_{BL}$ trajectories; the bottom shows trajectories in $r$-$z_{BL}$.  (The coordinates $x_{BL}$, $y_{BL}$, $z_{BL}$ are Cartesian-like representations of the Boyer-Lindquist coordinates: $x_{BL} = r\sin\theta\cos\phi$, etc.)  Increasing opacity of the trajectory curves denotes increasing time.  Panel (a) shows equatorial trajectories; for the blue (spinning-body) trajectory, the spin of the small body is aligned with the spin of the larger black hole.  The major difference in the trajectories in this case is the dephasing that occurs because spin-curvature coupling changes the timescales associated with orbital motions.  Panel (b) shows the same geodesic orbit as panel (a) but the spinning-body trajectory corresponds to a small body with its spin misaligned with its orbit.  Notice that the in-plane motion is similar to what we find in panel (a), at least over the time interval shown here, though the motion acquires an out-of-plane motion that is entirely absent from the geodesic case.  Note also the different scales used for the out-of-plane motion, versus the in-plane and radial motion: the out-of-plane motion is smaller by a factor $\sim 30$.  Panel (c) shows generic orbits for both cases.  In all panels, the parameters used are $a = 0.7 M$, $p = 10$, $e = 0.5$, $\varepsilon=0.1$, and $s = 1$.  In panels (b) and (c), we put $s_\parallel = 0.9s$ and $\phi_s = \pi/2$; in panel (c), we further put $x_I = 0.6967$.  Here and in many of the other plots, we have used a much less extreme mass ratio than is appropriate for these techniques in order to magnify the effect of spin-curvature coupling physics. } 
    \label{fig:trajectory}
\end{figure*}

We first consider equatorial orbits with aligned spin: $s = s_\parallel$, $s_\perp = 0$.  Spinning-body and geodesic orbits are quite similar in this case: motion is constrained to the plane $\theta = \pi/2$, and the radial motion is confined to an interval $r_2 \le r \le r_1$, where $r_2$ and $r_1$ are constants.  We show examples of equatorial non-spinning and spinning-body orbits with the same initial conditions in panel (a) of Fig.\ \ref{fig:trajectory}.  Differences emerge because the trajectories have different frequencies associated with both their radial and axial motions.

Qualitative differences become noticeable when $s_\perp \neq 0$.  When the small body's spin vector is misaligned, it precesses and the spinning body's orbit oscillates by an amount $\mathcal{O}(S)$ out of the equatorial plane.  For these ``nearly equatorial'' orbits, the radial motion remains constrained to the range $r_2 \le r \le r_1$, but the polar libration range is modified, with $\theta = \pi/2 + \delta \vartheta_S$.  The orbital plane precesses in response to the small body's spin precession, adjusting the turning points of the polar motion depending on the spin precession phase $\psi_s$.  This can be seen in panel (b) of Fig.\ \ref{fig:trajectory}: the orange (non-spinning) worldline is confined to the equatorial plane, while the blue (spinning-body) worldline oscillates about the equatorial plane.

Fully generic spinning-body orbits have eccentricity, are inclined with respect to the equatorial plane, and have an arbitrarily oriented small-body spin.  Functions evaluated along generic orbits have structure at harmonics of three frequencies: radial $\Omega_r$, polar $\Omega_{\theta}$, and spin-precessional $\Omega_s$.  We can use this to write functions evaluated along an orbit as a Fourier expansion of the form
\begin{equation}
    f[r, \theta, S^{\mu}] = \sum_{j = -1}^1 \sum_{k,n = -\infty}^\infty f_{jkn} e^{-ij\Omega_st}e^{-in\Omega_rt}e^{-ik\Omega_{\theta}t}\;,
\end{equation}
where $S^{\mu}$ is the small-body's spin vector.  Note the different index ranges in this sum: there are only three harmonics of the spin frequency $\Omega_s$, while in principle an infinite set of both polar and radial harmonics are present.  (In practice, these sums converge over a finite range, though one must study the system carefully to determine an appropriate truncation point \cite{Drummond2022_2}.)

The coupling of radial, polar and spin-precessional motions for generic spinning-body orbits causes the positions of the radial turning points to depend on $\theta$ and the spin-precession phase $\psi_s$. Similarly, the polar turning points depend on radial position and $\psi_s$, as derived in Ref.\ \cite{Witzany2019_2}.  Panel (c) of Fig.\ \ref{fig:trajectory} shows a generic geodesic (in orange) and spinning-body trajectory (in blue) with the same initial conditions.  The opacity of the curves increases as time advances; this illustrates how the trajectories diverge at late times, as the opacity increases.

\subsection{Spinning-body parameterizations}
\label{app:spinningbodyparams}

We have freedom in how we parameterize the motion of a spinning body, in the sense that we can construct various mappings between the triplet of constants $(p,e,x_I)$ which defines a ``reference" geodesic, to a specific spinning-body orbit. In Appendix A of \cite{Drummond2022_2}, three such mappings are discussed: (1) the turning points of the reference geodesic match those of the spinning-body orbit; (2) the initial conditions of the reference geodesic match those of the spinning-body orbit; and (3) the constants of motion $(\hat E, \hat L_z, \hat K)$ of the reference geodesic match the constants of the spinning-body orbit.  In this section, we will primarily discuss the parameterizations (1) and (2) and how to map between them.

References \cite{Drummond2022_1,Drummond2022_2} use parameterization (1): the turning points of the spinning-body orbit match those of a chosen reference geodesic defined by $(p,e,x_I)$.  Those references show how to compute the frequency corrections $\Upsilon_r^S(p,e,x_I)$, $\Upsilon_\theta^S(p,e,x_I)$, and  $\Upsilon_\phi^S(p,e,x_I)$ due to the small body's spin, relative to the frequencies of a reference geodesic with the same turning points.  Because of the additional harmonic complexity of spinning-body orbits relative to geodesics, the turning points of the non-spinning and spinning body orbits are matched in an orbit-averaged sense: the radial turning points of the ``purely radial'' piece of the spinning-body orbit are matched with the radial turning points of the geodesic, and likewise for the ``purely polar'' motion.  ``Purely radial'' means the contributions to the orbital motion that contains only harmonics of $\Upsilon_r$ or $\hat\Upsilon_r$; ``purely polar'' means contributions that contain only harmonics of $\Upsilon_\theta$ or $\hat\Upsilon_\theta$.  The equatorial spinning-body inspirals computed in \cite{Skoupy2022} also use this parameterization.

By construction, the perturbed motion found by solving the OG equations uses parameterization (2): the initial orbit coordinates and initial components of the four-velocity are the same for the spinning and non-spinning orbits.  We use this parameterization in this work, which was also used in \cite{Lynch2021}.  Parameterization (3), choosing the constants of motion $(E, L_z, K)$ of a spinning-body orbit to match those of a reference geodesic, is used in \cite{Witzany2019_2,Witzany2023}.

Because different parameterizations are used by different analyses, it is important to consider the mapping between the different choices, and to show that they describe the same orbits.  We begin by choosing a triplet $(p_{TP}, e_{TP}, x_{TP})$ that defines a geodesic with radial turning points $r_1 = p_{TP}/(1 - e_{TP})$ and $r_2 = p_{TP}/(1 + e_{TP})$, and with polar turning point $z_1 = \sqrt{1 - x_{TP}^2}$.  Using the approach of \cite{Drummond2022_1, Drummond2022_2}, we first compute the spinning-body trajectory that has the same turning points (on average) as this geodesic.  We next want to find the \textit{same} spinning-body orbit via the ``matched initial conditions'' parameterization, using the OG method presented in this paper.

To do this, we select initial values of ($r, z$) by choosing one of the radial and polar turning points of the spinning-body orbit we evaluated in the matched turning point parameterization.  We label these choices $r_{TP}$ and $z_{TP}$.  We use the subscript ``IC" to denote the triplet $(p_{IC}, e_{IC}, x_{IC})$ associated with a geodesic which has the same \textit{initial conditions} as the spinning-body orbit under consideration.  The geodesic orbit defined by $r_G$ and $z_G$ needs to initially have the same values of $r$ and $z$, so we equate $r_G(p_{IC}, e_{IC}, x_{IC}, q_{r0})$ and $z_G(p_{IC}, e_{IC}, x_{IC},q_{z0})$ as given in Eqs.\ (16)--(17) of Ref.\ \cite{vandeMeent2019}.  For convenience, we choose the spinning-body orbit to be at a turning point initially. The initial geodesic velocities must match, so we solve $R[r(p_{IC}, e_{IC}, x_{IC}, q_{r0})] = 0$ and $\Theta[z(p_{IC}, e_{IC}, x_{IC}, q_{z0}) = 0$ where the functions $R(r)$ and $\Theta(\theta)$ are given by equations (\ref{eq:geodr}) and (\ref{eq:geodtheta}).

We now have four equations and five unknowns, $(p_{IC},e_{IC},x_{IC},q_{r0},q_{z0})$.  To close this system, we find the initial value of $(d\phi/d\lambda)_{TP}$ of a spinning-body in the fixed turning point parameterization and equate it to $d\phi/d\lambda$ for a geodesic using $\Phi[r(p_{IC},e_{IC},x_{IC},q_{r0},q_{z0})]$, given in Eq.\ (\ref{eq:geodphi}). The final set of equations we solve is
\begin{align}
r_G(p_{IC},e_{IC},x_{IC},q_{r0}) &=r_{TP}\;, \\
z_G(p_{IC},e_{IC},x_{IC},q_{z0}) &=z_{TP}\;, \\
R[r(p_{IC},e_{IC},x_{IC},q_{r0})]&=0\;, \\
\Theta[r(p_{IC},e_{IC},x_{IC},q_{z0})]&=0\;, \\
\Phi[r(p_{IC},e_{IC},x_{IC},q_{r0},q_{z0})]&=\left( \frac{d\phi}{d\lambda}\right)_{TP}\;.
\end{align}
We solve the above equations to find the triplet $(p_{IC},e_{IC},x_{IC})$. We can then compute the spinning-body orbit corresponding to this choice of initial geodesic $(p_{IC},e_{IC},x_{IC})$. We now have a mapping between $(p_{TP},e_{TP},x_{TP})$ and $(p_{IC},e_{IC},x_{IC})$; this is how we compute the orbits in \ref{sec:spinviaforcedgeod}.

Note that the two parameterizations are not linearized in secondary spin in exactly the same way.  Feeding into the OG equations is the forcing term from the linearized MPD equations, Eq.\ (\ref{eq:spinforcelin}).  Beyond this point, the OG formulation does not assume the forcing term to be small and does not further linearize in spin. However, in the ``turning point matched'' prescription of Refs.\ \cite{Drummond2022_1,Drummond2022_2}, the expressions for the radial and polar trajectories, our Eqs.\ (\ref{eq:spintrajectoryr}) and (\ref{eq:spintrajectorytheta}), which have been explicitly divided into geodesic and secondary-spin pieces, are substituted into the MPD equations.  After this substitution, we then linearize the MPD equations.  This leads to a slight difference in the equations of motion between the two prescriptions at the $\mathcal{O}(S^2)$ level.  These two prescriptions are equivalent up to linear-order in secondary spin, but are not identical at $\mathcal{O}(S^2)$.  This is responsible for the slight drift seen after long integration times when comparing our methods for computing spinning-body orbits, discussed at the end of Sec.\ \ref{sec:spinviaforcedgeod}.

Note that we use the fact that we can evaluate the frequencies ($\Omega_r,\Omega_z,\Omega_\phi$) associated with a spinning-body orbit in both parameterizations in order to relate the reference geodesic triplets $(p_{TP},e_{TP},x_{TP})$ and $(p_{IC},e_{IC},x_{IC})$ in the two parameterizations. Explicitly, we find the mapping $(p_{TP},e_{TP},x_{TP})\rightarrow (p_{IC},e_{IC},x_{IC})$ by solving the equations:
\begin{align}
\Omega_r(p_{IC},e_{IC},x_{IC}) &=\Omega_r(p_{TP},e_{TP},x_{TP})\;, \\
\Omega_z(p_{IC},e_{IC},x_{IC}) &=\Omega_z(p_{TP},e_{TP},x_{TP}) \;, \\
\Omega_\phi(p_{IC},e_{IC},x_{IC}) &=\Omega_\phi(p_{TP},e_{TP},x_{TP})\;. 
\end{align}

\section{Forced motion via osculating geodesic orbital elements}
\label{app:oscelementframework}

In this appendix, we briefly discuss how to compute forced motion of a body in spacetime through a sequence of geodesic orbits, showing how the forcing terms lead to evolution of the orbital elements which characterize geodesics.  This synopsis is based on the discussion presented in Ref.\ \cite{Gair2011}.

Begin by writing the geodesic equation
\begin{equation}
    \frac{d^2x^\alpha}{d\tau^2}= - {\Gamma^\alpha}_{\beta\gamma}\frac{dx^\beta}{d\tau}\frac{dx^\gamma}{d\tau}\;
\end{equation}
in the form
\begin{equation}
\label{eq:ddotrgeo}
\ddot x^\alpha = a^\alpha_{\rm geo}\;,
\end{equation}
where overdot denotes $d/d\tau$.  As observed in Sec.\ \ref{sec:geodesics}, bound Kerr geodesics can be described by seven parameters:
\begin{equation}
   \mathcal{E}^A \doteq \{p, e, x_I, \chi_r^S, \chi_\theta^S, \phi_0, t_0\}\;.
\end{equation}
The capital Latin index introduced here ranges from $1$ to $7$; the symbol $\doteq$ means ``the components on the left-hand side are given by the elements of the set on the right-hand side.''  In this set, $p$, $e$, and  $x_I$ are the principal orbital elements describing the geometry of the orbit and $\chi_r^S$, $\chi_\theta^S$, $\phi_0$, and $t_0$ are the positional orbital elements that specify initial conditions.

The parameters $\mathcal{E}^A$ are strictly constant on a geodesic, and can be expressed as functions of spatial position and spatial velocity in an orbit.  In other words, we can write
\begin{equation}
    \mathcal{E}^A = \mathcal{E}^A(x^\alpha, \dot x^\alpha)\;.
\end{equation}
Using the chain rule, we write the rate of change of $\mathcal{E}^A$
\begin{equation}
\label{eq:dotI}
    \dot{\mathcal{E}}^A = \frac{\partial\mathcal{E}^A}{\partial x^\alpha}\dot x^\alpha + \frac{\partial\mathcal{E}^A}{\partial \dot x^\alpha}\ddot x^\alpha\;.
\end{equation}
Using Eq.\ (\ref{eq:ddotrgeo}) and requiring $\mathcal{E}^A$ to be constant on a geodesic, we obtain
\begin{equation}
\label{eq:dotIgeo}
    \dot{\mathcal{E}}^A = \frac{\partial\mathcal{E}^A}{\partial x^\alpha}\dot x^\alpha + \frac{\partial\mathcal{E}^A}{\partial \dot x^\alpha}a^\alpha_{\rm geo} = 0\;.
\end{equation}

Consider now forced motion.  In the presence of a perturbing force, the geodesic equation generalizes to
\begin{equation}
    \frac{d^2x^\alpha}{d\tau^2} + {\Gamma^\alpha}_{\beta\gamma}\frac{dx^\beta}{d\tau}\frac{dx^\gamma}{d\tau} = a^\alpha\;.
    \label{eq:forcedgeod}
\end{equation}
The non-geodesic acceleration $a^\alpha$ is subject to the constraint
\begin{equation}
    a^\alpha u_\alpha = 0\;.
\end{equation}
Equation (\ref{eq:forcedgeod}) can be written
\begin{equation}
\label{eq:forced}
\ddot{x}^\alpha = a^\alpha_{\rm geo} + a^\alpha\;.
\end{equation}
Our aim is to convert Eq.\ (\ref{eq:forced}) into a set of equations for the evolution of orbital elements $\mathcal{E}^A$.  This requires a mapping $\{x^\alpha, \dot x^\alpha \} \rightarrow \mathcal{E}^A$.  We assert that, at each moment along the worldline, a geodesic can be found with the same $(x^\alpha, \dot x^\alpha)$ as the accelerated body.  This assertion is known as the {\it osculation condition}.  Stated plainly, we assert that \cite{Pound2008}
\begin{align}
   x^\alpha(\tau) &= x_{\rm geo}^\alpha(\mathcal{E}^A,\tau)\;,
   \\
   \dot x^\alpha(\tau) &= \dot x_{\rm geo}^\alpha(\mathcal{E}^A,\tau)\;,\label{eq:secondeq}
\end{align}
where $a^\alpha(\tau)$ represents the coordinates of the true worldline, and $x^\alpha_{\rm geo}(\mathcal{E}^A,\tau)$ represents the coordinates of a geodesic worldline with orbital elements $\mathcal{E}^A$.  Note that the time derivative in Eq.\ (\ref{eq:secondeq}) holds $\mathcal{E}^A$ fixed.  Note also that the osculation condition involves 4 components of $x^\alpha$ and 4 components of $\dot x^\alpha$, one of which is constrained either by the condition $a^\alpha u_\alpha = 0$ or $u^\alpha u_\alpha = -1$.  The 8 components plus 1 constraint thus map to the 7 parameters $\mathcal{E}^A$, so the number of orbital elements matches the number of degrees of freedom \cite{Pound2008}.  

Under the influence of a perturbing force which accelerates the worldline by $a^\alpha$ relative to a geodesic, the parameters $\mathcal{E}^A$ do not remain constant.  We promote them to dynamical variables called \textit{osculating orbital elements}.  The accelerated trajectory $x^\alpha$ is then described by a sequence of geodesics with parameters
\begin{equation}
    \mathcal{E}^A(t) \doteq \{p(t), e(t), x_I(t), \chi_r^S(t), \chi_\theta^S(t), \phi_0(t), t_0(t)\}\;.
\end{equation}
Here $t$ is simply Boyer-Lindquist coordinate time along the inspiral, which we use as our parameter along the inspiral worldline.  Other parameter choices could be used (e.g., proper time $\tau$ along the inspiral, or Mino time $\lambda$).  Boyer-Lindquist time is particularly convenient, as it is the time measured by distant observers.  Note that we have written both $\phi_0$ and $t_0$ as though they are promoted to dynamical quantities; we will soon show that the equations governing them do not need to be evolved, and they can be left as constants.

What remains is to prescribe how to dynamically evolve these elements.  We again use the chain rule and Eq.\ (\ref{eq:forced}) to evaluate $\dot{\mathcal{E}}^A(\tau)$, yielding 
\begin{equation}
\label{eq:dotI2}
    \dot{\mathcal{E}}^A = \frac{\partial\mathcal{E}^A}{\partial x^\alpha}\dot x^\alpha + \frac{\partial\mathcal{E}^A}{\partial \dot x^\alpha}a^\alpha_{\rm geo} + \frac{\partial\mathcal{E}^A}{\partial \dot x^\alpha}a^\alpha\;.
\end{equation}
Taking advantage of Eq.\ (\ref{eq:dotIgeo}), we obtain
\begin{equation}
    \dot{\mathcal{E}}^A = \frac{\partial\mathcal{E}^A}{\partial \dot x^\alpha}a^\alpha\;.
\label{eq:oscelementevolforced}
\end{equation}

Multiplying both sides of Eq.\ (\ref{eq:dotIgeo}) by $\partial x_{\rm geo}^\beta/\partial\mathcal{E}^A$ and both sides of Eq.\ (\ref{eq:oscelementevolforced}) by $\partial \dot x_{\rm geo}^\beta/\partial\mathcal{E}^A$ yields a particularly useful form of these equations:
\begin{align}
\frac{\partial x^\beta_{\rm geo}}{\partial\mathcal{E}^A} \dot{\mathcal{E}}^A &= 0\;,
\label{eq:oscelementevol1} \\
\frac{\partial \dot x^\beta_{\rm geo}}{\partial\mathcal{E}^A}\dot{\mathcal{E}}^A &= a^\beta\;.
\label{eq:oscelementevol2}
\end{align}
To derive Eq.\ (\ref{eq:oscelementevol2}), note that Eq.\ (\ref{eq:secondeq}) implies
\begin{equation}
    \frac{\partial\dot x_{\rm geo}^\beta}{\partial\mathcal{E}^A}\frac{\partial\mathcal{E}^A}{\partial\dot x^\alpha} = {\delta^\beta}_\alpha\;.
\end{equation}
These expressions can be used to derive explicit equations for osculating orbital element evolution, and can be written in either contravariant or covariant form (see Secs.\ III D 1 and 2  of Ref.\ \cite{Gair2011}).

\subsection{Quasi-Keplerian evolution equations}
\label{sec:contraevol}

Following the approach used in Ref.\ \cite{Pound2008}, we use the contravariant formulation (see Sec.\ III D 2 of \cite{Gair2011}).  Expanding Eq.\ (\ref{eq:oscelementevol1}) yields
\begin{align}
    \frac{\partial r}{\partial p}p' + \frac{\partial r}{\partial e}e' + \frac{\partial r}{\partial I}I' + \frac{\partial r}{\partial \chi_r^S}\chi_r^{S\prime} + \frac{\partial r}{\partial \chi_\theta^S} \chi_\theta^{S\prime} &= 0\;,
    \label{eq:rgeodeq}\\
    \frac{\partial \theta}{\partial p}p' + \frac{\partial \theta}{\partial e}e' + \frac{\partial \theta}{\partial I}I' + \frac{\partial \theta}{\partial \chi_r^S}\chi_r^{S\prime} + \frac{\partial \theta}{\partial \chi_\theta^S}\chi_\theta^{S\prime} &= 0\;,
    \label{eq:thetageodeq}\\
    \frac{\partial \phi}{\partial p}p' + \frac{\partial \phi}{\partial e}e' + \frac{\partial \phi}{\partial x_I}x_I' + \frac{\partial \phi}{\partial \chi_r^S}\chi_r^{S\prime} + \frac{\partial \phi}{\partial \chi_\theta^S}\chi_\theta^{S\prime} + \phi_0' &= 0\;,
    \label{eq:phigeodeq}\\
    \frac{\partial t}{\partial p}p' + \frac{\partial t}{\partial e}e' + \frac{\partial t}{\partial x_I}x_I' + \frac{\partial t}{\partial \chi_r^S}\chi_r^{S\prime} + \frac{\partial t}{\partial \chi_\theta^S}\chi_\theta^{S\prime} + t_0'&=0\;.\label{eq:tgeodeq}
\end{align}
Prime represents differentiation with respect to the variable that parameterizes the trajectory, $t^{\rm i}$.


Equations (\ref{eq:phigeodeq}) and (\ref{eq:tgeodeq}), which govern the evolution of the axial offset $\phi_0$ and time offset $t_0$, contain elliptic integrals which are introduced due to terms like $\partial t/\partial p$.  Computing such integrals at each time step introduces additional computational expense.  Instead of evolving Eqs.\ (\ref{eq:phigeodeq}) and (\ref{eq:tgeodeq}), we find $\phi$ and $t$ along the worldline by using the geodesic expressions computed along the instantaneous orbit, as was done in Refs.\ \cite{Gair2011} and \cite{Pound2008}.  Rewriting Eqs.\ (\ref{eq:geodphi}) and (\ref{eq:geodt})), these equations are
\begin{align}
\frac{d\phi}{d\lambda}&=\Phi_r(r, E, L_z,Q )+\Phi_\theta(\theta, E, L_z, Q) \nonumber \\ 
&=\Phi_r[p(\lambda),e(\lambda),x_I(\lambda),\chi_r^S(\lambda)]\nonumber \\
&+\Phi_\theta[p(\lambda),e(\lambda),x_I(\lambda),\chi_\theta^S(\lambda)]\;, \label{eq:phiOGeqn}\\ 
\frac{d t}{d\lambda}&=T_r(r, E, L_z, Q)+T_\theta(\theta, E, L_z, Q) \nonumber\\
&=T_r[p(\lambda),e(\lambda),x_I(\lambda),\chi_r^S(\lambda)]  \nonumber 
\\&+T_\theta[p(\lambda),e(\lambda),x_I(\lambda),\chi_\theta^S(\lambda)]\;. \label{eq:tOGeqn}
\end{align}
Integrating up Eqs.\ (\ref{eq:phiOGeqn}) and (\ref{eq:tOGeqn}) for $\phi$ and $t$ along the inspiral is equivalent to solving (\ref{eq:phigeodeq}) and (\ref{eq:tgeodeq}).  Observe that Eqs.\ (\ref{eq:rgeodeq}) -- (\ref{eq:tgeodeq}) arise from Eq.\ (\ref{eq:oscelementevol1}), which in turn arises from (\ref{eq:dotIgeo}). Equation (\ref{eq:dotIgeo}) simply states that the geodesic equation $\ddot x^\alpha = a^\alpha_{\rm geo}$ holds when the osculating elements $\mathcal{E}^A$ are constant.  When $\{p, e, x_I, \chi_r^S, \chi_\theta^S\}$ are all constant, Eqs.\ (\ref{eq:phiOGeqn}) and (\ref{eq:tOGeqn}) yield geodesic solutions; when $\{p, e, x_I, \chi_r^S, \chi_\theta^S\}$ are evolving, we obtain the solution for forced motion.

We therefore need only consider Eqs.\ (\ref{eq:rgeodeq}) and (\ref{eq:thetageodeq}).  We rearrange these equations to obtain
\begin{align}
\chi_r^{S\prime} &= \frac{1}{\partial r/\partial\chi_r^S}\left(\frac{\partial r}{\partial p}p' + \frac{\partial r}{\partial e}e' + \frac{\partial r}{\partial x_I}x_I'\right) \equiv X_r^S(\mathcal{E}^A)\;,
\label{eq:psi0dash}\\
\chi_\theta^{S\prime} &= \frac{1}{\partial \theta/\partial\chi_\theta^S}\left(\frac{\partial \theta}{\partial p}p' + \frac{\partial \theta}{\partial e}e' + \frac{\partial \theta}{\partial x_I}x_I'\right) \equiv X_\theta^S(\mathcal{E}^A)\;.
\label{eq:chi0dash}
\end{align}

We next expand Eq.\ (\ref{eq:oscelementevol2}) just as we expanded (\ref{eq:oscelementevol1}):
\begin{align}
    \frac{\partial \dot r}{\partial p}p'+ \frac{\partial \dot  r}{\partial e}e'+ \frac{\partial \dot  r}{\partial x_I}x_I'+ \frac{\partial \dot  r}{\partial \chi_r^S}\chi_r^{S\prime}+ \frac{\partial \dot  r}{\partial \chi_\theta^S}\chi_\theta^{S\prime}&=a^r\tau'\;,
    \label{eq:rdoteq}\\
    \frac{\partial \dot  \theta}{\partial p}p'+ \frac{\partial \dot  \theta}{\partial e}e'+ \frac{\partial \dot \theta}{\partial x_I}x_I'+ \frac{\partial \dot \theta}{\partial \chi_r^S}\chi_r^{S\prime}+ \frac{\partial \dot \theta}{\partial \chi_\theta^S}\chi_\theta^{S\prime}&=a^\theta\tau'\;,
    \label{eq:thetadoteq}\\
    \frac{\partial \dot  \phi}{\partial p}p'+ \frac{\partial \dot  \phi}{\partial e}e'+ \frac{\partial \dot \phi}{\partial x_I}x_I'+ \frac{\partial \dot \phi}{\partial \chi_r^S}\chi_r^{S\prime}+ \frac{\partial \dot \phi}{\partial \chi_\theta^S}\chi_\theta^{S\prime}&=a^\phi\tau'\;,
    \label{eq:phidoteq}\\
     \frac{\partial \dot  t}{\partial p}p'+ \frac{\partial \dot  t}{\partial e}e'+ \frac{\partial \dot t}{\partial x_I}x_I'+ \frac{\partial \dot t}{\partial \chi_r^S}\chi_r^{S\prime}+ \frac{\partial \dot t}{\partial \chi_\theta^S}\chi_\theta^{S\prime}&=a^t\tau'\;, \label{eq:tdoteq}
\end{align}
Following Refs.\ \cite{Gair2011, Pound2008}, we use the condition $a^\alpha u_\alpha = 0$ to eliminate Eq.\ (\ref{eq:tdoteq}).  Following \cite{Gair2011}, we define the useful expression
\begin{equation}
    \mathcal{L}_b(c) \equiv \frac{\partial\dot c}{\partial b} - \frac{\partial r/\partial b}{\partial r/\partial \chi_r^S}\frac{\partial\dot c}{\partial \chi_r^S} - \frac{\partial \theta/\partial b}{\partial \theta/\partial \chi_\theta^S}\frac{\partial\dot c}{\partial \chi_\theta^S}\;,
\end{equation}
where $b$ denotes $p$, $e$ or $x_I$, and  where $c$ denotes $r$, $\theta$ or $\phi$.  This definition allows us to write Eqs.\ (\ref{eq:rdoteq}) -- (\ref{eq:phidoteq}) in the convenient form
\begin{widetext}
\begin{align}
p'& = \frac{\tau'}{D}\left((\mathcal{L}_e(\theta)\mathcal{L}_{x_I}(\phi) - \mathcal{L}_e(\phi)\mathcal{L}_{x_I}(\theta))a^r + (\mathcal{L}_{x_I}(r)\mathcal{L}_e(r) - \mathcal{L}_I(\phi)\mathcal{L}_e(r))a^\theta + (\mathcal{L}_e(r)\mathcal{L}_I(\theta) - \mathcal{L}_e(\theta)\mathcal{L}_I(r))a^\phi\right)\;,
\label{eq:pdash}\\
e'& = \frac{\tau'}{D}\left((\mathcal{L}_I(\theta)\mathcal{L}_p(\phi) - \mathcal{L}_I(\phi)\mathcal{L}_p(\theta))a^r + (\mathcal{L}_p(r)\mathcal{L}_{x_I}(r) - \mathcal{L}_p(\phi)\mathcal{L}_{x_I}(r))a^\theta + (\mathcal{L}_{x_I}(r)\mathcal{L}_p(\theta) - \mathcal{L}_{x_I}(\theta)\mathcal{L}_p(r))a^\phi\right)\;,
\label{eq:edash}\\
I'& = \frac{\tau'}{D}\left((\mathcal{L}_p(\theta)\mathcal{L}_e(\phi) - \mathcal{L}_p(\phi)\mathcal{L}_e(\theta))a^r + (\mathcal{L}_e(r)\mathcal{L}_p(r) - \mathcal{L}_e(\phi)\mathcal{L}_p(r))a^\theta + (\mathcal{L}_p(r)\mathcal{L}_e(\theta) - \mathcal{L}_p(\theta)\mathcal{L}_e(r))a^\phi\right)\;,
\label{eq:Idash}\\
D& = \mathcal{L}_p(r)\left(\mathcal{L}_e(\theta)\mathcal{L}_{x_I}(\phi) - \mathcal{L}_{x_I}(\theta)\mathcal{L}_e(\phi)\right) - \mathcal{L}_e(r)\left(\mathcal{L}_p(\theta)\mathcal{L}_I(\phi) - \mathcal{L}_{x_I}(\theta)\mathcal{L}_p(\phi)\right) + \mathcal{L}_{x_I}(r)\left(\mathcal{L}_p(\theta)\mathcal{L}_e(\phi) - \mathcal{L}_p(\phi)\mathcal{L}_e(\theta)\right)\;.
\end{align}
\end{widetext}
Equations (\ref{eq:pdash})--(\ref{eq:Idash}) tell us how to evolve the principal orbital elements, given non-geodesic accelerations $a^{r,\theta,\phi}$.

We further substitute these equations into Eqs.\ (\ref{eq:psi0dash})--(\ref{eq:chi0dash}) in order to obtain the evolution of the phase constants $\chi_r^S$ and $\chi_\theta^S$.  This gives us a closed system of ordinary differential equations which allow us to evolve $p$, $e$, $x_I$, $\chi_r^S$, and  $\chi_\theta^S$ given the non-geodesic accelerations $a^{r,\theta,\phi}$.  Augmenting with two auxiliary equations for $t$ and $\phi$, Eqs.\ (\ref{eq:phiOGeqn}) and (\ref{eq:tOGeqn}), yields a complete scheme for evolving the elements of our phase space, $\{p, e, x_I, \chi_r^S, \chi_\theta^S, \phi, t \}$.

\subsection{Action-angle evolution equations}

\label{sec:AAevol}

Action-angle coordinates are very useful for formulating near-identity transformations. In the action-angle picture, the OG equations of motion are given by \cite{Lynch2021}
\begin{align}
    \frac{dP_j}{d\lambda} &= F_j(\vec{P}, \vec q)\;,
    \\
    \frac{dq_i}{d\lambda} &= \hat \Upsilon_i(\vec{P}) + f_i^{(1)}(\vec{P}, \vec q)\;.
    \label{eq:AA3}
\end{align}
Here, $\vec{P}=\{p, e, x_I\}$ and $\vec q=\{q_r,q_z\}$.  We write the explicit forms for these equations below. The $F_j(\vec{P},\vec q)$ terms are given by
\begin{align}
\frac{dp}{d\lambda} &= \frac{2}{(r_1+r_2)^2}\left[r_2^2\frac{dr_1}{d\lambda} + r_1^2\frac{dr_2}{d\lambda}\right]^{-1} \equiv F_p\;,
\\
\frac{de}{d\lambda} &= \frac{2}{(r_1+r_2)^2}\left[r_2\frac{dr_1}{d\lambda} + r_1\frac{dr_2}{d\lambda}\right]^{-1} \equiv F_e\;,
\\
\frac{dx_I}{d\lambda} &= -\frac{z_-}{x_I}\frac{dz_-}{d\lambda} \equiv F_{x_I}\;.
\end{align}

The $f_i^{(1)}$ terms are given by
\begin{align}
\frac{dq_{i,0}}{d\lambda} = -\frac{1}{\partial x_G^i/\partial q_i}\left(\frac{\partial x^i_G}{\partial P_j}\frac{dP_j}{d\lambda}\right) \equiv f_i^{(1)}.
\end{align}
For detail about the derivation of these expressions, refer to Appendix C of Ref.\ \cite{Lynch2021}.

\section{Near-identity transformation details}
\label{app:NITdetails}

In this appendix, we describe in some detail the equations underlying the NIT.  Further details can be found in Refs.\ \cite{vandeMeent2018_2, Lynch2021, Lynch2022, Lynch2023}.

\subsection{Mino-time NIT derivation}
\label{app:NITMinoderiv}
\subsubsection{Inverse NIT}

The inverse transformations can be found for $P_k$ and $q_i$ by requiring that their composition with the transformations in Eqs.\ \eqref{eq:transformation} must give the identity transformation. Expanding order by order in $\varepsilon$, this gives us
\begin{subequations}\label{eq:inverse_trasformaiton}
	\begin{align}
		\begin{split}
			P_j &= \nit{P_j} - \varepsilon Y_j^{(1)}(\vec{\nit{P}},\vec{\nit{q}},\nit{\psi}_s) + \mathcal{O}(\varepsilon^2)\;,
		\end{split}\\
		\begin{split}
			q_i &= \nit{q_i} - \varepsilon X_i^{(1)}(\vec{\nit{P}},\vec{\nit{q}},\nit{\psi}_s)  + \mathcal{O}\varepsilon^2)\;,
		\end{split}\\
		\begin{split}
			\psi_s &= \nit{\psi}_s - W_s^{(0)}(\vec{\nit{P}},\vec{\nit{q}}) 
			\\& - \varepsilon \Biggl( W_s^{(1)}(\vec{\nit{P}},\vec{\nit{q}},\nit{\psi}_s)  - \PD{W_s^{(0)}(\vec{\nit{P}},\vec{\nit{q}})}{\nit{P_j}} Y_j^{(1)}(\vec{\nit{P}},\vec{\nit{q}}) 
			\\& - \PD{W_s^{(0)}(\vec{\nit{P}},\vec{\nit{q}})}{\nit{q_i}} X_i^{(1)}(\vec{\nit{P}},\vec{\nit{q}})
			\Biggr) + \mathcal{O}(\varepsilon^2)\;.
		\end{split}
	\end{align}
\end{subequations}
	
\subsubsection{Transformed equations of motion}

By taking the time derivative of the NIT \eqref{eq:transformation}, substituting the EMRI equations of motion \eqref{eq:Generic_EMRI_EoM} and inverse NIT \eqref{eq:inverse_trasformaiton}, and expanding in powers of $\mr$ we obtain the NIT transformed equations of motions
\begin{subequations}\label{eq:transformed_EoM_oscillating}
	\begin{align}
		\begin{split}
			\frac{d \nit{P}_j}{d \lambda} &=\varepsilon \nit{F}_j^{(1)}(\vec{\nit{P}},\vec{\nit{q}},\nit{\psi}_s) + \varepsilon^2 \nit{F}_j^{(2)}(\vec{\nit{P}},\vec{\nit{q}},\nit{\psi}_s) +  \mathcal{O}(\varepsilon^3)\;,
		\end{split}\\
		\begin{split}
			\frac{d \nit{q}_i}{d\lambda} &= \Upsilon_i^{(0)}(\vec{\nit{P}}) +\varepsilon \Upsilon_i^{(1)}(\vec{\nit{P}},\vec{\nit{q}},\nit{\psi}_s) + \mathcal{O}(\varepsilon^2)\;,
		\end{split}\\
		\begin{split}
			\frac{d \nit{\psi}_s}{d\lambda} &= \Upsilon_s^{(0)}(\vec{\nit{P}},\vec{\nit{q}},\nit{\psi}_s) +\varepsilon \Upsilon_s^{(1)}(\vec{\nit{P}},\vec{\nit{q}},\nit{\psi}_s) + \mathcal{O}(\varepsilon^2)\;,
		\end{split}\\
		\end{align}
\end{subequations}
where 
\begin{subequations}
\begin{align}
	\begin{split}
	    \Upsilon_s^{(0)} & = f^{(0)}_s + \PD{W_j^{(1)}}{\nit{q}_i} \Upsilon_i^{(0)}\;,
	\end{split}\\
	\begin{split}
		\nit{F}_j^{(1)} & = F^{(1)}_j + \PD{Y_j^{(1)}}{\nit{q}_i} \Upsilon_i^{(0)} + \PD{Y_j^{(1)}}{\nit{\psi}_s} \Upsilon_s^{(0)}\;,
	\end{split}\\
	\begin{split}
	    \Upsilon_s^{(1)} & = \PD{W_s^{(1)}}{\nit{q}_i} \Upsilon^{(0)}_i + \PD{W_s^{(1)}}{\nit{\psi}_s} \Upsilon_s^{(0)} 
	    \\& - \PD{f^{(0)}_s}{\nit{P}_j} Y_j^{(1)} -\PD{f^{(0)}_s}{\nit{q}_i} X_i^{(1)}\;,
	\end{split}\\
	\begin{split}
		\Upsilon_i^{(1)} & = f^{(1)}_i + \PD{X_i^{(1)}}{\nit{q}_k} \Upsilon^{(0)}_k + \PD{X_i^{(1)}}{\nit{\psi}_s} \Upsilon_s^{(0)} - \PD{\Upsilon^{(0)}_i}{\nit{P}_j} Y_j^{(1)}\;,
	\end{split}\\
	\begin{split}
	    \nit{F}_j^{(2)} & = F^{(2)}_j + \PD{Y_j^{(2)}}{\nit{q}_i} \Upsilon_i^{(0)} + \PD{Y_j^{(1)}}{\nit{q}_i} f_i^{(1)} +  \PD{Y_j^{(1)}}{\nit{P}_k} F_k^{(1)}
	    \\& -\PD{\nit{F}_j^{(1)}}{\nit{P}_k} Y_k^{(1)} - \PD{\nit{F}_j^{(1)}}{\nit{q}_i}X_i^{(1)} - \PD{\nit{F}_j^{(1)}}{\nit{\psi}_s} W_s^{(1)}\;.
	\end{split}
\end{align}
\end{subequations}
Note that all functions on the right hand side are evaluated at $\vec{\nit{P}}$, $\vec{\nit{q}}$, and $\nit{\psi}_s$ and that we have adopted the convention that all repeated roman indices are summed over.  Notice also that $\Upsilon_s^{(1)}$ will be suppressed by a factor of the mass-ratio: every term it appears in is proportional to secondary spin, and therefore will not contribute at 1PA order. We include these terms here for completeness, but only the 1PA contributions appear in Sec.\ \ref{sec:AveragedEoM}.
	
\subsubsection{Cancellation of oscillating terms at adiabatic order}
	
	We can recast the expression for $\nit{\Upsilon}^{(0)}_s$ as 
	
	\begin{align}
			\begin{split}
				\Upsilon_s^{(0)} & = f^{(0)}_s + \PD{W_s^{(1)}}{\nit{q}_i} \Upsilon_i^{(0)} \\
								&= \avg{f^{(0)}_s} + \sum_{\vec{\kappa} \neq \vec{0}} \left(f^{(0)}_{s,\vec{\kappa}} + i (\vec{\kappa} \cdot \vec{\Upsilon}^{(0)}) \osc{W}^{(0)}_{s,\vec{\kappa}}  \right) e^{i \vec{\kappa} \cdot \vec{q}}\;.
			\end{split}
		\end{align}
	As such, we can cancel the oscillatory pieces of $\nit{\Upsilon}^{(0)}_s$ by choosing the oscillatory part of $W_s^{(0)}$ to be
	\begin{equation}
		\osc{W}^{(0)}_{j,\vec{\kappa}} \equiv \frac{i}{\vec{\kappa} \cdot \vec{\Upsilon}^{(0)}} f^{(0)}_{s,\vec{\kappa}}(\vec{P}) = - (\psi_{sr}(q_r) + \psi_{sz}(q_z))\;.
	\end{equation}
	Conveniently, this is related to the oscillating pieces of the geodesic solution for the spin phase which is known analytically. Due to the separability of this solution, this transformation is always well defined, even in the presence of orbital resonances where $\vec{\kappa}_{\text{res}} = \{k_r,k_z \}$ where $k_r,k_z \in \mathbb{Z}$, such that $\vec{\kappa}_{\text{res}} \cdot \vec{\Upsilon}^{(0)} = k_r \Upsilon_r^{(0)} + k_z \Upsilon_z^{(0)} = 0$.
	
	We can continue with this analysis and recast the expression for $\nit{F}^{(1)}_j$ as
		\begin{align}
			\begin{split}
				\nit{F}^{(1)}_j &=  F^{(1)}_j + \PD{Y_j^{(1)}}{\nit{q}_i} \Upsilon^{(0)}_i + \PD{Y_j^{(1)}}{\nit{\psi}_s} \Upsilon^{(0)}_s = F^{(1)}_j + \PD{Y_j^{(1)}}{\nit{\mathcal{Q}}_k} \Upsilon^{(0)}_k\\
				&= \avg{F^{(1)}_j} + \sum_{\vec{\kappa} \neq \vec{0}} \left(F^{(1)}_{j,\vec{\kappa}} + i (\vec{\kappa} \cdot \vec{\Upsilon}^{(0)}) \osc{Y}^{(1)}_{j,\vec{\kappa}}  \right) e^{i \vec{\kappa} \cdot \vec{\mathcal{Q}}}\;.
			\end{split}
		\end{align}
	As such, we can cancel the oscillatory pieces of $\nit{F}_j^{(1)}$ by choosing the oscillatory part of $Y_j^{(1)}$ to be
	\begin{equation}
		\osc{Y}^{(1)}_{j,\vec{\kappa}} \equiv \frac{i}{\vec{\kappa} \cdot \vec{\Upsilon}^{(0)}} F^{(1)}_{j,\vec{\kappa}}(\vec{P})\;.
	\end{equation}
	Clearly, one can only make this choice so long as there is no $\vec{\kappa}_{\text{res}} = \{\kappa_r,\kappa_z,\kappa_s\}$ where $\kappa_r,\kappa_z ,\kappa_s\in \mathbb{Z}$, such that $\vec{\kappa}_{\text{res}} \cdot \vec{\Upsilon}^{(0)} = \kappa_r \Upsilon_r^{(0)} + \kappa_z \Upsilon_z^{(0)} + \kappa_s \Upsilon_s^{(0)} = 0$.  This is occasionally the case in the presence of resonances, where the radial and polar frequencies become commensurate or when the spin, the radial and/or the polar frequencies become commensurate.  We have carefully chosen our data grids so that we do not encounter such orbits in our study.
	
	\subsubsection{Cancellation of oscillating terms at post-adiabatic order}
 
	Using the above choice for $\osc{Y}_j^{(1)}$, the equation for $\Upsilon_i^{(1)}$ becomes
	\begin{align}
		\begin{split}
			\Upsilon_i^{(1)} &= f^{(1)}_i - \PD{\Upsilon^{(0)}_i}{\nit{P}_j} Y_j^{(1)} + \PD{X_i^{(1)}}{\nit{q}_k} \Upsilon^{(0)}_k + \PD{X_i^{(1)}}{\nit{\psi}_s} \Upsilon^{(0)}_s   
			\\ & = f^{(1)}_i - \PD{\Upsilon^{(0)}_i}{\nit{P}_j} Y_j^{(1)} + \PD{X_i^{(1)}}{\nit{\mathcal{Q}}_k} \Upsilon^{(0)}_k
			 \\ &= \avg{f^{(1)}_i} - \PD{\Upsilon^{(0)}_i}{\nit{P}_j} \avg{Y_j^{(1)}} 
			 \\ &+ \sum_{\vec{\kappa} \neq \vec{0}} \Biggl( f^{(1)}_{i,\vec{\kappa}} 
			 +   i (\vec{\kappa} \cdot \vec{\Upsilon}^{(0)} )\osc{X}^{(1)}_{i,\vec{\kappa}}
			 \\&- \frac{i}{\vec{\kappa} \cdot \vec{\Upsilon}^{(0)}} \PD{\Upsilon^{(0)}_i}{\nit{P}_j} F^{(1)}_{j,\vec{\kappa}} \Biggr) e^{i \vec{\kappa} \cdot \vec{\mathcal{Q}}}\;.
		\end{split}
	\end{align}
	As a result, we can remove the oscillating pieces of $\Upsilon_i^{(1)}$ by choosing
	\begin{equation}
		\osc{X}_{i, \vec{\kappa}}^{(1)} \equiv  \frac{i}{\vec{\kappa} \cdot \vec{\Upsilon}^{(0)}} f_{i,\vec{\kappa}}^{(1)} + \frac{1}{(\vec{\kappa} \cdot \vec{\Upsilon}^{(0)})^2} \frac{\partial \Upsilon_i^{(0)}}{\partial P_j}F_{j,\vec{\kappa}}^{(1)}\;.
	\end{equation}
	
	Similarly, looking at the equation for $\Upsilon_s^{(1)}$, we see that:
	\begin{align}
	\begin{split}
	    \Upsilon_s^{(1)} = & \PD{W_s^{(1)}}{\nit{q}_i} \Upsilon^{(0)}_i + \PD{W_s^{(1)}}{\nit{\psi}_s} \Upsilon_s^{(0)} - \PD{f^{(0)}_s}{\nit{P}_j} Y_j^{(1)} -\PD{f^{(0)}_s}{\nit{q}_i} X_i^{(1)}  
	    \\ = & \PD{W_s^{(1)}}{\nit{q}_k} \Upsilon^{(0)}_k - \PD{f^{(0)}_s}{\nit{P}_j} Y_j^{(1)} -\PD{f^{(0)}_s}{\nit{q}_i} X_i^{(1)}
	    \\ = &  \PD{W_s^{(1)}}{\nit{q}_k} \Upsilon^{(0)}_s - \avg{\PD{f^{(0)}_s}{\nit{P}_j} Y_j^{(1)}} - \avg{\PD{f^{(0)}_s}{\nit{q}_i} X_i^{(1)}} 
	    \\& -\left\{ \PD{f^{(0)}_s}{\nit{P}_j} Y_j^{(1)} \right\} - \left\{ \PD{f^{(0)}_s}{\nit{q}_i} X_i^{(1)} \right\}
	    \\ = & - \avg{\PD{f^{(0)}_s}{\nit{P}_j} Y_j^{(1)}} - \avg{\PD{f^{(0)}_s}{\nit{q}_i} X_i^{(1)}}
	    \\&+ \sum_{ \vec{\kappa} \neq \vec{0}} \Biggl( i (\vec{\kappa} \cdot \vec{\Upsilon}^{(0)}) \osc{W}^{(1)}_{s,\vec{\kappa}} 
	    \\& - \sum_{ \vec{\kappa}' \neq \vec{0}} \Biggl[
	    \PD{f^{(0)}_{s,\kappa'}}{\nit{P}_j} Y_{j, \vec{\kappa} - \vec{\kappa}'}^{(1)} +  \PD{f^{(0)}_{s,\vec{\kappa}'}}{\nit{q}_i} X_{i\vec{\kappa} - \vec{\kappa}'}^{(1)}
	     \Biggr] \Biggr) e^{i \vec{\kappa} \cdot \vec{\mathcal{Q}}}\;,
	\end{split}
	\end{align}
	where we introduced the additional notation $\{\cdot\}$ to denote the oscillatory part of a product of functions.
	From this we obtain
	\begin{align}
		\begin{split}
			\osc{W}^{(1)}_{s,\vec{\kappa}} = & \frac{i}{\vec{\kappa} \cdot \vec{\Upsilon}^{(0)}} 
			\Biggl(  \sum_{\vec{\kappa}' \neq \vec{0}} \Biggl[
	    \PD{f^{(0)}_{s,\kappa'}}{\nit{P}_j} Y_{j, \vec{\kappa} - \vec{\kappa}'}^{(1)} +  \PD{f^{(0)}_{s,\vec{\kappa}'}}{\nit{q}_i} X_{i\vec{\kappa} - \vec{\kappa}'}^{(1)}
	     \Biggr] \Biggr)\;.
		\end{split}
	\end{align}
	
	Using the above choice for $\osc{Y}_j^{(1)}$, we can express the oscillatory part of the expression for $\nit{F}_j^{(2)}$ as
	\begin{align}
		\begin{split}
		\osc{\nit{F}}^{(2)}_j = & \osc{F}_j^{(2)} + \PD{\osc{Y}_j^{(2)}}{\nit{\mathcal{Q}}_k} \Upsilon_k^{(0)} + \left\{\PD{Y_j^{(1)}}{\nit{q}_i} f_i^{(1)} \right\} 
		\\& +  \left\{\PD{Y_j^{(1)}}{\nit{P}_k} F_k^{(1)} \right\} - \PD{\avg{F_j^{(1)}}}{\nit{P}_k} \osc{Y}_k^{(1)}
		\\ =& \sum_{\vec{\kappa} \neq \vec{0}}  \Biggl( 
			 F_{j,\vec{\kappa}}^{(2)} + i (\vec{\kappa} \cdot \vec{\Upsilon}^{(0)}) \osc{Y}^{(2)}_{j,\vec{\kappa}}
			 + \PD{\avg{Y_j^{(1)}}}{\nit{P}_k} F^{(1)}_{k,\kappa} 
		\\ & - i \PD{\avg{F_j^{(1)}}}{\nit{P}_k} \frac{F_{k,\vec{\kappa}}}{\vec{\kappa} \cdot \vec{\Upsilon}^{(0)}} +   \sum_{\vec{\kappa}' \neq \vec{0}} \biggl(   i \frac{F^{(1)}_{k,\vec{\kappa} - \vec{\kappa} '}}{\vec{\kappa}' \cdot \vec{\Upsilon}^{(0)}} \biggl( \PD{F^{(1)}_{j,\vec{\kappa}'}}{\nit{P}_k}
		\\ &  - \frac{F^{(1)}_{j,\vec{\kappa} '}}{\vec{\kappa}' \cdot \vec{\Upsilon}^{(0)}} \PD{(\vec{\kappa}' \cdot \vec{\Upsilon}^{(0)}) } {\nit{P}_k}  \biggr) 
		 - \frac{\vec{\kappa}' \cdot \vec{f}^{(1)}_{\vec{\kappa} - \vec{\kappa}'}}{\vec{\kappa}' \cdot \vec{\Upsilon}^{(0)}} F^{(1)}_{j,\vec{\kappa}'}
		\biggr) \Biggr) e^{i \kappa \cdot \vec{\mathcal{Q}}}\;,
		\end{split}
	\end{align}
	Thus we can remove the oscillatory part of $\nit{F}_j^{(2)}$ by choosing
	\begin{align}
		\begin{split}
			\osc{Y}^{(2)}_{j,\vec{\kappa}} = & \frac{i}{\vec{\kappa} \cdot \vec{\Upsilon}^{(0)}} \Biggl( 
				F_{j,\vec{\kappa}}^{(2)} + \PD{\avg{Y_j^{(1)}}}{\nit{P}_k} F^{(1)}_{k,\vec{\kappa}} 
				\\ & - i \PD{\avg{F_j^{(1)}}}{\nit{P}_k} \frac{F^{(1)}_{k,\vec{\kappa}}}{\vec{\kappa} \cdot \vec{\Upsilon}^{(0)}}
			 +   \sum_{\vec{\kappa}' \neq \vec{0}} \biggl(   i \frac{F^{(1)}_{k,\vec{\kappa} - \vec{\kappa} '}}{\vec{\kappa}' \cdot \vec{\Upsilon}^{(0)}} \biggl( \PD{F^{(1)}_{j,\vec{\kappa}'}}{\nit{P}_k} 
            \\ &- \frac{F^{(1)}_{j,\vec{\kappa} '}}{\vec{\kappa}' \cdot \vec{\Upsilon}^{(0)}} \PD{(\vec{\kappa}' \cdot \vec{\Upsilon}^{(0)}) } {\nit{P}_k}  \biggr) 
		 - \frac{\vec{\kappa}' \cdot \vec{f}^{(1)}_{\vec{\kappa}- \vec{\kappa}'}}{\vec{\kappa}' \cdot \vec{\Upsilon}^{(0)}} F^{(1)}_{j,\vec{\kappa}'}
			\biggr) \Biggr)\;.
		\end{split}
	\end{align}
	
	\subsubsection{Freedom in the averaged pieces}
	With the oscillatory pieces of the NIT equations of motion removed, terms in the equations of motion become
	\begin{subequations}
		\begin{gather}
			\nit{F}_j^{(1)} = \avg{F^{(1)}_j}\;, \quad \Upsilon_s^{(0)} = \avg{f^{(0)}_s}\;, \tag{\theequation a-b}
		\end{gather}
		\begin{align}
		    \Upsilon_i^{(1)} = \avg{f^{(1)}_i} - \PD{\Upsilon^{(0)}_i}{\nit{P}_j}\;,
		\end{align}
	\end{subequations}
	and 
	\begin{align}
	\begin{split}
	\Upsilon_s^{(1)} = & - \avg{\PD{\osc{f}^{(0)}_s}{\nit{P}_j} \osc{Y}_j^{(1)}} - \avg{\PD{\osc{f}^{(0)}_s}{\nit{q}_i} \osc{X}_i^{(1)}}
	\\&- \PD{\avg{f^{(0)}_s}}{\nit{P}_j} \avg{Y_j^{(1)}}\;,
	\end{split}\\
	\begin{split}
		\nit{F}_j^{(2)} = & \avg{F^{(2)}_j} + \avg{\PD{\osc{Y}_j^{(1)}}{\nit{q}_i} \osc{f}_i^{(1)}} + \avg{\PD{\osc{Y}_j^{(1)}}{\nit{P}_k} F_k^{(1)}} 
		\\&+ \PD{\avg{Y_j^{(1)}}}{\nit{P}_k} \avg{F_k^{(1)}} -\PD{\avg{F_j^{(1)}}}{\nit{P}_k} \avg{Y_k^{(1)}}\;.
	\end{split}
	\end{align}
	 Note that we still have freedom to set the averaged pieces of the transformation functions $\avg{Y_j^{(1)}}$, $\avg{Y_j^{(2)}}$, $\avg{W_s^{(0)}}$, $\avg{W_s^{(1)}}$, and  $\avg{X_i^{(1)}}$ to be anything we choose.
	There are many valid and interesting choices that one could make that are explored in Refs.~\cite{vandeMeent2018_2,Lynch2021,Lynch2022}. For this work, we make use of the simplest choice: $\avg{Y_j^{(1)}}=\avg{Y_j^{(2)}}=\avg{W_s^{(0)}}=\avg{W_s^{(1)}}=\avg{X_i^{(1)}} = 0$, as this makes it easy to compare between OG and NIT inspirals. 
	It also has the benefit of drastically simplifying equations of motion to
	\begin{subequations}
		\begin{gather}
			\nit{F}_j^{(1)} = \avg{F^{(1)}_j}\;, \Upsilon_s^{(0)} = \avg{f^{(0)}_s}\;,
			\Upsilon_i^{(1)} = \avg{f^{(1)}_i}\;, \tag{\theequation a-c}
		\end{gather}
	\end{subequations}
	and 
	\begin{align}
	\begin{split}
	\Upsilon_s^{(1)} & = - \avg{\PD{\osc{f}^{(0)}_s}{\nit{P}_j} \osc{Y}_j^{(1)}} - \avg{\PD{\osc{f}^{(0)}_s}{\nit{q}_i} \osc{X}_i^{(1)}}\;,
	\end{split}\\
	\begin{split}
		\nit{F}_j^{(2)} & = \avg{F^{(2)}_j} + \avg{\PD{\osc{Y}_j^{(1)}}{\nit{q}_i} \osc{f}_i^{(1)}} + \avg{\PD{\osc{Y}_j^{(1)}}{\nit{P}_k} F_k^{(1)}}\;.
	\end{split}
	\end{align}
	
	\subsubsection{Evolution of extrinsic quantities}
 \label{section:NIT_extrinsic}
	Now we look to remove the oscillatory pieces of the evolution equations for the extrinsic quantities $\vec{\mathcal{X}}$:
	\begin{equation}
		\frac{d \mathcal{X}}{d \lambda} = f_k^{(0)} (\vec{P},\vec{q})\;.
	\end{equation}
	Since these terms do not depend directly on the spin phase $\psi$, this calculation goes through the same as in the non-spinning case.  Substituting the inverse transformation \eqref{eq:inverse_trasformaiton} and re-expanding in $\mr$ we can write this as an equation involving only the NIT variables $\vec{\nit{P}}$ and $\vec{\nit{q}}$,
	\begin{equation}
		\frac{d \mathcal{X}}{d \lambda} = f_k^{(0)} - \mr \left( \PD{f_k^{(0)}}{\nit{P}_j} Y_j^{(1)} + \PD{f_k^{(0)}}{\nit{q}_i} X_i^{(1)}  \right) + \HOT{2}\;,
	\end{equation}
	where all of the functions on the right hand side are now functions of $\vec{\nit{P}}$ and $\vec{\nit{q}}$.
	In order to remove the oscillatory pieces of the equations, we make use of a new set of extrinsic quantities $\vec{\nit{\mathcal{X}}}$ that are related to the original quantities by the following transformation:
	\begin{equation} \label{eq:extrinsic_transformation}
		\nit{\mathcal{X}}_k = \mathcal{X}_k + Z_k^{(0)} (\vec{\nit{P}},\vec{\nit{q}}) + \mr Z_k^{(1)} (\vec{\nit{P}},\vec{\nit{q}})\;.
	\end{equation}
	We note that since this transformation has a zeroth order in mass ratio term $Z_k^{(0)}$, it is not an near-identity transformation. 
	Thus when we produce waveforms it will be necessary to be able to calculate $Z_k^{(0)}$ explicitly.
	
	We then take the time derivative of \eqref{eq:extrinsic_transformation}, substitute the equations of motion for $\vec{\mathcal{X}}$ and expand order by order to obtain equations of motion for $\vec{\nit{\mathcal{X}}}$:
	\begin{equation} \label{eq:extrinsic_transformation2}
		\frac{d \nit{\mathcal{X}}_k}{d \lambda} = \Upsilon_k^{(0)} + \mr \Upsilon_k^{(1)} + \HOT{2}\;,
	\end{equation}
	 where  
	\begin{subequations}
	 	\begin{align}
	 		\begin{split}
	 			\Upsilon_k^{(0)} =& f_k^{(0)} + \PD{Z_k^{(0)}}{\nit{q}_i} \Upsilon_i^{(0)}\;,
	 		\end{split}\\
 		\begin{split}
 			\Upsilon_k^{(1)} = &
 			\PD{Z^{(0)}_k}{\nit{q}_i} \Upsilon_i^{(1)} + \PD{Z^{(0)}_k}{\nit{P}_j} \nit{F}_j^{(1)} + \PD{Z_k^{(1)}}{\nit{q}_i} \Upsilon_i^{(0)}  
 			\\ & - \PD{f_k^{(0)}}{\nit{P}_j} Y^{(1)}_j - \PD{f_k^{(0)}}{\nit{q}_i} X^{(1)}_i\;.
 		\end{split}
	 	\end{align}
	 \end{subequations}
 	We can now remove the oscillating pieces of the functions $\Upsilon_k^{(0)}$ by solving the equations
 	\begin{subequations}
 		\begin{align}
 			\begin{split}
 				0 = & \osc{f}_k^{(0)} + \PD{\osc{Z}_k^{(0)}}{\nit{q}_i} \Upsilon_i^{(0)}\;,
 			\end{split}\\
 			\begin{split}
 				 0  = &  \PD{\osc{Z}^{(0)}_k}{\nit{q}_i} \Upsilon_i^{(1)} + \PD{\osc{Z}^{(0)}_k}{\nit{P}_j} \nit{F}_j^{(1)}+ \PD{\osc{Z}_k^{(1)}}{\nit{q}_i} \Upsilon_i^{(0)}
 				 \\& - \left\{ \PD{f_k^{(0)}}{\nit{P}_j} Y^{(1)}_j \right\}  - \left\{ \PD{f_k^{(0)}}{\nit{q}_i} X^{(1)}_i \right\}\;,
 			\end{split}
 		\end{align}
 	\end{subequations}
 	for the oscillatory parts of the transformation $\osc{Z}_k^{(0)}$ and $\osc{Z}_k^{(1)}$. 
 	The first of these is satisfied by using the oscillating pieces for the analytic solutions for the geodesic motion of $t$ and $\phi$, 
 	\begin{equation}\label{eq:Z_solution}
 		\osc{Z}_k^{(0)} = - \osc{\mathcal{X}}_{k,r} (q_r) - \osc{\mathcal{X}}_{k,z} (q_z)\;.
 	\end{equation}
 	It is unclear whether the equation for $Z_k^{(1)}$ would yield analytic solutions, but it can be solved numerically. 
 	Since we only need to know the extrinsic quantities to $\mathcal{O}(\mr)$ to generate waveforms, we do not need to be able to calculate this explicitly and it is sufficient to know that a solution exists. 
 	
 	Now the forcing functions only depend only on $\vec{\nit{P}}$ and are given by
 	\begin{subequations}
 		\begin{align}
 			\begin{split}
 				\Upsilon_k^{(0)} =& \avg{f_k^{(0)}}\;,
 			\end{split}\\
 			\begin{split}
 				\Upsilon_k^{(1)} =& \PD{\avg{Z^{(0)}_k}}{\nit{P}_j} \nit{F}_j^{(1)} -\PD{\avg{f_k^{(0)}}}{\nit{P}_j}\avg{ Y^{(1)}_j} 
 				\\& - \avg{\PD{\osc{f}_k^{(0)}}{\nit{P}_j} \osc{Y}^{(1)}_j} - \avg{\PD{\osc{f}_k^{(0)}}{\nit{q}_i} \osc{X}^{(1)}_i}\;.
 			\end{split}
 		\end{align}
 	\end{subequations}
 	Again, we have freedom in the average pieces of the transformation functions which we use to simplify this problem further. 
 	As before, we chose the simplest option and set $\avg{Z_k^{(0)}} = 0$ which along with our previous choices simplifies the expression for $\Upsilon_k^{(1)}$ to be 
 	
 	\begin{equation}
 		\Upsilon_k^{(1)} = - \avg{\PD{\osc{f}_k^{(0)}}{\nit{P}_j} \osc{Y}^{(1)}_j} - \avg{\PD{\osc{f}_k^{(0)}}{\nit{q}_i} \osc{X}^{(1)}_i}\;.
 	\end{equation}

  \begin{figure}
\centerline{\includegraphics[scale=0.6]{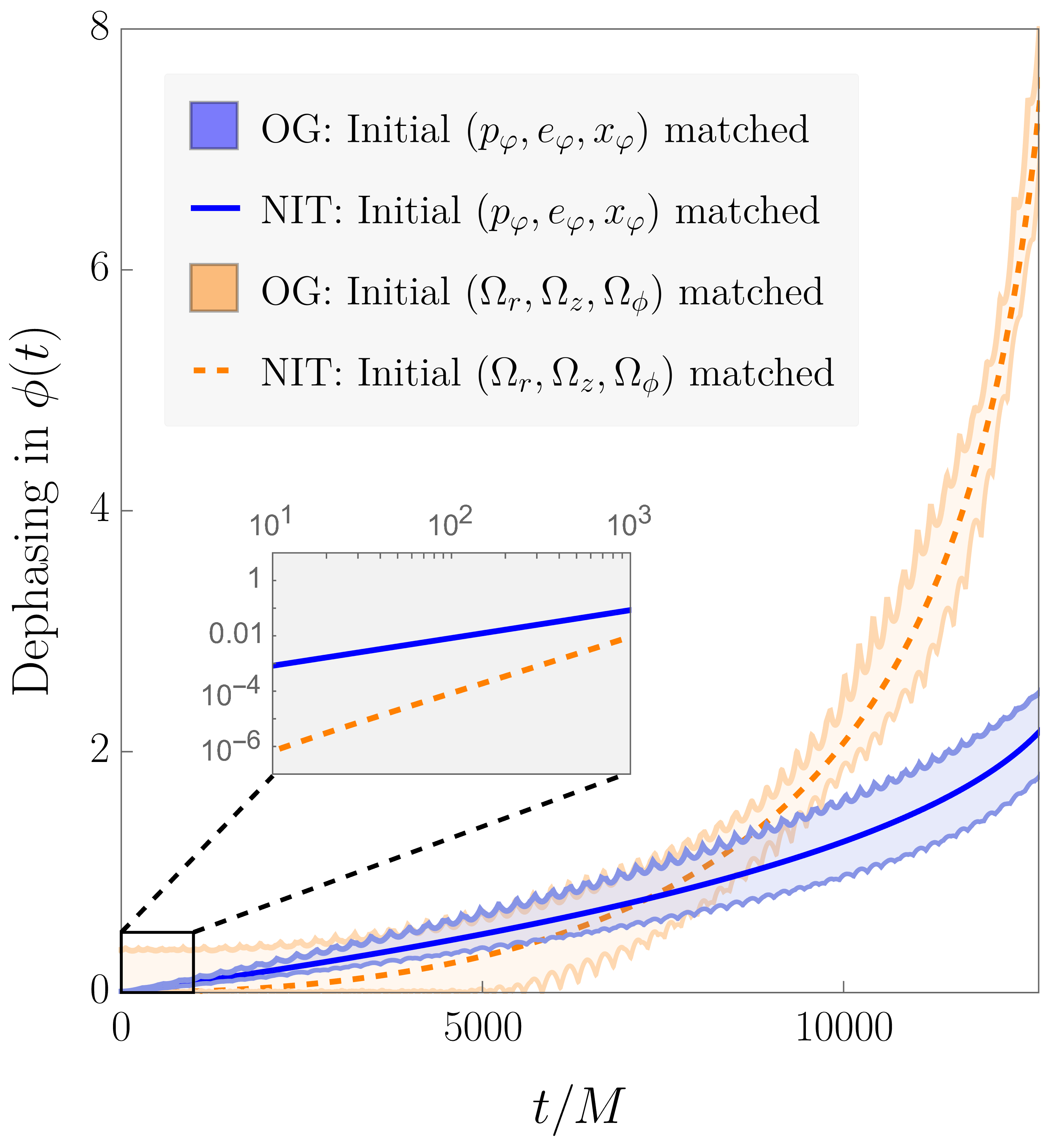}}
    \caption{Dephasing in $\phi(t)$ of a spinning-body orbit relative to a non-spinning body orbit for two different choices of initial conditions.  The system has mass ratio $\varepsilon=10^{-2}$ and the small body orbits a black hole with spin $a=0.7 M$. The magnitude and orientation of the small body's spin is specified by $s = 1$, $s_\parallel=s$.  The blue curves correspond to matching initial orbital elements $(p_\varphi,e_\varphi,x_\varphi)$ while the orange curves denotes matching the initial Boyer-Lindquist frequencies $(\Omega_r,\Omega_z,\Omega_\phi)$.  The solid curves show the averaged dephasing of $\phi(t)$, i.e., $\varphi_\phi^{RR+SCF}-\varphi_\phi^{RR}$ while the shading shows the dephasing of $\phi(t)$ given directly by the OG equations, i.e., $\phi^{RR+SCF}-\phi^{RR}$. Initially, $p=9.5$, $e=0.19$, and  $x_I=0.699$.}
     \label{fig:deltaphiIC}
\end{figure}

\begin{figure}
\centerline{\includegraphics[scale=0.55]{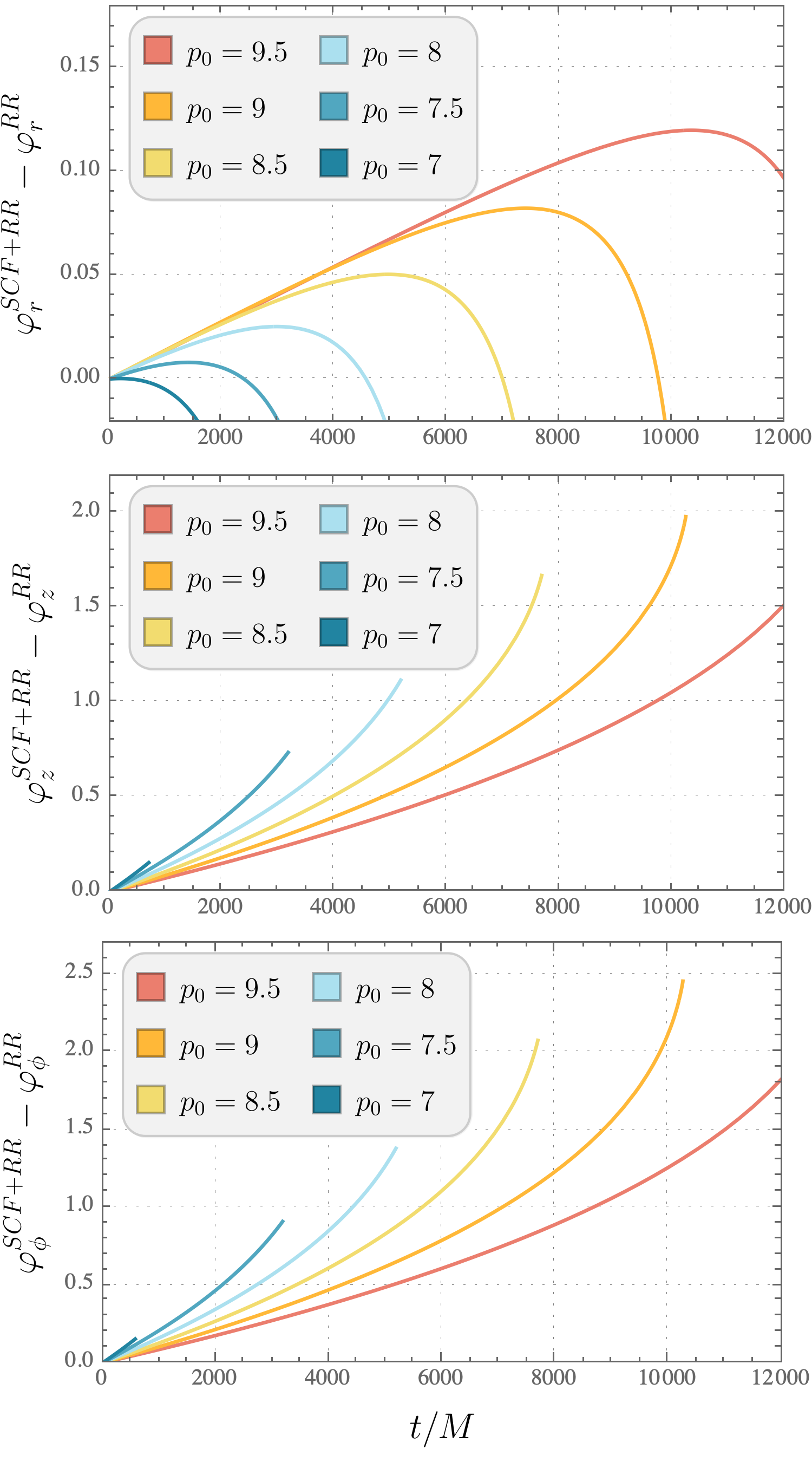}}
    \caption{Averaged dephasing in $q_r(t)$, $q_z(t)$, and  $\phi(t)$ for a spinning body relative to a non-spinning body with mass ratio $\varepsilon=10^{-2}$ orbiting a black hole with spin $a=0.7 M$; the small body's spin is given $s = 1$, $s_\parallel=s$.  The top panel shows $\varphi_r^{SCF+RR}-\varphi_r^{RR}$, the middle panel shows $\varphi_z^{SCF+RR}-\varphi_z^{RR}$ and the bottom panel shows $\varphi_\phi^{SCF+RR}-\varphi_\phi^{RR}$.  Different colors correspond to different initial $p$ values for the inspiral; duration of inspiral also correlates with initial $p$ (inspiral with $p_0 = 9.5$ is longest, that with $p_0 = 7$ is shortest, etc.).  For all panels, $e=0.2$, $x_I=0.7$, $q_r=0$, $q_z=0$, and $\phi=0$ initially.}
    \label{fig:pinitplot}
\end{figure}

\begin{figure*}
\centerline{\includegraphics[scale=0.55]{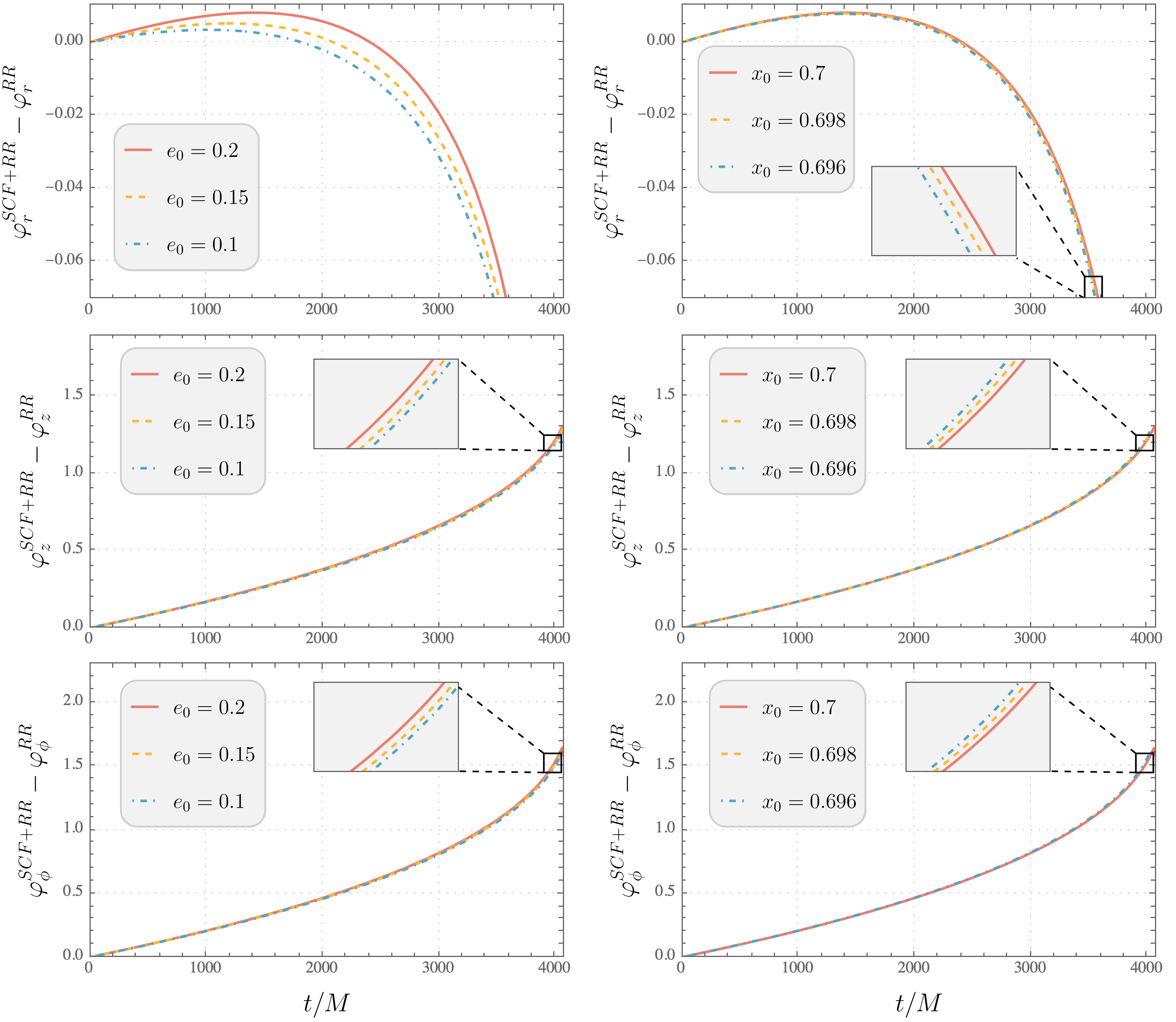}}
    \caption{Averaged dephasing in $q_r(t)$, $q_z(t)$, and $\phi(t)$ for a spinning body relative to a non-spinning body with mass ratio $\varepsilon=10^{-2}$ orbiting a black hole with spin $a=0.7 M$; the small body's spin has $s = 1$, $s_\parallel=s$.  The top row shows $\varphi_r^{SCF+RR}-\varphi_r^{RR}$, the middle row shows $\varphi_z^{SCF+RR}-\varphi_z^{RR}$ and the bottom row shows $\varphi_\phi^{SCF+RR}-\varphi_\phi^{RR}$. In the left column, the different colors correspond to different initial $e$ values for the inspiral; in the right column, the different colors correspond to different initial values of $x_I$. For all panels, $p=7.5$ and $x_I=0.7$ initially. For the left column, the initial value of $x_{I}$ is 0.7 and for the right column the initial value of $e$ is 0.2. }
    \label{fig:xeinitplot}
\end{figure*}

\subsection{Summary of Mino-time quantities}
\label{sec:summaryMinoNIT}
We chose the average pieces of the transformation terms to be $\avg{Y^{(1)}_j} = \avg{Y^{(2)}_j}= \avg{X^{(1)}_i}= \avg{W^{(0)}_s} = \avg{W^{(1)}_s} = \avg{Z^{(0)}_k} = \avg{Z^{(1)}_k} = 0$ and so the transformed forcing functions are related to the original functions by 
	\begin{subequations}
		\begin{gather}
			\nit{F}_j^{(1)} = \left<F_{j}^{(1)}\right>\;,\quad \Upsilon_{s}^{(0)} = \left<f_{s} ^{(0)}\right>\;, \tag{\theequation a-b}
		\end{gather}
		\begin{gather}
			\Upsilon_{i}^{(1)} = \left<f_{i}^{(1)}\right>\;, \quad \Upsilon_{k}^{(0)} = \left<f_{k} ^{(0)}\right>\;, \tag{\theequation c-d}
		\end{gather}
		\begin{equation}
		    \Upsilon_s^{(1)}  = - \avg{\PD{\osc{f}^{(0)}_s}{\nit{P}_j} \osc{Y}_j^{(1)}} - \avg{\PD{\osc{f}^{(0)}_s}{\nit{q}_i} \osc{X}_i^{(1)}}\;, \tag{\theequation e}
		\end{equation}
		\begin{equation}
			\nit{F}_j^{(2)} = \left<F_j^{(2)} \right> + \left<\frac{\partial \osc{Y}_j^{(1)}}{\partial \nit{q}_i} \osc{f}_i^{(1)} \right> + \left<\frac{\partial \osc{Y}_j^{(1)}}{\partial \nit{P}_k} \osc{F}_k^{(1)} \right>\;, \tag{\theequation f}
		\end{equation}
		\begin{equation}
			\Upsilon_k^{(1)} = - \left<\frac{\partial \osc{f}_k^{(0)}}{\partial \nit{P}_j} \osc{Y}_j^{(1)} \right> - \left<\frac{\partial \osc{f}_k^{(0)}}{\partial \nit{q}_i} \osc{X}_i^{(1)} \right>\;. \tag{\theequation g}
		\end{equation}
	\end{subequations}
	In deriving these equations of motion, we have constrained the oscillating pieces of the transformation functions to be
	\begin{equation}\label{eq:NIT_Y}
		\osc{Y}_j^{(1)} \equiv \sum_{\vec{\kappa} \neq \vec{0}} \frac{i}{\vec{\kappa} \cdot \vec{\Upsilon}} F_{j,\vec{\kappa}}^{(1)} e^{i \vec{\kappa} \cdot \vec{\mathcal{Q}}},
	\end{equation}
	
	\begin{equation}\label{eq:NIT_X}
		\osc{X}_i^{(1)} \equiv \sum_{\vec{\kappa} \neq \vec{0}}\left( \frac{i}{\vec{\kappa} \cdot \vec{\Upsilon}} f_{i,\vec{\kappa}}^{(1)} + \frac{1}{(\vec{\kappa} \cdot \vec{\Upsilon})^2} \frac{\partial \Upsilon_i}{\partial P_j}F_{j,\vec{\kappa}}^{(1)} \right) e^{i \vec{\kappa} \cdot \vec{\mathcal{Q}}}\;,
	\end{equation}
	\begin{align}\label{eq:NIT_W1}
		\osc{W}^{(1)}_{s,\vec{\kappa}}  &\equiv \sum_{\vec{\kappa} \neq \vec{0}}  \frac{i}{\vec{\kappa} \cdot \vec{\Upsilon}^{(0)}} \times
			\\ &\Biggl(  \sum_{\vec{\kappa}' \neq \vec{0}} \Biggl[
	    \PD{f^{(0)}_{s,\kappa'}}{\nit{P}_j} Y_{j, \vec{\kappa} - \vec{\kappa}'}^{(1)} +  \PD{f^{(0)}_{s,\vec{\kappa}'}}{\nit{q}_i} X_{i\vec{\kappa} - \vec{\kappa}'}^{(1)}
	     \Biggr] \Biggr) e^{i \vec{\kappa} \cdot \vec{\mathcal{Q}}}\;.
	\end{align}
	In order to generate waveforms, one only needs to know the transformations in Eq.~\eqref{eq:transformation} to zeroth order in the mass ratio so that the error is $\mathcal{O}(\mr)$ i.e.,
	\begin{subequations} \label{eq:leading_order_NIT}
		\begin{align}
			\begin{split}
				P_j &= \nit{P_j} + \mathcal{O}(\varepsilon)\;,
			\end{split}\\
			\begin{split}
				q_i &= \nit{q_i} + \mathcal{O}(\varepsilon)\;,
			\end{split}\\
			\begin{split}
				\psi_s &= \nit{\psi}_s -W_s^{(0)}(\vec{\nit{P}},\vec{\nit{q}} ) +  \mathcal{O}(\varepsilon)\;.
			\end{split}\\
			\begin{split}
				\mathcal{X}_k &= \nit{\mathcal{X}}_k -Z_k^{(0)}(\vec{\nit{P}},\vec{\nit{q}} ) +  \mathcal{O}(\varepsilon)\;.
			\end{split}
		\end{align}
	\end{subequations}
	where the zeroth order transformation term for the extrinsic quantities $\osc{Z}_k^{(0)}$ is known analytically
	\begin{equation}
	    \osc{W}^{(0)}_{j,\vec{\kappa}} = - \psi_{sr}(q_r) - \psi_{sz}(q_z)\;.
	\end{equation}
	
	Moreover, the zeroth order transformation term for the extrinsic quantities $\osc{Z}_k^{(0)}$ is known analytically as it related to the analytic solutions for the geodesic equations for $t$ and $\phi$ by
	\begin{equation}\label{eq:Z_solution2}
		\osc{Z}_k^{(0)} = - \osc{\mathcal{X}}_{k,r} (q_r) - \osc{\mathcal{X}}_{k,z} (q_z)\;.
	\end{equation}

 \subsection{NIT Operator}
 \label{sec:NITOperator}
 
For compact notation in the body of the text, we define the NIT operator $\mathcal{N}$ for generic spinning orbits by the following:
\begin{widetext}
    
\begin{align} \label{eq:gen_N_Operator}
			\begin{split}
				\mathcal{N}(A) & =  \sum_{(n,k,j) \neq (0,0,0)} \frac{i}{\Upsilon_{nkj}^{(0)}} \Biggl[
				i A_{knj}\left( n f_{r,-n-k-j}^{(1)} + k f_{z,-n-k-j}^{(1)} \right) 
				+\\& \PD{A_{nkj} }{\nit{p}} F_{p,-n-k-j}^{(1)} + \PD{A_{nkj} }{\nit{e} } F_{e,-n-k-j}^{(1)} + \PD{A_{nkj} }{\nit{x} } F_{x,-n-k-j}^{(1)}
				\\&  - \frac{A_{nkj}}{\Upsilon_{knj}^{(0)}} \Biggl( \PD{\Upsilon_{nkj}^{(0)}}{\nit{p}} F_{p,-n-k-j}^{(1)} 
				+\PD{\Upsilon_{nkj}^{(0)}}{\nit{e}} F_{e,-n-k-j}^{(1)} + \PD{\Upsilon_{nkj}^{(0)}}{\nit{x}} F_{x,-n-k-j}^{(1)} \Biggr) \Biggr]\;,
			\end{split}
		\end{align}

\end{widetext}
 where $\Upsilon_{nkj}^{(0)} = n\Upsilon_r^{(0)} + k \Upsilon_z^{(0)} +  j\Upsilon_s^{(0)}$.

Note that for the problem that we are solving in this work with only the radiation reaction driven only by the GW fluxes and the conservative effects comping only from the spin-curvature force, we find that $\mathcal{N}(F_p^{(1)})$, $ \mathcal{N}(F_e^{(1)})$, and $ \mathcal{N}(F_{x_I}^{(1)})$ are numerically consistent with zero. This is to be expected as there is no interplay between the modes of the dissipative and conservative forces. We would not expect this to hold if one were to include the first order conservative GSF needed for 1PA inspiral calculations. 
	
 \section{Initial conditions}
\label{app:initialconditions}

\subsection{OG and NIT}

To be able to directly compare between OG and NIT inspirals in Mino-time, we will need to match their initial conditions to sufficient accuracy.  To maintain an overall phase difference of $\mathcal{O}(\varepsilon)$ in the course of an inspiral, the initial values of the phases and extrinsic quantities need only be known to zeroth order in $\mr$. However, we need to know the initial values of the  orbital elements $\vec{P}$ to linear order in $\mr$ and so we use
	
	\begin{align} \label{eq:NIT_ICs}
	\begin{split}
	    \nit{P}_j (0) &\simeq P_j (0) 
     \\&+\mr  Y^{(1)}_j\bigl( \vec{P}(0), \vec{q}(0),\psi_s(0)-\osc{W}^{(0)}(\vec{P}(0),\vec{q}(0))\bigr)\;.
	\end{split}
	\end{align}

When comparing between OG inspirals and NIT inspirals that are  parameterized by Boyer-Lindquist time $t$, we set the initial conditions for the phases of the OG inspiral and match the initial conditions for the $\vec{\varphi}$ phases via:
\begin{align} \label{eq:Phase_ICs}
	\begin{split}
	    \varphi_\alpha (0) =&  \nit{\mathcal{Q}}_\alpha (0) + \Delta \varphi_\alpha(\vec{\nit{P}}(0) ,\vec{\nit{q}} (0)  ) + \mathcal{O}(\mr)\;,
	\end{split}
	\end{align}
	
where $\mathcal{Q}_\alpha(0)$ are given by Eqs.~\eqref{eq:leading_order_NIT} and $\nit{P}_j (0)$ is given by Eq.~\eqref{eq:NIT_ICs}.
However, we to maintain subradian accuracy in the phases, we need to know the initial conditions for the orbital elements to sub-leading order via:
\begin{align} \label{eq:Orbital_ICs}
	\begin{split}
	    \mathcal{P}_j (0) =&  \nit{P}_j (0) + \mr \Pi^{(1)}_j\bigl( \vec{P}(0), \vec{q}(0)\bigr) + \HOT{2}\;.
	\end{split}
	\end{align}

 \subsection{Adiabatic and post-adiabatic}
 
There are different approaches to matching initial conditions when comparing adiabatic and post-adiabatic inspirals. As discussed in Refs.\ \cite{Warburton2012,Osburn2016,Lynch2021}, matching the initial values of orbital parameters $(p_\varphi,e_\varphi,x_\varphi)$ between adiabatic and post-adiabatic inspirals leads to a linearly growing error in the orbital phases. By instead matching the initial Boyer-Lindquist frequencies $\Omega_r$, $\Omega_\phi$, and  $\Omega_z$, we will instead have quadratic growth in $t$. Explicitly, we can choose initial conditions $(p^{RR+SCF}_\varphi,e^{RR+SCF}_\varphi,x^{RR+SCF}_\varphi)$ for the inspiral that includes spin-curvature force and then find the initial conditions for the radiation-reaction-only inspiral $(p^{RR}_\varphi,e^{RR}_\varphi,x^{RR}_\varphi)$ by solving the simultaneous equations: \begin{align}
 &\Omega_r^{RR+SCF}(p^{RR+SCF}_\varphi,e^{RR+SCF}_\varphi,x^{RR+SCF}_\varphi)\nonumber \\ &= \Omega_r^{RR}(p^{RR}_\varphi,e^{RR}_\varphi,x^{RR}_\varphi)\;,\\
  &\Omega_z^{RR+SCF}(p^{RR+SCF}_\varphi,e^{RR+SCF}_\varphi,x^{RR+SCF}_\varphi)\nonumber \\ &= \Omega^{RR}_z(p^{RR}_\varphi,e^{RR}_\varphi,x^{RR}_\varphi)\;,\\
&\Omega_\phi^{RR+SCF}(p^{RR+SCF}_\varphi,e^{RR+SCF}_\varphi,x^{RR+SCF}_\varphi)\nonumber \\ &= \Omega^{RR}_\phi(p^{RR}_\varphi,e^{RR}_\varphi,x^{RR}_\varphi)\;.
 \end{align}
We explicitly demonstrate the difference in the choice of initial conditions for the post-adiabatic terms considered in this work in Fig.\  \ref{fig:deltaphiIC}. In Fig.\ \ref{fig:deltaphiIC}, the solid curves show the averaged dephasing of $\phi(t)$, i.e., $\varphi_\phi^{RR+SCF}-\varphi_\phi^{RR}$. The blue curve corresponds to the initial $(p_\varphi,e_\varphi,x_\varphi)$ values matching between the radiation-reaction only and the radiation-reaction plus spin-curvature inspirals. The orange curve corresponds to the initial $(\Omega_r,\Omega_z,\Omega_\phi)$ values matching between the radiation-reaction only and the radiation-reaction plus spin-curvature inspirals. The blue curve grows linearly with $t$ while the orange one grows quadratically with $t$. This can be seen clearly in the inset of Fig.\ \ref{fig:deltaphiIC}; on a log-log scale, the slopes of the orange line is twice that of the blue line. Note that, in the results presented in this article, we match initial orbital parameters $(p_\varphi,e_\varphi,x_\varphi)$ between adiabatic and post-adiabatic inspirals. 

 \begin{figure}[!]
\centerline{\includegraphics[scale=0.52]{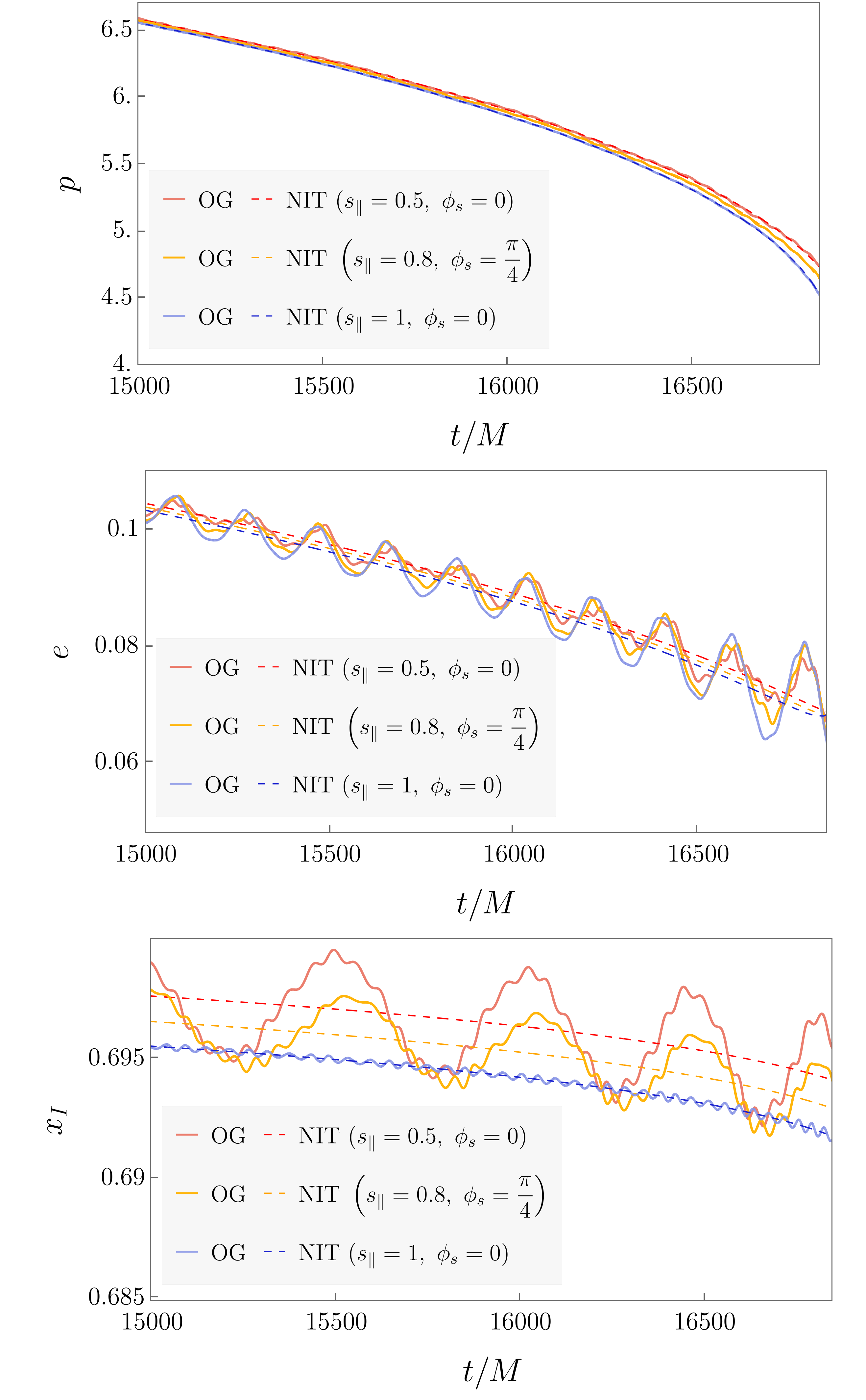}}
\caption{Orbital evolution of a small body with a misaligned spin vector. The top panel shows $p(t)$ (solid) and $p_\varphi(t)$ (dashed); middle shows $e(t)$ and $e_\varphi(t)$; and the bottom panel shows $x_I(t)$ and $x_\varphi(t)$.  In all the panels, we plot orbital element evolution for three values of spin alignment $s_\parallel=1$ (blue), $s_\parallel=0.8$ (orange) and $s_\parallel=0.5$ (red) using the NIT equations of motion (dashed) and the OG equations of motion (solid).  Especially in the middle and bottom panels, the different OG tracks can also be distinguished by the magnitude of the oscillations, which scale with $s_\perp$ and are thus smallest for $s_\parallel = 1$ and largest for $s_\parallel = 0.5$.  (The NIT tracks pass roughly the centers of the OG oscillations.)  The magnitude of the small body's spin is $s = 1$; $\phi_s$ is zero except for the orange curve which has $\phi_s=\pi/4$. The small body has mass ratio $\varepsilon=10^{-2}$ and is orbiting a black hole with spin $a=0.7 M$. For all panels, $p=10$, $e=0.2$, $x_I=0.7$, $q_r=0$, $q_z=0$, and $\phi=0$ initially.}
\label{fig:precessplot2}
\end{figure}

\subsection{Varying initial conditions}
\label{sec:varyinitialconditions}

Figure \ref{fig:pinitplot} depicts the dephasing of the radial, polar and axial phases due to spin-curvature force during an inspiral. As in Fig.\ \ref{fig:OGNITphases}, the top, middle and bottom panels display $\varphi_\alpha^{SCF+RR}-\varphi_\alpha^{RR}$ with $\alpha\in \{r,z,\phi\}$ respectively. Different-colored curves correspond to different initial $p$ values: Red corresponds to a larger initial $p$ value while blue corresponds to a smaller initial $p$ value that is closer to the LSO. Because the monotonic evolution of the dephasing of the polar and axial phases (middle and bottom panels), the curves that begin closer to the LSO do not accumulate as much dephasing before the plunge. However, for the case of the radial phase the initial value of $p$ will affect where the maximum of the dephasing will occur, because the evolution is not monotonic.  

Figure \ref{fig:xeinitplot} depicts the dephasing of the radial, polar and axial phases due to spin-curvature force with different curves on the same plot corresponding to different initial $e$ (left column) and $x_I$ (right column) values. The red curves correspond to a larger initial $e$ or $x_I$ value, yellow is an intermediate value and blue is the smallest value. As in Fig.\ \ref{fig:pinitplot}, the top, middle and bottom panels display $\varphi_\alpha^{SCF+RR}-\varphi_\alpha^{RR}$ with $\alpha\in \{r,z,\phi\}$ respectively. The initial $e_0$ values we plot are evenly spaced by $\Delta e=0.05$ and initial $x_I$ values are evenly spaced by $\Delta x=0.002$. Consider the insets of the two plots in the middle row; these curves show the evolution of $\varphi_\alpha^{SCF+RR}-\varphi_\alpha^{RR}$. Observe that even separation in $e$ does not correspond to even separation in $\varphi_\alpha^{SCF+RR}-\varphi_\alpha^{RR}$-space (middle left) while even separation in $x_I$ does correspond to roughly even separation in $\varphi_\alpha^{SCF+RR}-\varphi_\alpha^{RR}$-space (middle right).

In Fig.\ \ref{fig:precessplot2}, the solid lines show the the evolution of the orbital elements $(p,e,x_I)$ under the OG equations of motion, while the dashed lines show the averaged evolution of the orbital elements $(p_\varphi,e_\varphi,x_\varphi)$ under the NIT equations of motion.  The oscillations depicted by the solid curves exhibit harmonics of several frequencies: The spin-aligned case ($s_\parallel=1$, blue curve) contains harmonic of $\Omega_r$ and $\Omega_z$ while the spin-misaligned cases ($s_\parallel\neq1$, orange and red curves) contain harmonics of three  frequencies  $\Omega_r$,   $\Omega_z$and $\Omega_s$. The additional harmonic structure introduced by spin-precession is most evident in the evolution of $e$ shown in middle right panel. 

The effect of the perpendicular spin component $s_\perp$ is most evident in the evolution of $x_I$ in the bottom right panel. We can clearly see that the amplitude of the oscillations in $x_I$ increase with increasing $s_\perp$, i.e., decreasing $s_\parallel$. In addition, when there is a non-zero initial spin-precession phase ($\phi_s\neq0$, orange curve), we can see that the oscillations in $x_I$ are out-of-phase with the $\phi_s\neq0$ (red) curve. Because the initial conditions we use for the NIT equations of motion are determined by the oscillations present in the OG equations (as described in Appendix \ref{app:initialconditions}), the NIT (dotted) curve for the misaligned spin cases (red and orange curves) have slightly different initial conditions and evolution compared to the aligned spin curve (blue).

\normalem
\bibliographystyle{unsrt}

\bibliography{SpinningBodyInspiral}
\end{document}